\documentclass[11pt]{article}

\usepackage{jheppub} % for details on the use of the package, please
\usepackage[T1]{fontenc} % if needed

\usepackage{graphicx}
\usepackage{mathrsfs}                                                                                                                                                       
\usepackage{latexsym}
\usepackage{amsmath}
\usepackage{amssymb}
\usepackage{color}
\usepackage{subfig}
\usepackage{array}
\usepackage{multirow}
\usepackage{hhline}
\usepackage[table]{xcolor}
\usepackage{float}
\usepackage{subfig}
\usepackage{booktabs}

\DeclareCaptionLabelFormat{andtable}{#1~#2  \&  \tablename~\thetable}

\usepackage{xcolor}
\definecolor{plum}{rgb}{0.36078, 0.20784, 0.4}
\definecolor{chameleon}{rgb}{0.30588, 0.60392, 0.023529}
\definecolor{cornflower}{rgb}{0.12549, 0.29020, 0.52941}
\definecolor{scarlet}{rgb}{0.8, 0, 0}
\definecolor{brick}{rgb}{0.64314, 0, 0}

\newcommand{\ba}{\begin{eqnarray}}
\newcommand{\ea}{\end{eqnarray}}
\newcommand{\be}{\begin{equation}}
\newcommand{\ee}{\end{equation}}
\newcommand{\bd}{\begin{displaymath}}
\newcommand{\ed}{\end{displaymath}}

\newcommand{\tcr}{\textcolor{red}}
\newcommand{\rmd}{{\rm d}}

\begin{document} 

\begin{flushright}
CTPU-PTC-24-08
\end{flushright}

\title{\boldmath Scalar Field Perturbation of Hairy Black Holes in EsGB theory}

\author[a]{Young-Hwan Hyun,}
\author[b]{Boris Latosh,}
\author[b,1]{Miok Park\note{Corresponding author.}}

\affiliation[a]{Korea Astronomy and Space Science Institute, 776 Daedeok-daero, Yuseong-gu, Daejeon 34055, Republic of Korea,}
\affiliation[b]{Particle Theory  and Cosmology Group, Center for Theoretical Physics of the Universe, Institute for Basic Science (IBS), Daejeon, 34126, Republic of Korea}

\emailAdd{younghwan.hyun@gmail.com}
\emailAdd{latosh.boris@ibs.re.kr}
\emailAdd{miokpark76@ibs.re.kr}

%\date{\today}

%{\let\newpage\relax{\maketitle}}

\abstract{We investigate scalar field perturbations of the hairy black holes involved with spontaneous symmetry breaking of the global U(1) symmetry in Einstein-scalar-Gauss-Bonnet theory for asymptotically flat spacetimes. We consider the mechanism that black holes without hairs become unstable at the critical point of the coupling constant and undergo a phase transition to hairy black holes in the symmetry-broken phase driven by spontaneous symmetry breaking. This transition occurs near the black hole horizon due to the diminishing influence of the Gauss-Bonnet term at infinity. To examine such process, we introduce a scalar field perturbation on the newly formed background spacetime. We solve the linearized perturbation equation using Green's function method. We begin by solving the Green's function, incorporating the branch cut contribution. This allows us to analytically investigate the late-time behavior of the perturbation at both spatial and null infinity. We found that the late-time behavior only differs from the Schwarzschild black hole by a mass term. We then proceed to calculate the quasinormal modes (QNMs) numerically, which arise from the presence of poles in the Green's function. Our primary interest lies in utilizing QNMs to investigate the stability of the black hole solutions both the symmetric and symmetry-broken phases. Consistent with the prior study, our analysis shows that hairy black holes in the symmetric phase become unstable when the quadratic coupling constant exceeds a critical value for a fixed value of the quartic coupling constant. In contrast, hairy black holes in the symmetry-broken phase are always stable at the critical value. These numerical results provide strong evidence for a dynamical process that unstable black holes without hairs transition into stable hairy black holes in the symmetry-broken phase through the spontaneous symmetry breaking.}

\maketitle

%%%%%%%%%%%%%%%%%%%%%%%%%%%%%%%%%%%%%%%%%%%%%%%%
\section{Introduction}
%%%%%%%%%%%%%%%%%%%%%%%%%%%%%%%%%%%%%%%%%%%%%%%%

Detecting gravitational waves from the merger of binary black holes first observed by the Laser Interferometer Gravitational-Wave Observatory (LIGO)\cite{LIGOScientific:2016aoc} in 2015 was a significant physics breakthrough in recent decades. General relativity predicted the existence of gravitational waves, but their high frequency and weak nature made it challenging to confirm their existence experimentally, which took nearly a century. On the other hand, electromagnetic waves were observed just twenty years after their prediction from electromagnetic theory and have significantly impacted our lives. One of the primal objectives of LIGO is to test general relativity since it fails to account for dark matter, dark energy, and inflationary expansion \cite{Clowe:2006eq,Linde:1983gd,Starobinsky:1980te}.

Numerous alternative theories of gravity have been proposed to enhance general relativity \cite{Lovelock:1971yv,Kobayashi:2011nu,Copeland:2012qf,Babichev:2022djd,Dorlis:2023qug}. New terms or non-minimal coupling are introduced to modify the well-known Einstein-Hilbert Lagrangian. Models containing the Gauss-Bonnet term are of particular interest. First, the Gauss-Bonnet term is a total derivative in four-dimensional spacetime \cite{tHooft:1974toh}. Second, although the gravity action contains higher curvature terms, it does not introduce higher derivative terms into the equation of motion, even when it is non-minimally coupled to a scalar field \cite{Lovelock:1971yv,Kobayashi:2011nu,Copeland:2012qf}. Finally, the presence of the Gauss-Bonnet term is predicted from various string theory considerations \cite{Zwiebach:1985uq,Gross:1986mw}.

This paper focuses on the Einstein-scalar-Gauss-Bonnet theory, where a scalar field couples to the Gauss-Bonnet term. In this theory, the evasion of the no-hair theorem was first studied in \cite{Antoniou:2017acq} (generalising results obtained in \cite{Kanti:1995vq}) and soon after in \cite{Lee:2018zym} based on Bekenstein's argument \cite{Bekenstein:1972ny,Bekenstein:1995un}. Paper \cite{Antoniou:2017acq} claimed that a positive coupling function is necessary for the evasion, but \cite{Lee:2018zym} showed that evasion is possible for both signs of the coupling function. These inconsistent results caused confusion and led to the conclusion that there is a privileged way to validate the evasion of the no-hair theorem. This would imply that the theorem loses its universal nature. The complete derivation for the evasion of the no-hair theorem was done in \cite{Papageorgiou:2022umj}. It found that a surface term was omitted in the previous studies \cite{Antoniou:2017acq,Lee:2018zym}, resulting in inconsistent and incorrect results. The no-hair theorem is extended to spherical black holes in shift-symmetric scalar-tensor theory in \cite{Capuano:2023yyh}.

Coinciding with the discovery of hairy black holes in ESGB theory \cite{Antoniou:2017acq}, the mechanism of spontaneous scalarization was proposed to explain how black holes grow their scalar hair \cite{PhysRevLett.120.131104,Doneva:2017bvd,Doneva:2022ewd}, which claims the tachyonic instability is responsible for this transition. However, it is found that the newly generated hairy black holes are dynamically unstable under the radial perturbation of scalar fields \cite{Blazquez-Salcedo:2018jnn}. This implies that this hairy black hole may not be the final stage in the evolution of a black hole. Alternative approaches were proposed later to guarantee the stability of the hairy black holes \cite{Minamitsuji:2018xde, Macedo:2019sem, Minamitsuji:2023uyb,Antoniou:2022agj,Kleihaus:2023zzs}.

One promising approach in this regard utilizes spontaneous symmetry breaking (SSB) \cite{Latosh:2023cxm}. The model contains a single complex scalar field with a global U(1) symmetry, which couples to the Gauss-Bonnet term: 
\begin{align}
  V = - f(\varphi)\, \mathcal{G}, \qquad \mathcal{G} = R_{\mu \nu \rho \sigma} R^{\mu \nu \rho \sigma} - 4 R_{\mu \nu} R^{\mu \nu} +R^2.
\end{align}
Here, $\mathcal{G}$ is the Gauss-Bonnet term, $f$ is the coupling function, and $V$ is commonly called the interaction potential. We only discuss the coupling function of the following form
\begin{align}
  f(\varphi) = \alpha \, \varphi^{*}(r) \varphi(r) - \lambda \, \big( \varphi^{*}(r) \varphi(r) \big)^2 .
\end{align}
The interaction potential resembles the conventional scalar field potential in curved spacetime, with the Gauss-Bonnet term acting as a position-dependent magnitude. In this model, the Schwarzschild black hole becomes unstable under scalar field perturbation when the coupling constant reaches a critical value denoted as $\alpha_{\textrm{Sch.}}$. It was argued that this instability may lead to the formation of stable hairy black holes in symmetry-broken phases. Here the spontaneous symmetry breaking is involved in forming or evolving stable hairy black hole spacetimes from their counterparts without hairs.  The Gauss-Bonnet term in the interaction potential peaks near the horizon and decreases rapidly as distance increases. Since the magnitude of the Gauss-Bonnet term is non-vanishing near the horizon, it is responsible for producing a non-vanishing expectation value for the scalar field that spontaneously breaks the $U(1)$ symmetry. On the contrary, the contribution of the non-minimal coupling is negligible at infinity and so this process is irrelevant there. In the symmetry-broken phase, the equation of motion for Goldstone bosons is decoupled from other equations and due to the flux conservation only trivial solutions are physically acceptable. 

This application of spontaneous symmetry breaking to black hole systems is particularly intriguing, because it indicates that this mechanism, considered fundamental in our understanding of nature, might play a role in the formation and stability of hairy black holes. In the Standard Model, elementary particles gain mass through the Higgs mechanism and in condensed matter physics, spontaneous symmetry breaking is responsible for superconductivity. In cosmology, dark matter and dark energy may be composed of Goldstone bosons, created when symmetry is broken \cite{Amendola:2005ad,Frigerio:2011in,Alanne:2014kea}. 

While the previous work in \cite{Latosh:2023cxm} identified a sufficient condition for (in)stability by analyzing the effective potential of the perturbed scalar field, this approach is not enough to prove the dynamical evolution of hairy black holes from their counterparts without hairs. To gain a more rigorous understanding of the stability for hairy black hole solutions, a detailed examination of their dynamical behavior is necessary. Calculating the quasinormal modes(QNMs) provides a powerful tool for such an investigation. 

We aim to comprehensively analyze the scalar field perturbation on the fixed background of the hairy black hole solutions in the symmetric and symmetry-broken phases. We especially focus on calculating the QNMs to examine the instability of the hairy black holes. First, we solve the perturbed equation with Green's function. The function consists of three parts. $G_{\textrm{QNM}}$ describes the contribution of simple poles to scalar perturbations. $G_{\textrm{B}}$ describes the influence of the branch cut. $G_{\textrm{F}}$ contains all the other contributions coming from the part of the integration contour, which is two-quarter circles at infinity. We analytically obtained the solutions for $G_{\textrm{B}}$ in the low-frequency limit at spatial infinity ($r \rightarrow \infty$) and null infinity ($v \rightarrow \infty$) and demonstrated their solutions. As our background solutions are numerically generated, we utilized the numerical methods such as Chebysheve polynomials to compute the QNMs to calculate $G_{\textrm{QNM}}$. We proved that the hairy black holes in the symmetric phase become unstable at a certain value of $\alpha$ (denoted as $\alpha_{\textrm{crit.}}$), yielding the positive value for the imaginary part of the frequency, while the hairy black holes in the symmetry-broken phase are stable for all valid values of $\alpha$, yielding the negative value for the imaginary part of the frequency. As discussed in \cite{Latosh:2023cxm}, we also found $\alpha_{\textrm{Sch.}} \approx \alpha_{\textrm{crit.}}$. These numerical results suggest that black holes without hairs undergo a transition to hairy black holes in the symmetry-broken phase.

This paper is organized as follows. We introduce the gravity model and hairy black hole solutions in Section 2. Section 3 delves into the Green's function for scalar field perturbations on the hairy black hole background. We discuss the full solution structure of the scalar field perturbation, demonstrate some properties of QNMs, and show the late time behavior by calculating the branch cut in Green's function. Next, we discuss the numerical method employed to calculate QNMs in Section 4. Sections 5 and 6 focus on calculating the QNMs for hairy black holes. Section 5 demonstrates the hairy black holes in the symmetric phase and examines their QNMs (Section 5.3). Section 6 follows the same arguments for the hairy black holes in the symmetry-broken phase. Finally, Section 7 presents a summary of the key findings.

%%%%%%%%%%%%%%%%%%%%%%%%%%%%%%%%%%%%%%%%%%%%%%%%
\section{Hairy Black Holes by Spontaneous Symmetry Breaking}
%%%%%%%%%%%%%%%%%%%%%%%%%%%%%%%%%%%%%%%%%%%%%%%%

We consider the Einstein-scalar-Gauss-Bonnet theory in four-dimensional asymptotically flat spacetime as follows:
\begin{align}
  S =& \int \textrm{d}^4 x \sqrt{-g} \bigg[ \frac{1}{2 \kappa^2}R - \nabla_{\alpha} \varphi^{*} \nabla^{\alpha} \varphi + f(\varphi^*, \varphi) \mathcal{G}  \bigg], \label{eq:action1}\\
  \mathcal{L}_{\varphi} =& - \nabla_{\alpha} \varphi^{*} \nabla^{\alpha} \varphi + f(\varphi^*, \varphi) \mathcal{G} = T - V, \qquad V = - f(\varphi^*, \varphi) \mathcal{G}  \label{eq:action2}
\end{align}
where $\kappa^2 = 8 \pi G$ the following scalar field coupling function is employed
\begin{align*}
  f(\varphi^*, \varphi) = \alpha \, \varphi^{*}(r) \varphi(r) - \lambda \, \big(\varphi^{*}(r) \varphi(r) \big)^2
\end{align*}
where $\varphi(r)$ is a complex field and $\alpha$ and $\lambda$ are coupling constants. We assume that $\alpha$ can have real values, but $\lambda$ only takes positive ones. This Lagrangian respects the global $U(1)$ symmetry
\begin{align}
  \varphi(r) \rightarrow e^{i \chi} \varphi(r) ,
\end{align}
where $\chi$ is a constant.

We are specially interested in forming stable hairy black holes from black holes without hairs, suggested in \cite{Latosh:2023cxm} where the transition mechanism is realized by the spontaneous symmetry breaking. This will happen if the potential of the scalar field is invariant under a certain symmetry and possesses local minima. We treat the scalar field coupling with the Gauss-Bonnet term as an interacting potential $V= - f(\varphi^*, \varphi) \mathcal{G}$. Depending on the values of the parameters in $f$, the interacting potential can have either a single-well or double-well shape at every point in the spacetime. This allows us to investigate hairy black holes in two distinct phases: symmetric and symmetry-broken phases. These phases are determined by the sign of $\alpha$ or the vacuum value of the interacting potential. The \textit{symmetric phase} occurs when the scalar field near the horizon resides at either the ``global" minimum (when $\alpha< 0$) or the ``local" maximum (when $\alpha > 0$) of the interacting potential. In contrast, the \textit{symmetry-broken phase} is characterized by the scalar field near the horizon settling at the ``global" minimum ($\alpha > 0$) of the interacting potential. In later sections, we will construct hairy black hole solutions for both the symmetric and symmetry-broken phases.

We use the spherically symmetric and diagonal form of the metric ansatz
\begin{align}
  \rmd s^2 = - A(r) \rmd t^2 + \frac{1}{B(r)} \rmd r^2 + r^2 (\rmd \theta^2 + \sin^2 \theta \rmd \phi^2). \label{eq:metric}
\end{align} 
The equations of motion take the following form
\begin{align}
  \frac{1}{2\, \kappa^2} \bigg(R_{\mu \nu} - \frac{1}{2}\, R \, g_{\mu \nu} \bigg) =&  - \frac{1}{2} (\nabla_{\alpha} \varphi^* \nabla^{\alpha} \varphi)g_{\mu \nu} + \frac{1}{2}(\nabla_{\mu} \varphi^* \nabla_{\nu} \varphi + \nabla_{\mu} \varphi \nabla_{\nu} \varphi^*) \nonumber\\
  & - \frac{1}{2} (g_{\rho \mu} g_{\lambda \nu} + g_{\lambda \mu} g_{\rho \nu}) \, \eta^{\kappa \lambda \alpha \beta} \tilde{R}^{\rho \gamma}{}_{\alpha \beta} \nabla_{\gamma} \nabla_{\kappa} f(\varphi^*, \varphi), \label{eq:EE}\\
  \nabla^{\alpha} \nabla_{\alpha} \varphi + \frac{\partial f(\varphi^*, \varphi)}{\partial \varphi^*} \, \mathcal{G} =& 0, \qquad \nabla^{\alpha} \nabla_{\alpha} \varphi^{*} + \frac{\partial f(\varphi^*, \varphi)}{\partial \varphi} \, \mathcal{G} = 0 . \label{eq:KG}
\end{align}
Here $\tilde{R}^{\rho \gamma}{}_{\alpha \beta} = \eta^{\rho \gamma \sigma \tau} R_{\sigma \tau \alpha \beta} = \frac{\epsilon^{\rho \gamma \sigma \tau}}{\sqrt{-g}} R_{\sigma \tau \alpha \beta}$ and it is expanded as
\begin{align}
\frac{1}{2} (g_{\rho \mu} g_{\lambda \nu} + g_{\lambda \mu} g_{\rho \nu}) \, \eta^{\kappa \lambda \alpha \beta} \tilde{R}^{\rho \gamma}{}_{\alpha \beta} \nabla_{\gamma} \nabla_{\kappa} f(\varphi^*, \varphi) = 2 R \nabla_{\mu} \nabla_{\nu} f(\varphi^*, \varphi) - 4 R^{\rho}{}_{\nu} \nabla_{\rho} \nabla_{\mu} f(\varphi^*, \varphi) \nonumber\\
+ 4 R_{\mu \nu} \nabla^2 f(\varphi^*, \varphi) - 4 R_{\mu}{}^{\lambda} \nabla_{\nu} \nabla_{\lambda} f(\varphi^*, \varphi) + 4 R_{\sigma \nu \mu \lambda} \nabla^{\sigma} \nabla^{\lambda} f(\varphi^*, \varphi). 
\end{align}
A complex scalar field can be decomposed into two real scalar fields as follows :
\begin{align}
  \varphi(r) = \frac{1}{\sqrt{2}} \big(\varphi_1(r) + i \, \varphi_2(r) \big). \label{eq:decmpCS}
\end{align}
Thus, we replace the complex field with the real scalars. 

We only consider the case of asymptotically flat spacetime. Consequently, near infinity, the metric behaves as
\begin{align}
  &A(r) \sim 1+\frac{A_1}{r} -\frac{A_1 \left((\varphi_{1,1})^2+ (\varphi_{2, 1})^2\right)}{24\, r^3} + \cdots  \label{eq:AexpInf} \\
  &B(r) \sim 1+\frac{A_1}{r}+\frac{(\varphi_{1,1})^2+ (\varphi_{2,1})^2}{4 \, r^2}-\frac{A_1 \left((\varphi_{1,1})^2+ (\varphi_{2,1})^2\right)}{8 \, r^3} + \cdots \label{eq:BexpInf} \\
  &\varphi_i(r) \sim \varphi_{i\infty} + \frac{\varphi_{i,1}}{r} -\frac{A_1 \varphi_{i,1}}{2\, r^2}  -\frac{\left( (\varphi_{1,1})^2+ (\varphi_{2,1})^2-8 A_1^2 \right) \,\varphi_{i,1}}{24\, r^3}+ \cdots  \label{eq:vpexpInf}
\end{align}
where $i,j = 1,2$. The coefficient $A_1 = - 2 M$ is identified with the black hole mass $M$. The scalar charge is defined as follows
\begin{align}
D_i = -\frac{1}{4 \pi} \int_{S^2} \textrm{d}^2 \Sigma^{\mu} \nabla_{\mu} \varphi_i .
\end{align}
The definition immediately results in $\varphi_{i,1} = D_i$. For simplicity, we consider $\varphi_1 = \varphi_2$ hereafter. 

%%%%%%%%%%%%%%%%%%%%%%%%%%%%%%%%%%%%%%%%%%%%%%%%
\section{Green's function for Scalar Field Perturbation}
%%%%%%%%%%%%%%%%%%%%%%%%%%%%%%%%%%%%%%%%%%%%%%%%

The hairy black hole solutions satisfying the equations of motion (\ref{eq:EE}) - (\ref{eq:KG}) are hard to obtain analytically. However, they can be established with numerical methods. On these hairy black hole background, presented in Section 5 and 6, we consider scalar field perturbations and understand their solutions using Green's function.

\subsection{Scalar Field Perturbations}

Let us consider a perturbation of the scalar field described by (\ref{eq:KG})
\begin{align} 
\varphi(r) \rightarrow \varphi(r) + \delta \varphi(r), \qquad \varphi^*(r) \rightarrow \varphi^*(r) + \delta \varphi^*(r) .
\end{align}
The equation takes the following form
\begin{align}
  \bigg( \nabla_{\alpha} \nabla^{\alpha}  + f_{\varphi^* \varphi}\, \mathcal{G} \bigg) \delta \varphi (r)+ f_{\varphi^* \varphi^*} \, \mathcal{G} \, \delta \varphi^*(r)  = 0
\end{align}
where the subscript of $f$ indicates the variation with respect to those variables. We decompose the complex scalar with (\ref{eq:decmpCS}) and employ the tortoise coordinate
\begin{align}
\rmd x = \frac{1}{\sqrt{A(r)B(r)}} \rmd r . \label{eq:xtor}
\end{align}
The perturbation equation becomes 
\begin{align}
&\bigg[\partial_{x}^2 - \partial_{t}^2 - V_{\textrm{eff}}(r) \bigg]  \delta \varphi_1(t,x) = 0,  \label{eq:perturbRS} \\
&V_{\textrm{eff}}(r) \!=\! A(r) \bigg[ \frac{l(l+1)}{r^2} \!+\! \frac{B(r)}{2\,r} \bigg(\frac{A'(r)}{A(r)} + \frac{B'(r)}{B(r)} \bigg)\! - \frac{1}{2}f_{\varphi_1 \varphi_1}  \, \mathcal{G}  \bigg]. \label{eq:Veff}
\end{align}

To solve the equation for $\delta \varphi_1(t,x)$, we impose boundary conditions reflecting the classical nature of black holes: an incoming condition at the horizon ($x \rightarrow -\infty$) and an outgoing condition at infinity ($x \rightarrow \infty$). These conditions imply that a black hole in asymptotically flat spacetime acts as an open system, and energy carried by $\delta \varphi_1(t,x)$ dissipates. Consequently, the system is non-Hermitian, meaning its energy eigenvalues are complex rather than real. In Hermitian systems, energy is conserved and energy eigenvalues are real, leading to normal modes forming a complete set. In contrast, an open system has complex energy eigenvalues, and the corresponding solutions yield QNMs. They do not form a complete set and experience exponential decay or divergence over time depending on the imaginary part of the frequency.

To search for solutions of (\ref{eq:perturbRS}) one can use the Laplace transformation on $\delta \hat{\varphi}_1 (s,x)$ for real $s$ and positive $t$. 
\begin{align}
 \delta \hat{\varphi}_1 (s,x) = \int^{\infty}_{0} \rmd t ~e^{-s t} \delta \varphi_1(t,x).
\end{align}
If $\delta \varphi_1(t,x)$ is bounded, one can analytically continue $\delta \hat{\varphi}_1 (s,x) $ into the complex half-plane Re$(s)>0$. After the transformation, the perturbation equation takes the following form
\begin{align}
&\bigg[\partial_{x}^2 - s^2 - V_{\textrm{eff}}(r) \bigg]  \delta \hat{\varphi}_1(s, x) = \mathcal{J}(s, x),  \label{eq:LPperturbRS}\\
&\mathcal{J}(s, x) = - \partial_t \delta \varphi(t, x)|_{t=0} - s \delta \varphi(t, x)|_{t=0} . \label{eq:GreensJ}
\end{align}
The source term $\mathcal{J}(s, x)$ is determined by the initial conditions at $t=0$. The equation is an inhomogeneous ordinary second-order differential equation that can be solved using the Green's function. By definition, the Green's function is a solution of the following equation:
\begin{align}
\bigg[\partial_{x}^2 - s^2 - V_{\textrm{eff}}(r) \bigg]  \hat{G}(s, x, x') = \delta(x - x')
\end{align}
where $x'$ and $x$ are the distances to the source and the observer from the horizon, respectively. Once the Green's function is obtained and the initial conditions are given, one obtains the following solution of the perturbation equation
\begin{align}
\delta \hat{\varphi}_1 (s, x) = \int^{\infty}_{- \infty} \rmd x' ~\hat{G}(s, x, x')  \mathcal{J}(s, x'). \label{eq:Greenssol}
\end{align}
Here the Green's function is constructed by the solutions of the homogenous equations $\hat{\varphi}_{(\pm)}$ with $\mathcal{J}(s, x) = 0$ from (\ref{eq:LPperturbRS})  as follows 
\begin{align}
\bigg[\partial_{x}^2 - s^2 - V_{\textrm{eff}}(r) \bigg]  \hat{\varphi}_{(\pm)}(s, x) = 0  \label{eq:PertbHmEq} .
\end{align}
Then the Green's function takes the form
\begin{align}
\hat{G}(s, x, x') &= \frac{1}{\mathcal{W}_x [\varphi_{-}, \varphi_{+} ]} \bigg[ \theta(x-x') \hat{\varphi}_-(s,x') \hat{\varphi}_+(s,x) + \theta(x'-x) \hat{\varphi}_-(s,x) \hat{\varphi}_+(s,x')  \bigg], \label{eq:GreensTheta} \\
\mathcal{W}_x[\varphi_{-}, \varphi_{+} ] &\equiv \hat{\varphi}_-(s,x) \partial_x \hat{\varphi}_+(s,x) - \hat{\varphi}_+(s,x) \partial_x \hat{\varphi}_-(s,x) \label{eq:Wx}.
\end{align}
where  $\mathcal{W}_x(s)$ indicates the Wronskian computed in the $x$-coordinate.

To investigate the general properties of $\delta \hat{\varphi}_1(s, x)$, we temporarily neglect the contribution of the initial condition to the source term $\mathcal{J}(s, x')$ and instead, focus on the solution of the Green's function. We obtain the Green's function in $t$-plane by performing an inverse Laplace transformation 
\begin{align}
G(t, x, x') = \int^{\epsilon + i \infty}_{\epsilon - i \infty} \frac{\rmd s}{2 \pi i} e^{s t} ~\hat{G}(s,x,x').
\end{align}
To calculate this integral, one shall construct the closed contour integral in the complex $s$-plane. Since $t \geq 0$, the integral diverges for $\operatorname{Re}(s)>0$, which is the right half of the $s$-plane. Thus, we make a closed contour integral for negative Re$(s)$, which is the left half of the $s$-plane. One should also consider a branch cut for real negative $s$, so the integration contour should not cross the branch cut. Consequently, the Green's function consists of three parts:
\begin{align}
G(t, x, x') =G_{\textrm{QNM}}(t, x, x') + G_{\textrm{B}}(t, x, x') + G_{\textrm{F}}(t, x, x') ,
\end{align}
where
\begin{align}
&G_{\textrm{QNM}}(t, x, x') = \lim_{R \rightarrow \infty} \oint_{\gamma_{R}} \frac{\rmd s}{2 \pi i} e^{s t} \hat{G}(s, x, x'), \label{eq:GreenQNM}\\
&G_{\textrm{B}}(t, x, x') = \int^{0}_{-\infty} \frac{\rmd (s e^{2 \pi i})}{2 \pi i} e^{se^{2 \pi i} t} \hat{G}(se^{2 \pi i}, x, x') + \int^{-\infty}_{0} \frac{\rmd s}{2 \pi i} e^{s t} \hat{G}(s, x, x'), \label{eq:GreenB}\\
&G_{\textrm{F}}(t, x, x') = \lim_{R \rightarrow \infty} \bigg[ \int^{\pi}_{\phi = \frac{\pi}{2}} \frac{\rmd (s=R e^{i \phi})}{2 \pi i } e^{s t} \hat{G}(s, x, x') + \int^{-\frac{\pi}{2}}_{\phi = -\pi} \frac{\rmd (s=R e^{i \phi})}{2 \pi i } e^{s t} \hat{G}(s, x, x') \bigg]. 
\end{align}
The first term $G_{\text{QNM}}(t, x, x')$ describes the contribution of poles within the contour $\gamma_{R}$, the second term $G_{\text{B}}(t, x, x')$ arise from the branch cut, and the last term $G_{\text{F}}(t, x, x')$ is originated from the two quarter-circles at infinity.

When a scalar perturbation is introduced to the background spacetime, it generates an initial wave that propagates both inward toward the black hole and outward toward a distant observer. If the initial perturbation carries significantly higher energy compared to the black hole's effective potential, denoted by $V_{\text{eff}}(r)$ in (\ref{eq:Veff}), the outward propagating wave can effectively ignore the influence of this potential. Consequently, this high-energy component traveling outward reaches the observer at a time approximately equal to $t \simeq x - x'$. This solution is described by the term $G_{\text{F}}(t, x, x')$. However, while this high-frequency signal might offer insights into the early response of the black hole to the perturbation, calculating the exact behavior of the wave interacting with the potential is highly complex and remains an open problem.

At $t \simeq x + x'$, the part of the low-frequency scalar perturbation traveling towards the black hole bounces off the effective potential $V_{\text{eff}}(r)$ near the horizon and propagates back to the observer. This reflected signal is described by the term $G_{\text{QNM}}(t, x, x')$, and their basic properties are discussed in Section 3.2. There are several methods known to compute the QNMs \cite{1985ApJ291L33S,PhysRevD.35.3621,PhysRevD.35.3632,Chandrasekhar:1975zza,Leaver:1985ax,Jansen:2017oag}. Because only numerical background solutions are known, we use numerical methods to compute their QNMs. We discuss this in detail in Section 4. We present the numerical results of these QNMs in Sections 5 and 6, enabling a rigorous investigation of stability for our black hole solutions.

At late time $ t \gg x + x'$, the QNMs gradually fade away, and only the decaying modes of the signal persist. This late-time behavior arises from the outgoing wave that traveled to a large distance and then scattered back towards the observer due to the spacetime curvature at large distances. Consequently, the behavior at these late times depends on the asymptotic form of the effective potential in (\ref{eq:Veff}) at infinity. This signal can be seen in $G_{\textrm{B}}(t, x, x')$, and we analytically calculate this quantity for our hairy black hole solutions in section 3.3. \\

\subsection{Some Properties of Quasinormal Modes}

In this subsection, we closely investigate the physical properties of the QNMs in (\ref{eq:GreenQNM})
\begin{align*}
G_{\textrm{QNM}}(t, x, x') = \lim_{R \rightarrow \infty} \oint_{\gamma_{R}} \frac{\rmd s}{2 \pi i} e^{s t} \hat{G}(s, x, x')
\end{align*}
which is the contour integral that is only contributed by the poles.

We assume that the observer is far away from the source ($x > x'$), and so our Green's function in (\ref{eq:GreensTheta}) is written as
\begin{align}
G(t, x, x') = \lim_{R \rightarrow \infty} \oint_{\gamma_{R}} \frac{\rmd s}{2 \pi i} e^{s t} \frac{1}{\mathcal{W}_x [\varphi_{-}, \varphi_{+} ]} \bigg[ \hat{\varphi}_-(s,x') \hat{\varphi}_+(s,x)  \bigg].
\end{align}
Since $\hat{\varphi}_{+}$ and $\hat{\varphi}_{-}$ are analytic within the contour $\gamma_R$, poles can only appear because of the Wronskian zeros. In other words, the function has poles in points where $\mathcal{W}_x = 0$. 

We suppose that $\mathcal{W}_x$ only has a set of simple poles ($s=s_n$) and expand $\mathcal{W}_x$ near these poles
\begin{align}
\mathcal{W}_x (s) = \frac{\rmd \mathcal{W}_x(s)}{\rmd s} \bigg|_{s=s_n} (s - s_n) + \mathcal{O}(s-s_n)^2.
\end{align}
The residue theorem yields both Green's function and the scalar field perturbations to take the sums over all poles
\begin{align}
G(t, x, x') &= \sum_{n} e^{s_n t} \frac{1}{\mathcal{W}_x'(s_n)} \bigg[ \hat{\varphi}_-(s_n,x') \hat{\varphi}_+(s_n,x)  \bigg], \\
\delta \varphi_1 (t, x) &=\sum_{n} e^{s_n t} \frac{1}{\mathcal{W}_x'(s_n)} \hat{\varphi}_+(s_n,x) \mathcal{C}_-(s_n) \equiv \sum_{n} a_n \varphi_n(t, x) , \label{eq:Greenssol}
\end{align}
where
\begin{align}
\mathcal{C}_-(s_n) = \int^{\infty}_{- \infty} \rmd x' \hat{\varphi}_-(s_n,x') \mathcal{J}(s_n, x'),
\end{align}
and
\begin{align}
a_n = \frac{\mathcal{C}_-(s_n)}{\mathcal{W}_x'(s_n)}, \qquad \varphi_n(t, x) \equiv e^{s_n t} \hat{\varphi}_+(s_n,x). 
\end{align}
At the same time, the condition $\mathcal{W}_x = 0$ in (\ref{eq:Wx}) also lead its integration to yield
\begin{align}
\hat{\varphi}_{+} (s, x) = c_0(s) \hat{\varphi}_{-}(s, x) ,
\end{align}
where $c_0$ is a function depending only on $s$. This implies that $\hat{\varphi}_{+}$ and $\hat{\varphi}_{-}$ are dependent. 

Another distinct characteristic of QNMs is their non-normalizability under the Klein-Gordon inner product:
\begin{align}
\langle \phi_n, \phi_m \rangle = \frac{i}{2} \int_{\Sigma} \bigg[ (D_{\mu} \phi_n)^* \phi_m - \phi_n^* (D_{\mu} \phi_m) \bigg] ~ \rmd \Sigma^{\mu}.
\end{align}
Expanding (\ref{eq:LPperturbRS}) near infinity and changing $s = - i \omega$, we obtain
\begin{align}
\partial_r^2 \delta \hat{\varphi}_1(\omega,r)+ \bigg(\omega^2 - \frac{2 \omega^2 A_1}{r} \bigg)\delta \hat{\varphi}_1(\omega,r) \simeq 0,
\end{align}
where $x \sim r $ from (\ref{eq:xtor}) is used, the source term is neglected since the variable change to $\omega$ indicates the Fourier transformation. Separating variables as $\delta \hat{\varphi}_1(\omega,r) = e^{- i \omega t} \delta \hat{\varphi}_1(r)$, the asymptotic solution reads
\begin{align}
\delta \hat{\varphi}_1(\omega,r) =& r e^{-i \omega  (t-r)} \bigg[ U\left(i \omega  A_1+1,2,-2 i r \omega \right) c_1 +  \, _1F_1\left(i \omega  A_1+1;2;-2 i r \omega \right) c_2 \bigg]  \\
\simeq & (-2 i \omega)^{-1-i \omega A_1} e^{- i \omega(t-r)} r^{- i \omega A_1} c_1 \nonumber\\
&+ \bigg[ \frac{(2i\omega)^{-1-i \omega A_1}}{\Gamma(1-i \omega A_1)} e^{- i \omega(t-r)} r^{- i \omega A_1} + \frac{(-2i\omega)^{-1+i \omega A_1}}{\Gamma(1+i \omega A_1)} e^{- i \omega(t+r)} r^{i \omega A_1}  \bigg] c_2
\end{align}
where $c_1$ and $c_2$ are integration constants, the second line is expanded for large $r$. In this solution, terms $e^{- i \omega(t-r)} r^{- i \omega A_1} = e^{- i \omega (t-r+A_1 \log r)}$ and $e^{- i \omega(t+r)} r^{i \omega A_1} = e^{- i \omega (t+r-A_1 \log r)}$ describe an outgoing and incoming waves respectively. We require the outgoing boundary condition at infinity, so we set $c_2 = 0$. Since the frequency is discretized, the solution is written as
\begin{align}
\delta \hat{\varphi}_1(\omega_n,r) = a_n e^{- i \omega_n (t-r)} r^{- i \omega_n A_1}, \qquad a_n = (-2 i \omega_n)^{-1-i \omega_n A_1}  c_1.
\end{align}
Plugging this to the Klein-Gordon inner product, we obtain
\begin{align}
\langle \delta \hat{\varphi}_1(\omega_n,r), \delta \hat{\varphi}_1(\omega_m,r) \rangle = \frac{1}{2} \int_{\Sigma_t} (\omega_n^* +  \omega_m) a_n^* a_m e^{i (\omega_n^* - \omega_m)(t-r)} r^{i A_1(\omega_n^* - \omega_m)} \rmd \Sigma^t 
\end{align}
where $\Sigma_t$ is a hypersurface with a timelike normal vector $n_t$. Since we are working with asymptotic solutions, we employed $\rmd \Sigma^t = \rmd \Sigma~ n^t = \sqrt{-g^{tt} g_{rr}}~ r^2 \sin \theta \rmd r \rmd \theta \rmd \phi \simeq r^2 \sin \theta \rmd r \rmd \theta \rmd \phi$ by substituting the asymptotic expansions from (\ref{eq:AexpInf})-(\ref{eq:vpexpInf}). When $n=m$, the inner product becomes
\begin{align}
\langle \delta \hat{\varphi}_1(\omega_n,r), \delta \hat{\varphi}_1(\omega_n,r) \rangle =& - \textrm{Re}(\omega_n) |a_n|^2 \frac{1}{2} \int_{\Sigma_t}  e^{i (\omega_n^* - \omega_n)(t-r)} r^{i A_1(\omega_n^* -  \omega_n)} \rmd \Sigma^t \nonumber\\
\simeq & 4 \pi \textrm{Re}(\omega_n) |a_n|^2 \bigg(\frac{e^{2~ \textrm{Im}(\omega_n)(t-r)} r^{2 (1+ \textrm{Im}(\omega_n) A_1)}}{2~ \textrm{Im}(\omega_n)} \bigg) 
\end{align}
where we made a large $r$ expansion at the second line after integration. As mentioned above, $A_1$ is the black hole mass, with $2M = -A_1 > 0$. For stable black holes, the imaginary part of the quasinormal mode frequency, denoted by $\textrm{Im}(\omega_n)$, is negative (i.e., $\textrm{Im}(\omega_n) < 0$). Consequently, the Klein-Gordon inner product diverges at the boundary of infinity when evaluated for QNMs. This divergence implies that QNMs are non-normalizable and do not form a complete set.

\subsection{Branch Cut}

Here, we compute the part of the Green's function (\ref{eq:GreenB}) that accounts for the branch cut, following the approach in \cite{Andersson:1996cm}. This term takes the following form:
\begin{align*}
G_{B}(t,x,x') = \int^{-\infty}_{0} \frac{\rmd s}{2 \pi i} e^{s t} \bigg[\hat{G}(s e^{2 \pi i}, x, x') - \hat{G}(s, x, x') \bigg].
\end{align*}
Let us return to the $r$-coordinate and introduce a new variable $\hat{\psi}(s,r)$
\begin{align}
\hat{\varphi}(s, x) \; \; \rightarrow \; \; \hat{\varphi}(s, r) = f(r) \hat{\psi}(s,r),\qquad f(r) = \sqrt[4]{A(r) B(r)}.
\end{align}
In this variable the equation (\ref{eq:PertbHmEq}) reads:
\begin{align}
\hat{\psi}^{(0,2)}(s,r) - \left[\frac{1}{A(r) B(r)} \bigg(s^2 + V_{\textrm{eff}}(r) \bigg) + \frac{f''(r)}{f(r)} \right] \hat{\psi} (s,r) = 0. \label{eq:psiEq1}
\end{align}
Expanding (\ref{eq:psiEq1}) to the large $r$ regime by using (\ref{eq:AexpInf})-(\ref{eq:vpexpInf}), it yields
\begin{align}
\hat{\psi}^{(0,2)}(s,r) - \left[s^2 - \frac{2 A_1 s^2}{r} + \frac{l_1(l_1+1)}{r^2} \right] \hat{\psi} (s,r) = 0 \label{eq:psiEq2}.
\end{align}
Here, $l_1$ accounts for both the angular momentum, coming from the spherical harmonic contribution $Y_l^m$ and the leading order contribution from the metric and the scalar coupling function:
\begin{align}
l_1(l_1+1) = l(l+1) + 3 A_1^2 s^2 - \frac{\varphi_{1,1}^2 s^2 }{2}.
\end{align}
Whittaker functions $M$ and $W$ solve the equation(\ref{eq:psiEq2})
\begin{align}
\hat{\varphi}_{+} (s,r) &= \sqrt[4]{A(r) B(r)} ~ M_{A_1 s, l_1 + \frac{1}{2}}(2 r s), \\
\hat{\varphi}_{-} (s,r) &= \sqrt[4]{A(r) B(r)} ~ W_{A_1 s, l_1 + \frac{1}{2}}(2 r s)
\end{align}
and they have properties of 
\begin{align}
M_{k,\mu}(z e^{2 \pi i m}) &= (-1)^{m} e^{2 \pi i \mu m} M_{k, \mu}(z), \\
W_{k, \mu}(z e^{2 \pi i m}) &= (-1)^{m} e^{- 2 \pi i \mu m} W_{k, \mu}(z) + \frac{(-1)^{m+1} 2 \pi i \sin(2 \pi \mu m)}{\Gamma(\frac{1}{2} - \mu - k) \Gamma(1+2 \mu) \sin(2 \pi \mu)} M_{k,\mu}(z).
\end{align}
In turn, the Wronskians take the following form
\begin{align}
&\mathcal{W}_{z}[M_{k,\mu}(z), W_{k,\mu}(z)] = M_{k,\mu}(z) \partial_z W_{k,\mu}(z) - W_{k,\mu}(z) \partial_z M_{k,\mu}(z) = - \frac{\Gamma(1+2 \mu)}{\Gamma(\frac{1}{2} + \mu - k)}, \\
&\mathcal{W}_{z}[M_{k,\mu}(z e^{2 \pi i m}), W_{k,\mu} (z e^{2 \pi i m})]  = \mathcal{W}_{z}[M_{k,\mu}(z), W_{k,\mu}(z)].
\end{align}
Using these properties, the Green's function in the $s$-space is computed as
\begin{align}
\hat{G}(s e^{2 \pi i}, x, x') = \hat{G}(s, x, x') + \frac{1}{\mathcal{W}_x[\varphi_+, \varphi_-]} \frac{2 \pi i ~e^{2 \pi i l_1}}{\Gamma(2l_1+2) \Gamma(-l_1-A_1 s)} \hat{\varphi}_{+}(s,x) \hat{\varphi}_{+}(s,x') ,
\end{align}
where the Wronskian is
\begin{align}
\mathcal{W}_x[\varphi_+, \varphi_-] &= \hat{\varphi}_{+}(s,x) ~ \partial_{x} \hat{\varphi}_{-}(s,x) - \hat{\varphi}_{-}(s,x)~ \partial_{x} \hat{\varphi}_{+}(s,x) \nonumber\\
&= - 2 s \sqrt{A(x)B(x)} \frac{\Gamma(2l_1+2)}{\Gamma(1+l_1 - A_1 s)}.
\end{align}
Consequently, the branch cut contribution to the Green's function reads
\begin{align}
G_{B}(t,x,x') &= - \frac{e^{2 \pi i l_1}}{ \{(2l_1+1)! \}^2} \frac{1}{2 \sqrt{A(x)B(x)}}\int^{-\infty}_{0} \frac{\rmd s}{2 \pi i} e^{s t} \frac{1}{s} \frac{\Gamma(1+l_1-A_1s)}{\Gamma(-l_1-A_1 s)}\hat{\varphi}_{+}(s,x) \hat{\varphi}_{+}(s,x').
\end{align}
In terms of the $r$ coordinate, this expression becomes
\begin{align}
G_{B}(t,r,r') &= - \frac{e^{2 \pi i l_1}}{ \{(2l_1+1)! \}^2} \frac{1}{2 A(r)B(r)}\int^{-\infty}_{0} \frac{\rmd s}{2 \pi i} e^{s t} \frac{1}{s} \frac{\Gamma(1+l_1-A_1s)}{\Gamma(-l_1-A_1 s)}\hat{\varphi}_{+}(s,r) \hat{\varphi}_{+}(s,r') ,
\end{align}
where we used the Wronskian computed in the $r$ coordinate 
\begin{align}
\mathcal{W}_r[\varphi_+, \varphi_-] = \sqrt{AB}~\mathcal{W}_x[\varphi_+, \varphi_-].
\end{align}
Since we consider the low-frequency regime ($s \sim i \omega$ for Fourier modes), we take the limit $A_1 s \rightarrow 0$ and $\varphi_{1,1} s \rightarrow 0$. This leads the Green's function to be
\begin{align}
G_{B}(t,r,r') \simeq (-1)^{-l} \frac{A_1 2^{2 l+1} \Gamma (l+1)^2 r^{l+1} (r')^{l+1} }{\Gamma (2 l+2)^2} \int^{-\infty}_{0} \rmd s  ~e^{s (t-r-r')} s^{2 (l+1)} {_1}F_1(2 r s)~ {_1}F_1(2 r' s),
\end{align}
where ${_1}F_1(2 r s) = {_1}F_1(l+1;2 l+2; 2 r s)$ is the confluent hypergeometric function of the first kind. 

Firstly, we consider the low-frequency limit $r s \ll 1$ and  $r' s \ll 1$. In the limit $t \rightarrow \infty$ with $r$ fixed, we obtain
\begin{align}
G_{B}(t,r,r') \simeq (-1)^{l+1} \frac{2 (-A_1) (2 l+2)!}{((2 l+1)!!)^2} \frac{r^{l+1} (r')^{l+1}}{t^{2 l+3}}.
\end{align}
Our late-time solution aligns with the leading order term of the Schwarzschild solution at large $r$ in \cite{Andersson:1996cm,Gundlach:1993tp}, when we substitute $-A_1$ for $2M$.

In the limit $v \rightarrow \infty$ ($t \rightarrow \infty$ with $t/r$ fixed), we can take $r \rightarrow \infty$ and $r' s \ll 1$. Consequently, we obtain the following expression
\begin{align}
G_{B}(t,r,r') &\simeq - \frac{A_1 (r')^{l+1}}{(2 l+1)!!} \int^{-\infty}_{0} \rmd s ~ e^{s (t-r)} s^{l+1}+ \frac{A_1 (-1)^l (r')^{l+1}}{(2l+1)!} \int^{-\infty}_{0} \rmd s ~ e^{s (r+t)} s^{l+1}, \\
&= \frac{(l+1)!}{(2 l+1)!!} \bigg[ \frac{(-A_1) (-1)^{l+1} }{u^{l+2}}-\frac{A_1}{v^{l+2}} \bigg] (r')^{l+1} \simeq  (-1)^{l+1}  (-A_1) \frac{(l+1)!}{(2 l+1)!!}  \frac{ (r')^{l+1}}{u^{l+2}}  
\end{align}
where $v = t + x \sim t \tcr{+} r$ and $u = t -x \sim t - r$ and $v \rightarrow \infty$ is taken at the last step. This result also agrees with the radiative case for no initial static field at null infinity in \cite{Gundlach:1993tp} by replacing the mass of hairy black holes.

%%%%%%%%%%%%%%%%%%%%%%%%%%%%%%%%%%%%%%%%%%%%%%%%
\section{Numerical Method for Quasinormal Modes}
%%%%%%%%%%%%%%%%%%%%%%%%%%%%%%%%%%%%%%%%%%%%%%%%

Since the background black hole solutions are numerical, we use a numerical method, namely, the Chebyshev spectral method, to calculate the QNMs. This subsection discusses the employed numerical methods initially introduced in \cite{Jansen:2017oag}. We present results for black holes in the symmetric and the symmetry-broken phases in Sections 5 and 6 correspondingly.

One needs to set up a grid to realize a space to use numerical calculations. Here we employ Chebyshev polynomials $T_n (y)$ defined by 
\begin{align}
\cos n \theta = T_n (\cos \theta).
\end{align}
which introduces unequispacing grids \cite{dolapcci2004chebyshev, XU201617}. If one sets $\cos \theta = y$, then the polynomials become
\begin{align}
&T_0(y) = 1, \qquad T_1(y) = y, \qquad T_2(y) = 2 y^2 -1, \qquad T_3(y) = 4 y^3 - 3 y, \qquad \cdots. 
\end{align}
The Chebyshev polynomials form a complete basis, so any function defined on an interval $[-1,1]$ admits the Chebyshev expansion:
\begin{align}
f(y) = \sum^{\infty}_{n=0} C_n T_n (y) .
\end{align}
The polynomials are also orthogonal with respect to the following inner product 
\begin{align}
\langle f,g \rangle \equiv \int^{1}_{-1} \frac{f(y)g(y) \rmd y}{\sqrt{1-y^2}},
\end{align}
implying
\begin{align}
\langle T_n, T_m \rangle = 
 \left\{ 
  \begin{array}{rl} 
   0 & \; \; \; \textrm{if} \; \; n \neq m \\ 
   \pi  &  \; \; \;  \textrm{if} \; \; n=m=0 \\ 
   \frac{\pi}{2} &  \; \; \;  \textrm{if} \; \; n = m \neq 0
  \end{array} 
 \right. 
 \end{align}

Due to the limit of the numerical capability, we approximate the Chebyshev polynomial $T_n (y)$ to $N$-th degree polynomial
\begin{align}
f(y) = \sum^{N}_{n=0} C_n T_n (y) = T C .
\end{align}
We generate $(N+1)$ Chebyshev points $y_n = \cos(\frac{n \pi}{N})$, $n = 0,1, \cdots, N$ accordingly. The derivative of the function of $f(y)$  is expressed as
\begin{align}
f'(y) = \sum^{N}_{n=0} C_n T^{(1)}_n (y) = T^{(1)} C = T (2 M C) \equiv T C^{(1)}.
\end{align}
where $M$ is a matrix that is made up with the coefficients from the derivative of the Chebyshev polynomial with respect to $y$ on the basis of the Chebyshev polynomial  $T_n(y)$. It takes a form of

\begin{align}
M_{(n=\textrm{even})} =
\begin{bmatrix}
    0 & \frac{1}{2} & 0 & \frac{3}{2} & 0 & \frac{5}{2} & \dots  & \frac{N}{2} \\
    0 & 0 & 2 & 0 & 4 & 0 & \dots  & 0 \\
    0 & 0 & 0 & 3 & 0 & 5 & \dots  & N \\
    \vdots & \vdots & \vdots & \vdots & \vdots & \vdots & \vdots & \vdots \\
    0 & 0 & 0 & 0 & 0 & 0 & \dots  & N \\
    0 & 0 & 0 & 0 & 0 & 0 & \dots  & 0 \\
\end{bmatrix}, \; \; \;
M_{(n=\textrm{odd})} =
\begin{bmatrix}
    0 & \frac{1}{2} & 0 & \frac{3}{2} & 0 & \frac{5}{2} & \dots  & 0 \\
    0 & 0 & 2 & 0 & 4 & 0 & \dots  & N \\
    0 & 0 & 0 & 3 & 0 & 5 & \dots  & 0 \\
    \vdots & \vdots & \vdots & \vdots & \vdots & \vdots & \vdots & \vdots \\
    0 & 0 & 0 & 0 & 0 & 0 & \dots  & N \\
    0 & 0 & 0 & 0 & 0 & 0 & \dots  & 0 \\
\end{bmatrix}.
\end{align}
The higher derivatives of $f(y)$ are obtained by 
\begin{align}
f^{(m)}(y) = T^{(m)} C = T (2 M)^{m} C \equiv T C^{(m)}. 
\end{align}
For example, if the perturbed equation has a form of 
\begin{align}
A_2 \varphi''(y) + (A_1 + A_{1\omega} \omega) \varphi'(y) + (A_0 + A_{0 \omega} \omega^2) \varphi(y) = 0,
\end{align}
we can express the functions as follows
\begin{align}
A_2 T (4 M^2 C) + (A_1 + A_{1\omega} \omega) T (2 M C) + (A_0 + A_{0 \omega} \omega^2) T C= 0.
\end{align}
We rearrange the above expression as
\begin{align}
&T \bigg(4 A_2 M^2 + 2 (A_1 + A_{1\omega} \omega) M + (A_0 + A_{0 \omega} \omega^2) I \bigg) C= T (\alpha + \beta_1 \omega + \beta_2 \omega^2 )C = 0 .
\end{align}
Here $I$ is the $(N+1) \times (N+1)$ identity matrix and
\begin{align}
\alpha = 4 A_2 M^2 + 2 A_1 M+ A_0 I, \qquad \beta_1 = A_{1\omega} M, \qquad \beta_2 = A_{0 \omega} I.
\end{align}
The subscript of $\beta$ indicates the power of $\omega$. Then we redefine $\alpha$, $\beta_1$, and $\beta_2$ to take a form of
\begin{align}
\tilde{\alpha}+ \tilde{\beta}~  \omega = 0
\end{align}
where 
\begin{align}
	\tilde{\alpha} = \begin{pmatrix}
	\alpha & \beta_1 \\
	0 & I
	\end{pmatrix}, \qquad 
	\tilde{\beta} = \begin{pmatrix}
	0 & -\beta_2 \\
	I & 0
	\end{pmatrix}.
\end{align}
For the case of equations containing higher orders of $\omega$, we can generalize $\tilde{\alpha}$ and $\tilde{\beta}$ as 
\begin{align}
	\tilde{\alpha} = \begin{pmatrix}
	\alpha & \beta_1 & \cdots & \beta_{n_{max}-1} \\
	0 & I & \cdots & 0 \\
	0 & 0 & \ddots & 0 \\
	0 & 0 & \cdots & I
	\end{pmatrix}, \qquad \tilde{\beta} = \begin{pmatrix}
	0 & 0 & \cdots & -\beta_{n_{max}} \\
	I & 0 & \cdots & 0 \\
	0 & I & \ddots & 0 \\
	0 & 0 & 1 & \cdots
	\end{pmatrix},
\end{align}
where $n_{max}$ is the highest power of $\omega$.

By using the command ``Eigenvalues$[\{\tilde{\alpha}, \tilde{\beta} \}]$" in Mathematica, we can compute the quasinormal frequencies $\omega$.  We here employ the Chebyshev zero points
\begin{align}
y_j = \cos \frac{(j-1/2) \pi}{N}, \qquad (1 \leq j \leq N)
\end{align}
which generate $N$ Chebyshev points \cite{XU201617}. Then we obtain $N$ values of $\omega$. Because of the numerical errors we repeat the numerical calculation with different numbers of grids and changing the accuracy and precision. If these calculations produce the same values of $\omega$ within the tolerance of $10^{-2}$, we consider that those values of $\omega$ are reliable candidates. (In our numerical calculation, the tolerance ranges from $10^{-35}$ to $10^{-2}$.) Then, we check the eigenfunctions for the selected $\omega$ in the previous step. If they satisfy our boundary conditions, we consider those values of $\omega$ to be correct.\\

We first describe our numerical solutions on the grids based on these properties and numerical techniques. Since the Chebyshev polynomials are defined for  $-1\lesssim y \lesssim 1$, we transform the coordinate $r$, which ranges from the black hole horizon to infinity, into a finite interval ($z_h \leq z \leq z_{\infty}$) using the substitution $r = 1/z$. Subsequently, we discretize $z$ into $z_n$ by rescaling $y_n$
\begin{align}
z_n = \frac{z_h-z_{\infty}}{2 ~ z_h} y_n + \frac{z_h+ z_{\infty}}{2 ~ z_h} ,
\end{align}
where $z_{\infty} = 0$. Then, the Chebyshev points are shifted to $0 \lesssim z_n \lesssim1$.  

To incorporate the incoming and outgoing boundary conditions into our numerical calculations, we use ingoing Eddington-Finkelstein (EF) coordinates. These coordinates have a distinct advantage: when solving the perturbation equation near the boundaries (both horizon and infinity), the asymptotic solutions naturally separate into incoming and outgoing modes. As we will see in more detail later, the incoming modes are generally normalizable and exhibit smooth behavior. Conversely, the outgoing modes are non-normalizable and oscillate rapidly as they approach the horizon or infinity. Because of such rapid oscillations, the wavelength of the outgoing mode becomes much smaller than the grid spacing. Hence, our numerical calculations will not be able to capture it effectively. This limitation is beneficial near the horizon because we aim to impose the incoming boundary condition there. However, it becomes problematic at infinity, where we require the outgoing boundary condition that involves the non-normalizable and rapidly oscillating modes. To make the outgoing modes numerically treatable, we can redefine the outgoing mode to become normalizable. This transformation will automatically render the incoming mode non-normalizable and rapidly oscillating. By strategically redefining the field in this manner, we can effectively impose the boundary conditions for both incoming and outgoing QNMs on our numerical calculations.

%%%%%%%%%%%%%%%%%%%%%%%%%%%%%%%%%%%%%%%%%%%%%%%%
\section{Quasinormal Modes for Hairy Black Holes in the Symmetric Phase}
%%%%%%%%%%%%%%%%%%%%%%%%%%%%%%%%%%%%%%%%%%%%%%%%

We numerically generate hairy black hole solutions in the symmetric phase for both positive and negative values of $\alpha$. Based on prior theoretical expectations \cite{Latosh:2023cxm}, hairy black holes beyond a certain positive value of $\alpha$ are anticipated to be unstable. To verify this, we calculate their QNMs and demonstrate that they indeed become unstable under scalar field perturbations at a critical value of $\alpha$, denoted as $\alpha_{\textrm{crit.}}$.

\subsection{Background solution}

In order to use our numerical method, we transform it to the ingoing EF coordinate as follows
\begin{align}
\rmd s^2 = -A(z) \rmd v^2 - \frac{2}{z^2} \sqrt{\frac{A(z)}{B(z)}} \rmd v \rmd z + \frac{1}{z^2} (\rmd \theta^2 + \sin^2 \theta \rmd \phi^2) ,
\end{align}
where $z=1/r$ and $v$ is defined as
\begin{align}
v = t + z_*, \qquad z_* = - \int \frac{1}{z^2} \frac{1}{\sqrt{AB}} \rmd z. \label{eq:vzs}
\end{align}
Then, the Gauss-Bonnet term is written as
\begin{align}
  \mathcal{G} =& R_{\mu \nu \rho \sigma} R^{\mu \nu \rho \sigma} - 4 R_{\mu \nu} R^{\mu \nu} +R^2 \nonumber\\
  =& 2 z^6 B\bigg[\frac{(3 B-1) A' B'}{A B}+(B-1) \left(\frac{2 A''}{A}-\frac{A'^2}{A^2}+\frac{4 A'}{z A}\right)\bigg],
\end{align}
where $'$ is the derivative with respect to $z$. Equations of motions (\ref{eq:EE})-(\ref{eq:KG}) are expressed in terms of $\varphi_1$ and $\varphi_2$ as follows
%\begin{align}
%&\frac{-z B'+B-1}{z^4 B}  + \kappa^2 \sum^{2}_{i=1} \bigg[ \frac{1-16 z^2 (B-1) \left(x-2 \lambda  \varphi_i^2\right)}{2 z^2}  \varphi_i'^2  -\frac{4 x \left(z (3 B-1) B' +4 (B-1) B \right)}{z B} \varphi_i \varphi_i' \label{eq:SMeom1} \nonumber\\
%&\qquad \qquad -8 x (B-1) \varphi_i \varphi_i '' \bigg] + 32 \kappa^2 \lambda  (B -1) \prod^2_{i=1} \varphi_i \varphi_i'  = 0, \\
% &\frac{1}{z^3} \bigg(\frac{A'}{A}-\frac{B'}{B} \bigg)+ \kappa^2 \sum^{2}_{i=1}\bigg[ \frac{1-8 z^2 (B-1) \left(x-2 \lambda  \varphi_i^2\right)}{z^2} \varphi_i'^2 + 4 x \bigg( \frac{4 (1-B)}{z}  + \frac{(3 B -1) \left(B A' -A B' \right)}{A B} \bigg) \varphi_i \varphi_i' \label{eq:SMeom2}\nonumber\\
%& \qquad \qquad - 8 x (B-1) \varphi_i \varphi_i '' \bigg] + 32 \kappa^2 \lambda  (B-1) \prod^{2}_{i=1} \varphi_i \varphi_i' = 0, \\
%&\frac{A''}{2 z^3 A}+\frac{1}{4 z^4}\left(2-\frac{z A'}{A}\right) \left(\frac{A'}{A}-\frac{B'}{B}\right) + \kappa^2 \sum^{2}_{i=1} \bigg[ \frac{ \left(A + 8 z^3 B A' \left(x-2 \lambda  \varphi_i^2\right)\right)}{2 z^3 A} \varphi_i'^2 + \frac{4 x B A'}{A} \varphi_i \varphi_i''  \label{eq:SMeom3} \nonumber\\
%&\qquad \qquad + 2 x B \bigg(\frac{2 A''}{A}+\frac{3 A' B'}{A B}-\frac{A'^2}{A^2}+\frac{8 A'}{z A} \bigg) \varphi_i \varphi_i' \bigg] - \frac{16 \kappa ^2 \lambda  B A'}{A}  \prod^{2}_{i=1} \varphi_i \varphi_i' = 0,\\
%&\varphi_i'' + \frac{1}{2}  \left(\frac{A'}{A}+\frac{B'}{B}\right) \varphi_i' +2 x z \bigg[\frac{z (3 B-1) A' B'}{A B}+(B-1) \left(\frac{2 z A''}{A}-\frac{z A'^2}{A^2}+\frac{4 A'}{A}\right) \bigg] \varphi_i  = 0 , \label{eq:SMeom4}
%\end{align}
\begin{align}
&\frac{-z B'+B-1}{z^4 B}  + \kappa^2 \sum^{2}_{i=1} \bigg[ \frac{1-16 z^2 (B-1) \left(x-2 \lambda  \varphi_i^2\right)}{2 z^2}  \varphi_i'^2  -8 x (B-1) \varphi_i \varphi_i ''  \label{eq:SMeom1} \nonumber\\
&\qquad  -\frac{4 x \left(z (3 B-1) B' +4 (B-1) B \right)}{z B} \varphi_i \varphi_i' \bigg] + 32 \kappa^2 \lambda  (B -1) \prod^2_{i=1} \varphi_i \varphi_i'  = 0, \\
 &\frac{1}{z^3} \bigg(\frac{A'}{A}-\frac{B'}{B} \bigg)+ \kappa^2 \sum^{2}_{i=1}\bigg[ \frac{1-8 z^2 (B-1) \left(x-2 \lambda  \varphi_i^2\right)}{z^2} \varphi_i'^2 - 8 x (B-1) \varphi_i \varphi_i '' \label{eq:SMeom2}\nonumber\\
& \qquad + 4 x \bigg( \frac{4 (1-B)}{z}  + \frac{(3 B -1) \left(B A' -A B' \right)}{A B} \bigg) \varphi_i \varphi_i'  \bigg] + 32 \kappa^2 \lambda  (B-1) \prod^{2}_{i=1} \varphi_i \varphi_i' = 0, \\
&\frac{A''}{2 z^3 A}+\frac{1}{4 z^4}\left(2-\frac{z A'}{A}\right) \left(\frac{A'}{A}-\frac{B'}{B}\right) + \kappa^2 \sum^{2}_{i=1} \bigg[ \frac{ \left(A + 8 z^3 B A' \left(x-2 \lambda  \varphi_i^2\right)\right)}{2 z^3 A} \varphi_i'^2 + \frac{4 x B A'}{A} \varphi_i \varphi_i''  \label{eq:SMeom3} \nonumber\\
&\qquad + 2 x B \bigg(\frac{2 A''}{A}+\frac{3 A' B'}{A B}-\frac{A'^2}{A^2}+\frac{8 A'}{z A} \bigg) \varphi_i \varphi_i' \bigg] - \frac{16 \kappa ^2 \lambda  B A'}{A}  \prod^{2}_{i=1} \varphi_i \varphi_i' = 0,\\
&\varphi_i'' + \frac{1}{2}  \left(\frac{A'}{A}+\frac{B'}{B}\right) \varphi_i' +2 x z \bigg[\frac{z (3 B-1) A' B'}{A B}+(B-1) \left(\frac{2 z A''}{A}-\frac{z A'^2}{A^2}+\frac{4 A'}{A}\right) \bigg] \varphi_i  = 0 , \label{eq:SMeom4}
\end{align}
where
\begin{align}
  x = \alpha - \lambda \, \varphi_1^2 - \lambda \, \varphi_2^2 \, .
\end{align}
Asymptotic flatness is required at infinity. Under this condition, the metric and scalar fields are expanded as follows:
\begin{align}
  &A(z) \sim 1 + A_1 z  -\frac{1}{24} A_1 z^3 \left(\varphi_{1,1}^2+\varphi_{2,1}^2\right)  + \cdots  \label{eq:AexpInfz} \\
  &B(z) \sim 1+A_1 z +\frac{1}{4} z^2 \left(\varphi_{1,1}^2+\varphi_{2,1}^2 \right) -\frac{1}{8} A_1 z^3 \left(\varphi_{1,1}^2+\varphi_{2,1}^2 \right)+ \cdots \label{eq:BexpInfz} \\
  &\varphi_i(z) \sim \varphi_{i, \infty }+\varphi_{i,1} z -\frac{1}{2} A_1 \varphi_{i,1} z^2  -\frac{1}{24} \varphi_{i,1} z^3 \left(\varphi_{1,1}^2+\varphi_{2,1}^2 -8 A_1^2 \right)+ \cdots  \label{eq:vpexpInfz}
\end{align}
where all coefficients are constant. As before, we identify the coefficient $A_1$ as ADM mass of the black hole such that $A_1 = - 2M$, and $\varphi_{i,1}$ is the scalar charge, $\varphi_{i,1}=Q_i$. Hereafter we set $\varphi_1 = \varphi_2$ for simplicity. \\

\begin{figure}
\begin{center}
	\includegraphics[scale=0.28]{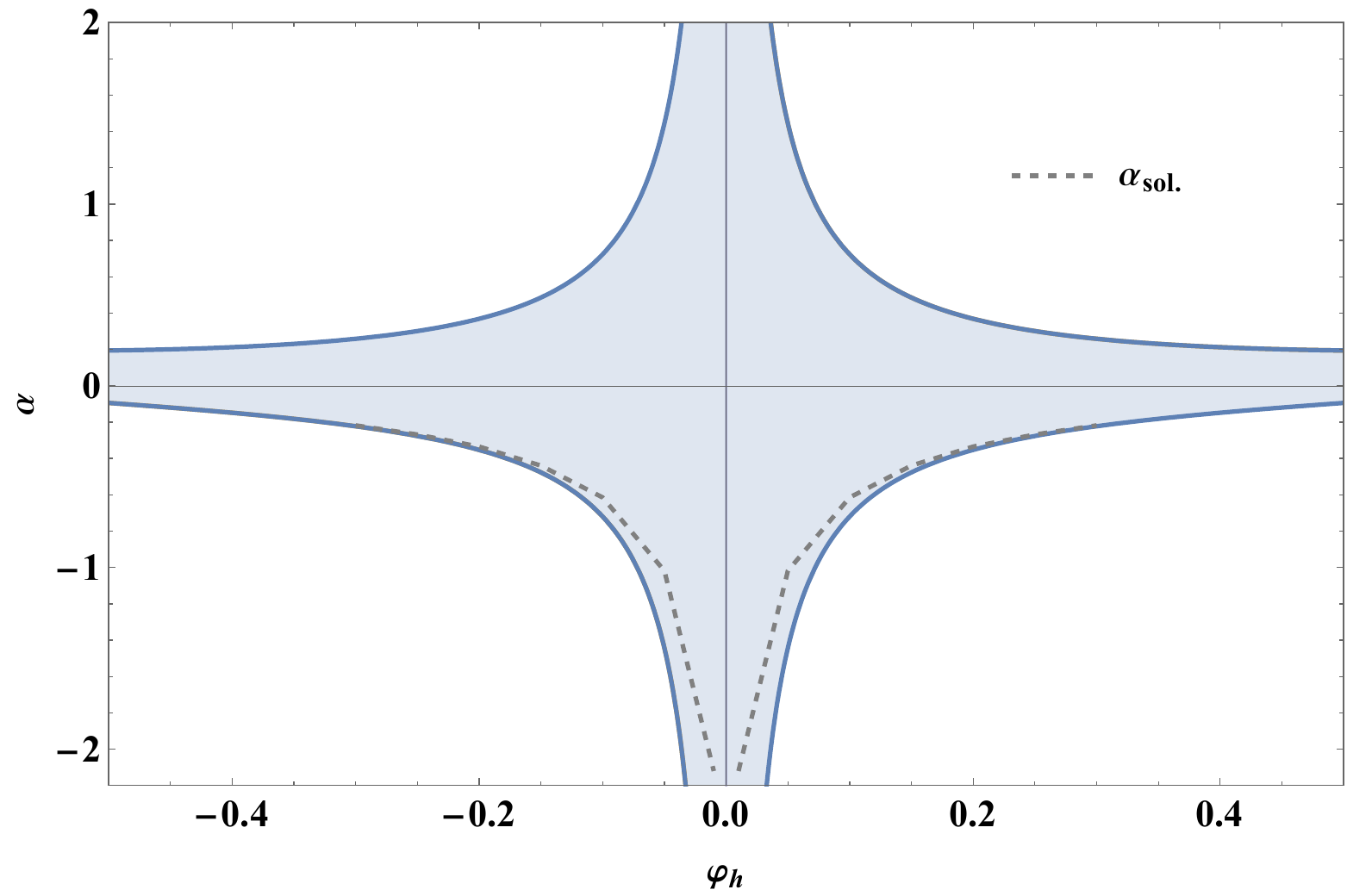}
	\caption{Phase space of $\alpha$ and $\varphi_h$ for $\lambda = 0.1$ in the symmetric phase}
	\label{fig:regularCD}
\end{center}
\end{figure}

For a regular black hole to exist, specific boundary conditions must be met near the horizon:
\begin{align}
&A(z) \sim A_h \epsilon + \mathcal{O}(\epsilon^2) , \; \; \; B(z) \sim B_h \epsilon  + \mathcal{O}(\epsilon^2), \; \; \; \varphi_i(z) \sim \varphi_{ih} + {\varphi_{ih,1}} \epsilon  + \mathcal{O}(\epsilon^2) ,  \label{eq:NHexp}  
\end{align}
where $\epsilon = z_h - z$ is the expansion parameter  and  $A_h, B_h, \varphi_{i h}$ and $\varphi_{i h,1}$ ($i=1,2$) are constants. We also set $\kappa^2 = 1/2$. The following relations between these constants ensure the regularity of the metric and scalar fields:
\begin{align}
  B_h&= \frac{1-\sqrt{1-96 z_h^4 \left(\varphi _{1h}^2+\varphi _{2h}^2\right) \left(\alpha -\lambda  \left(\varphi _{1h}^2+\varphi _{2h}^2\right)\right){}^2}}{48  z_h^5 \left(\varphi _{1h}^2+\varphi _{2h}^2\right) \left(\alpha -\lambda  \left(\varphi _{1h}^2+\varphi _{2h}^2\right)\right)^2}, \label{eq:Bh} \\
  {\varphi_{i h,1}} &= -\frac{\varphi _i \left(1-\sqrt{1-96 z_h^4  \left(\varphi _{1h}^2+\varphi _{2h}^2\right) \left(\alpha -\lambda  \left(\varphi _{1h}^2+\varphi _{2h}^2\right)\right){}^2}\right)}{4 z_h^3 \left(\varphi _{1h}^2+\varphi _{2h}^2\right)  \left(\alpha - \lambda  \left(\varphi _{1h}^2+\varphi _{2h}^2\right) \right)} . \label{eq:dvarphi}
\end{align}
As seen above, the near horizon expansion of the fields is determined uniquely by two boundary values, which are $A_h$ and $\varphi_{1h} = \varphi_{2h} = \varphi_h$. We discovered that $\varphi''(r_h)$ diverges when the value inside the root becomes zero. To avoid this, we impose the regularity conditions:
\begin{align}
\left(\varphi _{1h}^2+\varphi _{2h}^2\right) \left(\alpha -\lambda  \left(\varphi _{1h}^2+\varphi _{2h}^2\right)\right)^2 < \frac{1}{96 z_h^4}. \label{eq:constraint2}
\end{align}
Thus, this regularity condition constrains the values of parameters ($\alpha, \lambda$) and the scalar field value near the horizon ($\varphi_h$). In Figure \ref{fig:regularCD}, we present the phase space for these parameters at a fixed value of $\lambda$. The shaded regions enclosed by the solid lines represent valid parameters for hairy black hole solutions. Points lying on the solid lines themselves are not considered valid. 

\begin{figure}
\subfloat[The metric solutions of $A$ and $B$ with $\alpha=-0.8$]{\includegraphics[scale=0.25]{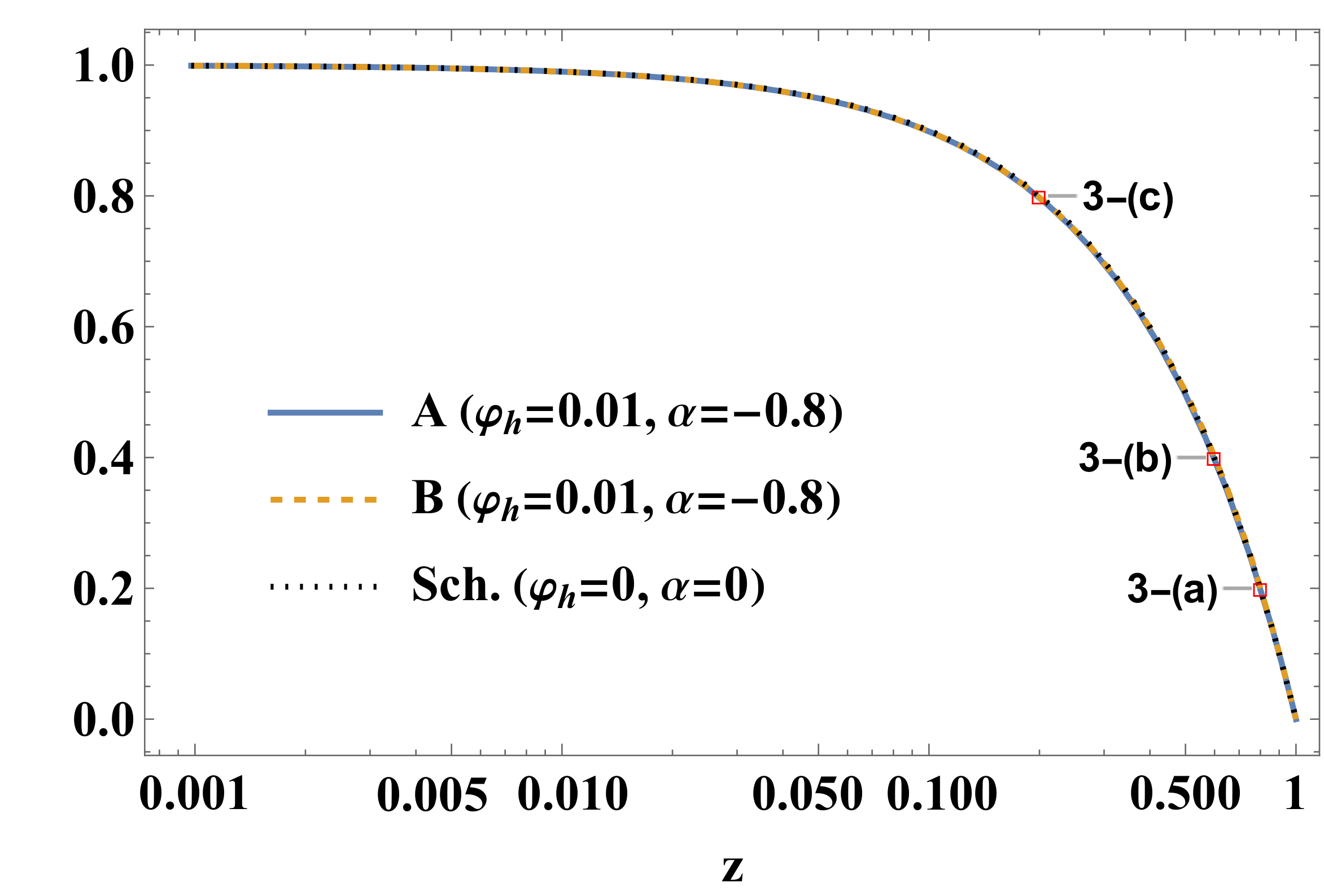}} $\;$
\subfloat[The metric solutions of $A$ and $B$ with $\alpha=3.5$]{\includegraphics[scale=0.25] {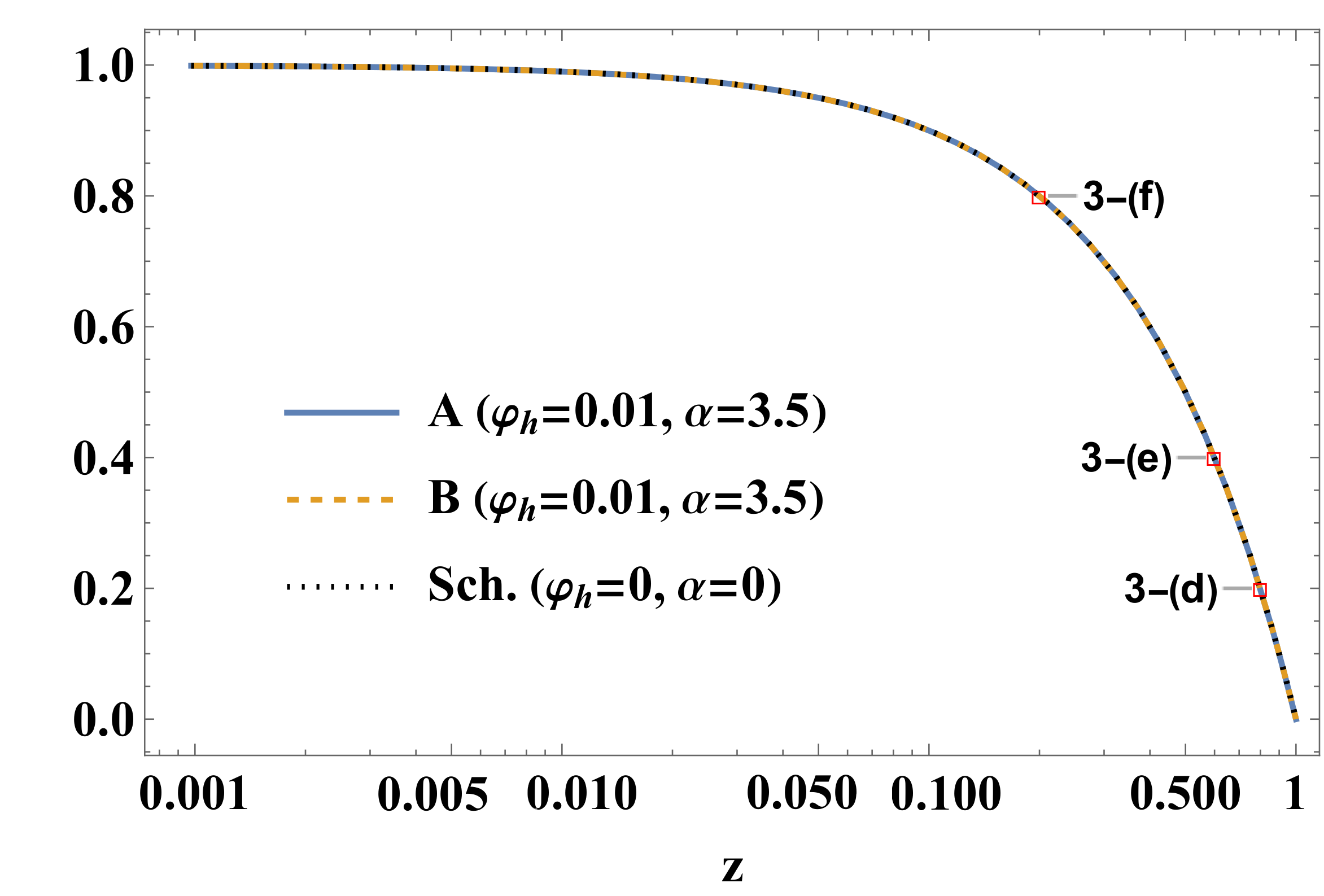}} $\;$
\subfloat[The scalar field solutions $\varphi$ for various values of $\alpha$]{\includegraphics[scale=0.25]{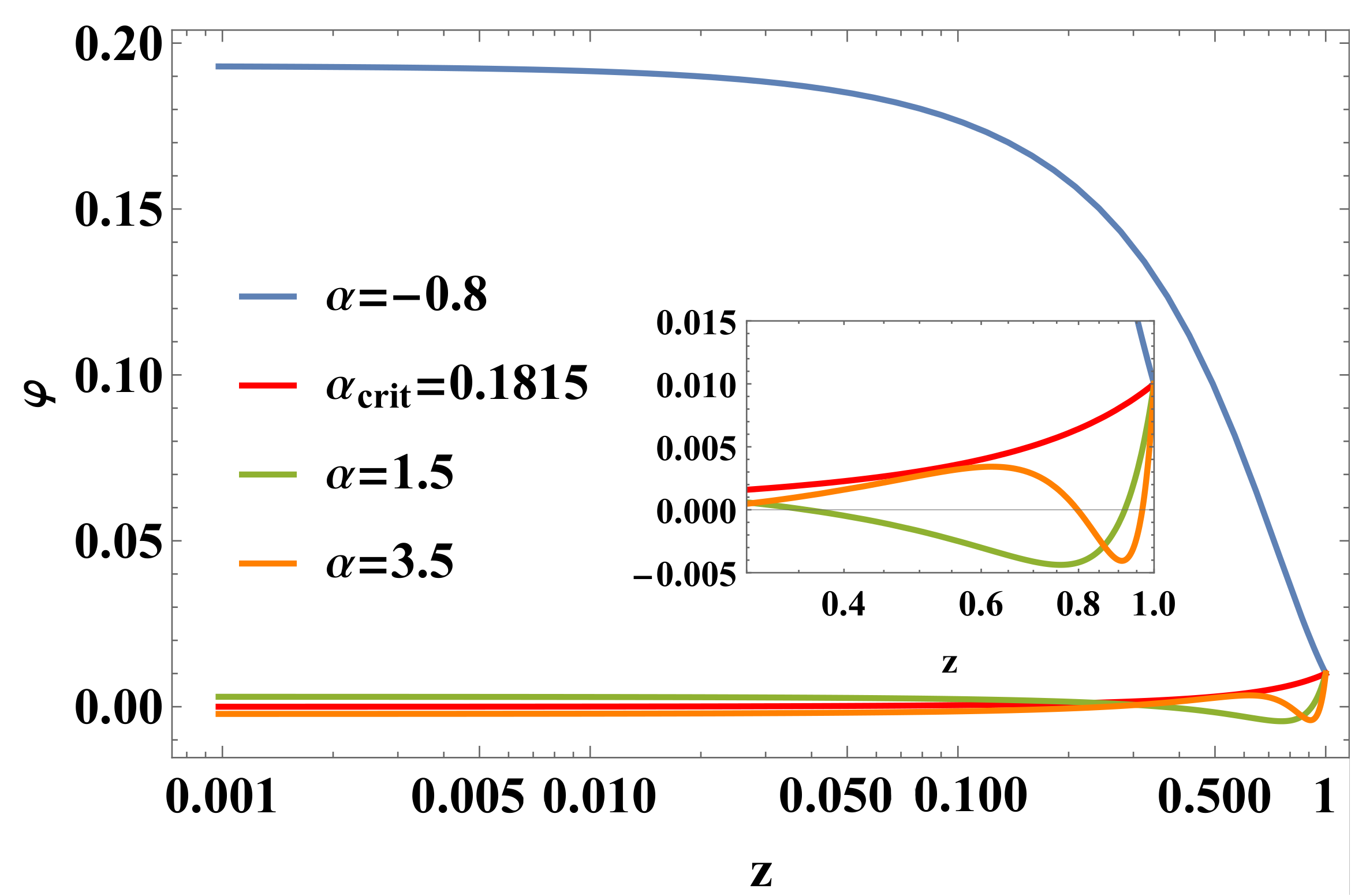}}
\caption{Hairy black hole solutions with $\varphi_h = 0.01$ for various values of $\alpha$. The black dotted line denotes the solutions of Schwarzschild black hole. The small regimes with the callout index (a)$\sim$(f) are plotted in Figure~\ref{fig:SM001sub}.
}
	\label{fig:SM001}

\vspace{2cm}

\subfloat[Case of $\alpha=-0.8$]{\includegraphics[scale=0.27]{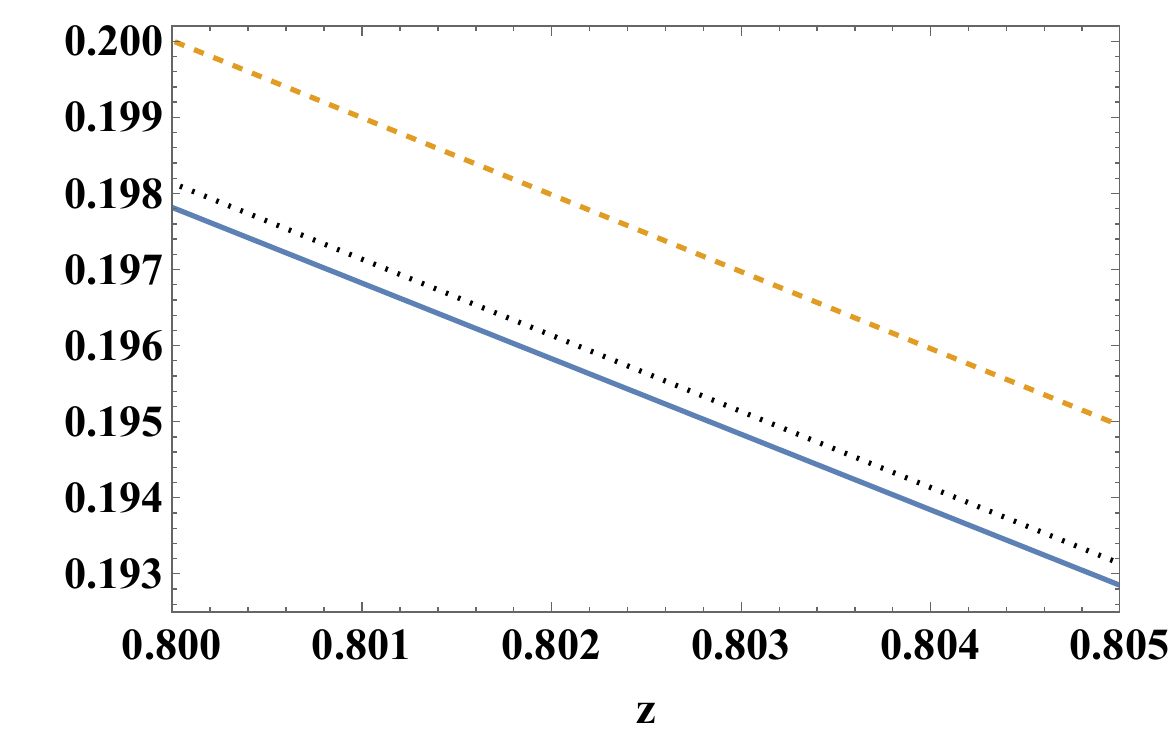}}
\subfloat[Case of $\alpha=-0.8$]{\includegraphics[scale=0.27]{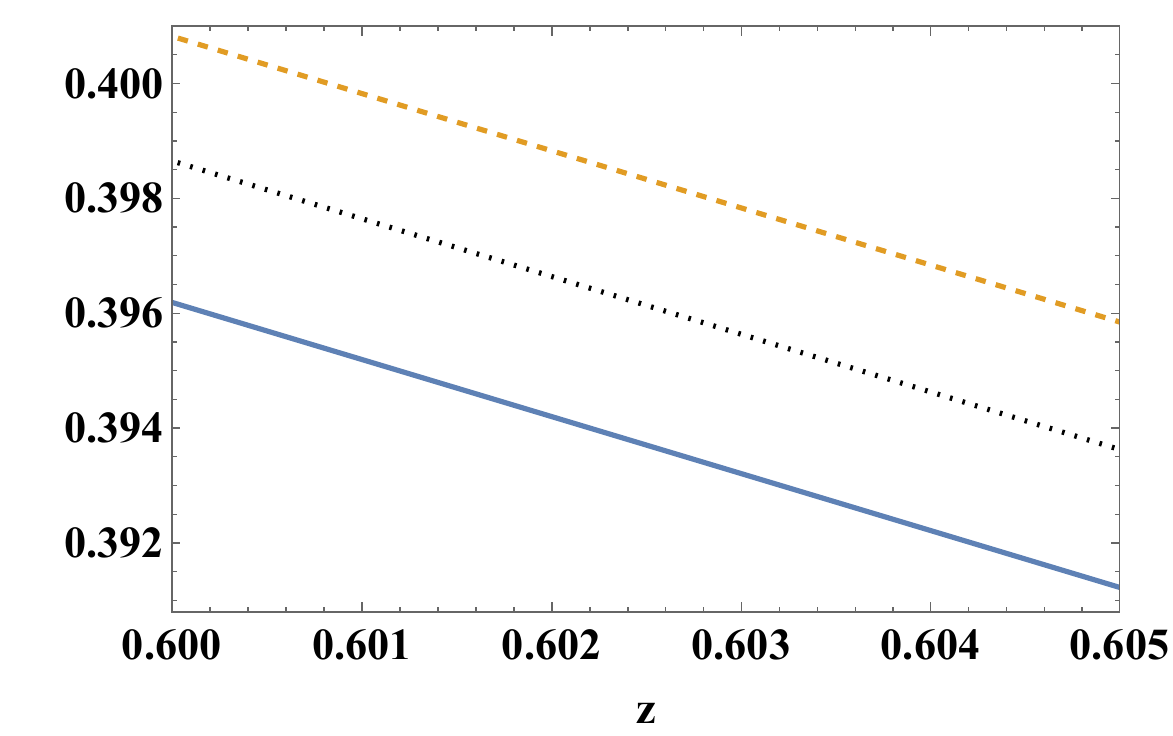}}
\subfloat[Case of $\alpha=-0.8$]{\includegraphics[scale=0.27]{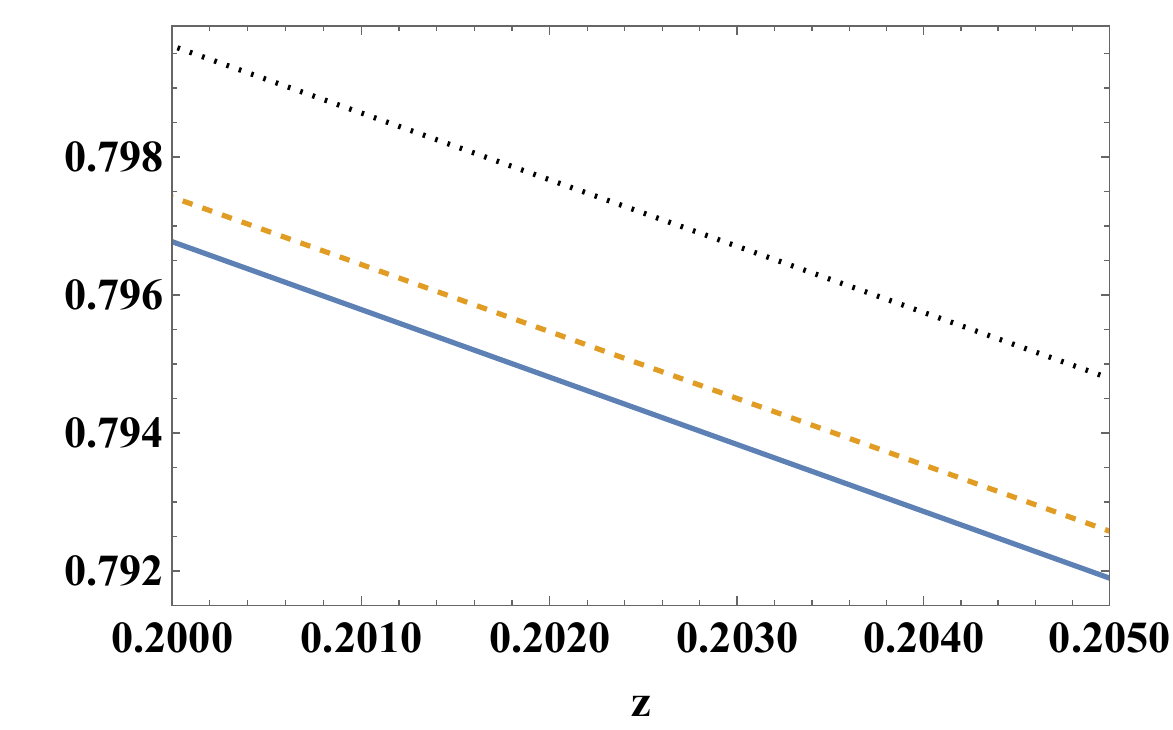}}

\subfloat[Case of $\alpha=3.5$]{\includegraphics[scale=0.27]{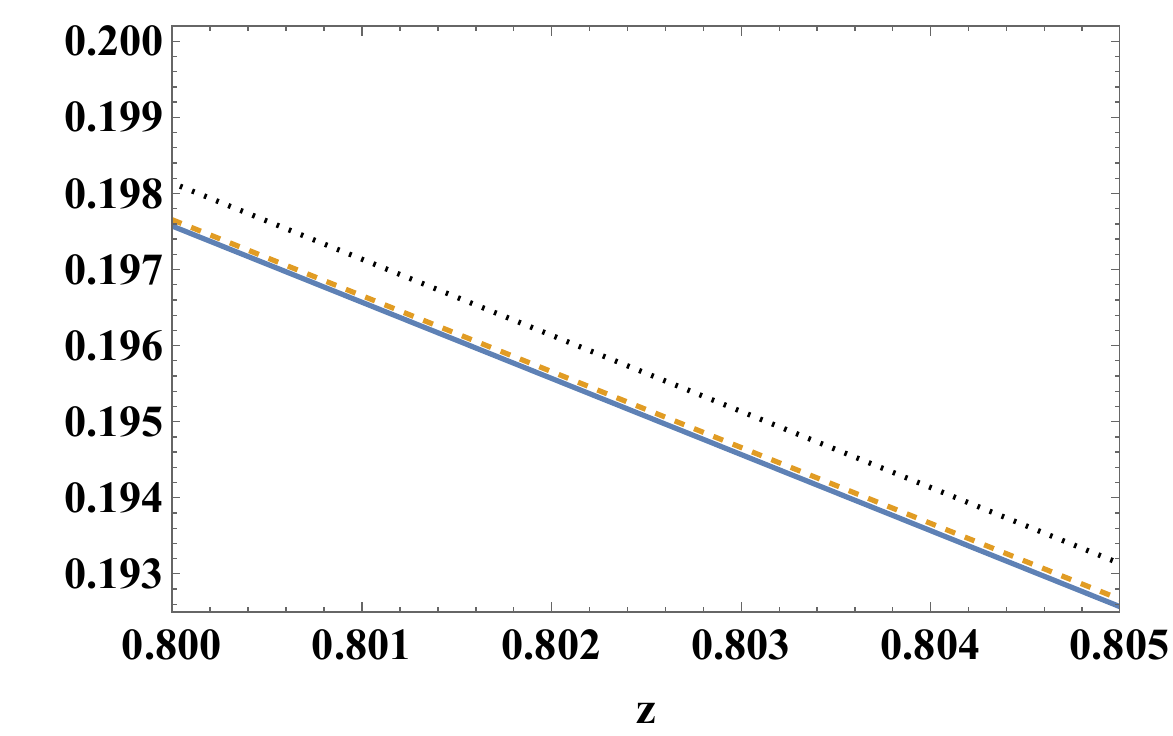}}
\subfloat[Case of $\alpha=3.5$]{\includegraphics[scale=0.27]{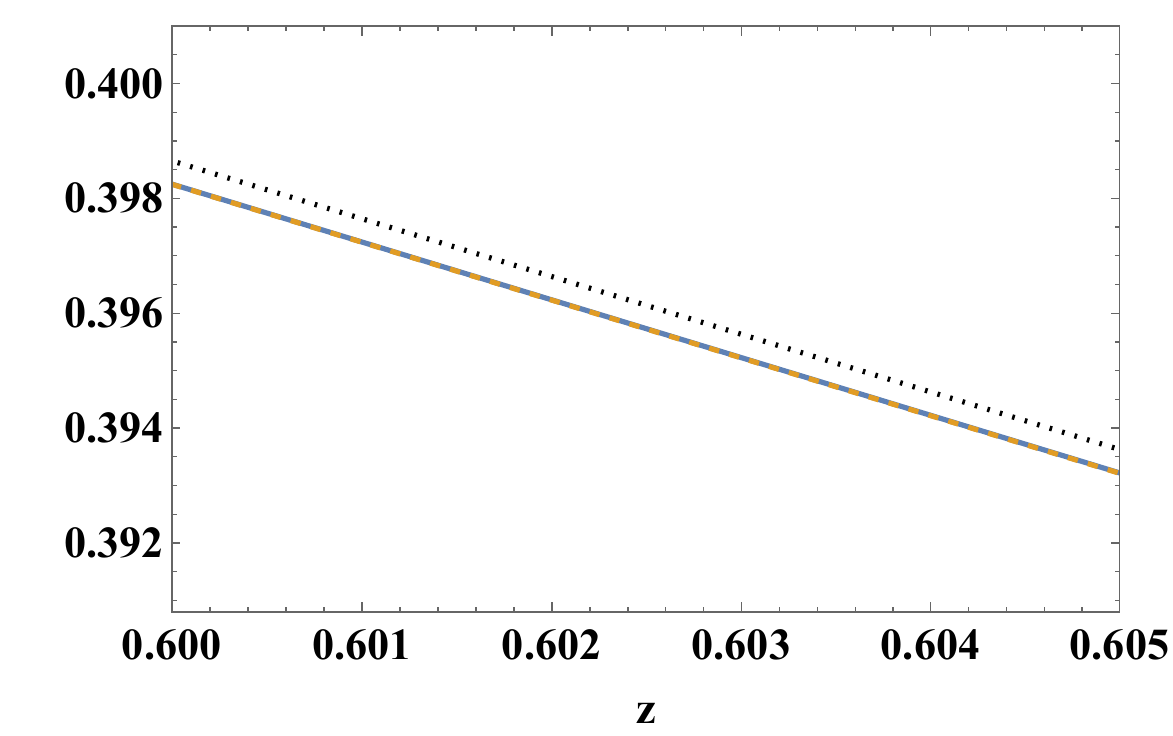}}
\subfloat[Case of $\alpha=3.5$]{\includegraphics[scale=0.27]{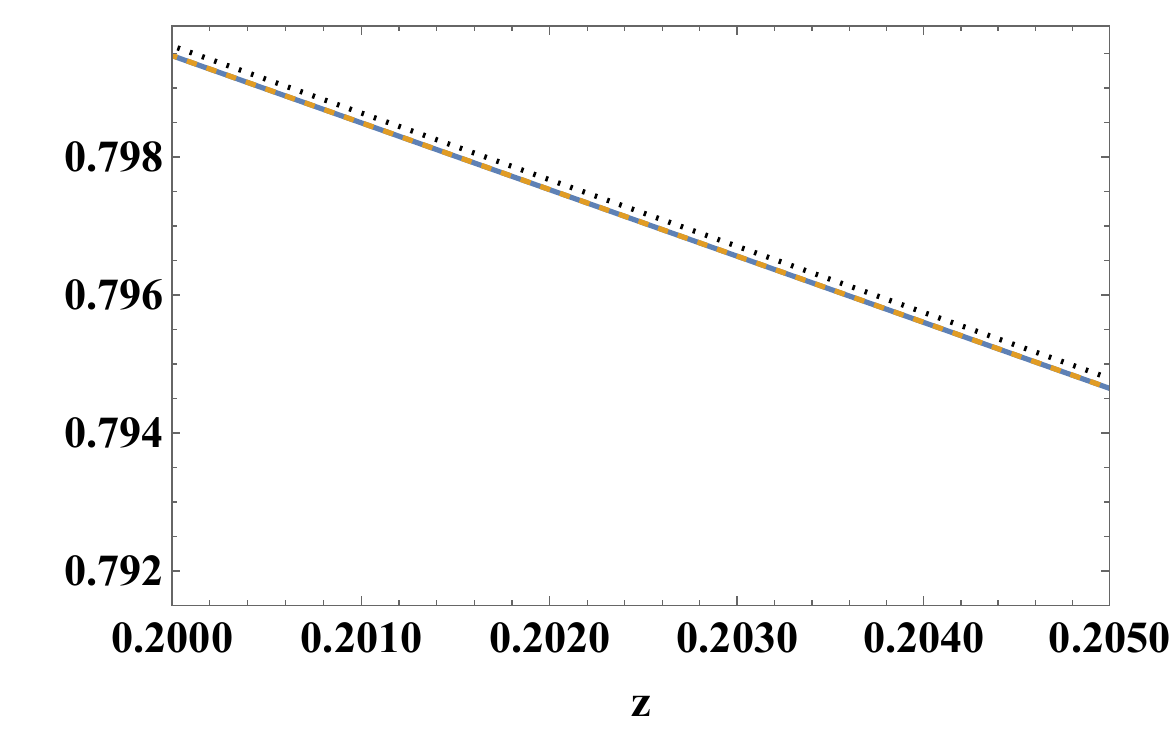}}
\caption{The comparison of metric solutions for hairy black holes and the Schwarzschild black hole. When $\alpha = -0.8$, hairy black holes exhibit a slight departure from the Schwarzschild black holes which shows the isotropic property ($A/B=1$) through the radial distance of $z$. This deviation, though subtle, leads to noticeably distinct behavior in their QNMs. Conversely, for $\alpha = 3.5$, the hairy black hole solutions show a much weaker deviations from the Schwarzschild metric across the radial distance.}
	\label{fig:SM001sub}
\end{figure}

To obtain the solutions, we impose the boundary value of $\varphi_h$ for given values of $\alpha$ and $\lambda$ and numerically solve the equations of motion (\ref{eq:SMeom1})-(\ref{eq:SMeom4}) from the near horizon ($z=z_h - \epsilon$) to infinity ($z=\epsilon$). Here, $A_h$ is determined by the boundary conditions at infinity. To perform numerical calculation, we fixed $z_h=1$ and $\epsilon=10^{-50}$ and used ``NDSolve" implemented in Mathematica with precision up to $300$. High precision and accuracy in constructing the background solution are crucial for reliable computation of quasinormal frequencies. This is because numerical errors in the background solution directly propagate to the calculated frequencies. As mentioned in \cite{Latosh:2023cxm}, numerical solutions were not found in the entire valid parameter space but found beyond the certain value of $\alpha$, which is denoted as $\alpha_{\textrm{sol.}}$ in Figure~\ref{fig:regularCD}. The absence of solutions in the region $\alpha < \alpha_{\textrm{sol.}}$ is discussed in \cite{Latosh:2023cxm}. We present the numerically generated hairy black hole solutions in the symmetric phase for two different scalar field values at the horizon ($\varphi_h$). Given the value of $\lambda = 0.1$, Figure~\ref{fig:SM001} shows the numerical solutions for the metric and scalar field with $\varphi_h = 0.01$, comparing with the Schwarzschil black holes. To highlight the contrasting behavior of the solutions, Figure~\ref{fig:SM001sub} zooms in on a smaller region of the $z$-axis. This magnified view clearly reveals  a significantly more pronounced anisotropy ($A/B \neq 1$) in the metric for large negative $\alpha$ values compared to positive $\alpha$ values.

\subsection{Scalar Field Perturbation}

We consider a linearized perturbation of the complex scalar field around the background solution
\begin{align}
\varphi(z) \sim \varphi(z)+ \delta \varphi(v,z,\theta,\phi),
\end{align}
so the linearized equation becomes
\begin{align}
\bigg(\nabla_{\alpha} \nabla^{\alpha}  + f_{\varphi^* \varphi}\, \mathcal{G} \bigg) \delta \varphi (z) + f_{\varphi^* \varphi} \, \mathcal{G} \delta \varphi^*(z) = 0. \label{eq:PerbEqvz}
\end{align}
Employing the following ansatz for scalar field perturbation
\begin{align}
\delta \varphi (v,z,\theta,\phi) = \sum_{l,m} \Phi (z) Y_{lm}(\theta, \phi ) e^{-i \omega v}, 
\end{align}
and using
\begin{align}
\frac{1}{Y} \frac{1}{\sin \theta} \frac{\partial}{\partial \theta} \bigg( \sin \theta \frac{\partial Y}{\partial \theta} \bigg) + \frac{1}{Y} \frac{1}{\sin^2 \theta} \frac{\partial^2 Y}{\partial \phi^2}  = -l(l+1) =-\gamma, \label{eq:sphericalY}
\end{align}
the linearized scalar field equation (\ref{eq:PerbEqvz}) is written as
\begin{align}
&z^2 A(z) B(z) \Phi''(z) + \frac{1}{2}  \bigg( z^2 B(z) A'(z)+z^2 A(z) B'(z)+4 i \omega  B(z) \sqrt{\frac{A(z)}{B(z)}} \bigg) \Phi'(z) \nonumber\\
&+ \frac{1}{z^2} \bigg( A(z) \left( \mathcal{G} \tilde{f}_{\varphi^* \varphi}(\varphi)-\gamma  z^2\right)-2 i \omega  z B(z) \sqrt{\frac{A(z)}{B(z)}} \bigg) \Phi(z) = 0 \label{eq:PerbEq}
\end{align}
and 
\begin{align}
\tilde{f}_{\varphi^* \varphi} =   \alpha - 6 \lambda \varphi_1^2. 
\end{align}
Let us investigate the asymptotic solution of (\ref{eq:PerbEq}) at both boundaries. Near the horizon, the perturbation equation is expanded as
\begin{align}
&z^3 \left(z-z_h\right) \Phi''(z) + \left[ z^3-  \frac{2 i \omega z}{\sqrt{A_h B_h}}\right] \Phi'(z) \nonumber\\
&+ \bigg[\frac{\gamma}{B_h} z -4 z^4 \left(B_h \left(3 z-2 z_h\right)+2\right) (f''\left(\varphi _h\right) + \varphi _{h,1} \left(z_h-z\right) f^{(3)}\left(\varphi _h\right))+\frac{2 i \omega}{\sqrt{A_h B_h}}\bigg] \Phi(z) = 0,
\end{align}
and plugging $\Phi(z) \sim (z_h -z)^p = \epsilon^p$ into above it expands
\begin{align}
\epsilon ^p \left[\frac{p}{\epsilon } \left(-p z_h^2+  \frac{2 i \omega}{\sqrt{A_h B_h}}\right)+\frac{1}{z_h}   \frac{2 i \omega}{\sqrt{A_h B_h}}-4 z_h^3 \left(B_h z_h+2\right) f''\left(\varphi _h\right)+\frac{\gamma}{B_h}+2 p^2 z_h\right] \sim 0. 
\end{align}
For the equation not to be singular, the first term should vanish. This requires $p$ to take values:
\begin{align}
p=0 \qquad \textrm{or} \qquad p=\frac{1}{z_h^2}   \frac{2 i \omega}{\sqrt{A_h B_h}} \label{eq:NHp}
\end{align}
which yield
\begin{align}
&\delta \varphi(v,z) = e^{- i \omega v} \textrm{const.}  \\
\textrm{or} \qquad &\delta \varphi(v,z) = e^{- i \omega v}(z_h-z)^{\frac{2 i \omega}{z_h^2 \sqrt{A_h B_h}}} = e^{-i \omega \big(v - \frac{2}{z_h^2 \sqrt{A_h B_h}} \log(z_h - z) \big)} . \label{eq:NHsol2}
\end{align}
The first solution corresponds to the incoming mode. The second one is written as
\begin{align}
\delta \varphi(v,z) = e^{-i \omega (v - 2 z_* + 2 c)} = e^{-i \omega (u + 2 c)} ,
\end{align} 
where the outgoing null vector $u = t - z_*$ is used and $z_*$ is expanded near the horizon such as
\begin{align}
z_* \simeq \frac{1}{z_h^2 \sqrt{A_h B_h}} (1 + i \pi - \log(z_h) + \log(z_h-z)) = \frac{1}{z_h^2 \sqrt{A_h B_h}} \log(z_h-z) + c ,
\end{align}
where $c$ indicates the first three terms in the second equality. This indicates that the second solution in (\ref{eq:NHp}) corresponds to the outgoing mode. Also, as shown in (\ref{eq:NHsol2}), the phase of this outgoing solution diverges and oscillates very fast when $z$ approaches $z_h$ in $(v,z)$-coordinate. Due to the limitations of numerical calculations on a grid, the rapid oscillations of the outgoing modes are not captured. Consequently, we can only impose the incoming boundary condition at the horizon, effectively excluding the outgoing modes from the numerical calculation. 

Near infinity the perturbation equation (\ref{eq:PerbEq}) is expanded
	\begin{align}
	\Phi ''(z) +\frac{ 2 i \omega}{z^2}\Phi '(z)  -\frac{2 i \omega }{z^3} \Phi (z) \simeq 0 , \label{eq:PerbEqIF}
	\end{align}
	and the solution takes the following form
	\begin{align}
	\Phi (z) = c_1 z e^{\frac{2 i \omega }{z}}-\frac{i c_2}{2 \omega} z. 
	\end{align}
To rigorously check this solution, we generalize the form to be
	\begin{align}
	\Phi (z) = c_1 z^p e^{\frac{2 i \omega }{z}}-\frac{i c_2}{2 \omega} z^p. 
	\end{align}
and expand (\ref{eq:PerbEqIF}) near $z \sim 0$. Then we obtain 
		\begin{align}
		&\frac{z^{p-3} e^{\frac{2 i \omega }{z}}}{A_1 z+1}  \bigg[A_1 \left(p^2 z^2-4 i p \omega  z+2 i \omega  (z+2 i \omega )\right)+p^2 z-p (z+2 i \omega )+2 i \omega -\gamma  z \bigg]c_1 \nonumber\\
		&-\frac{i}{2 \omega} \frac{z^{p-3} }{A_1 z+1} \bigg[A_1 p^2 z^2+p^2 z+2 i p \omega -p z-2 i \omega -\gamma  z \bigg]c_2 \simeq 0. 
		\end{align}
Taking $z \rightarrow 0$ limit, the first and second line require $p =1+2 i A_1 \omega$ and $p=1$ respectively and they yield
\begin{align}
\delta \varphi(v,z) \sim c_1 e^{- i \omega v} e^{\frac{2 i \omega }{z}} (z^{1+2 i A_1 \omega} + \cdots) + c_2 e^{- i \omega v} (z + \cdots).
\end{align}
The second term having $c_2$ describes the incoming mode and is normalisable. The first term having $c_1$ is rewritten as 
\begin{align}
\delta \varphi(v,z) = e^{- i \omega (v - \frac{2}{z} - 2 A_1 \log z)} + \cdots = e^{- i \omega (v - 2 z_*)} + \cdots = e^{- i \omega u} + \cdots ,
\end{align}
where $u = t - z_*$ is an outgoing null vector and $z_*$ in (\ref{eq:vzs}) is expanded in $z \rightarrow 0$ limit 
\begin{align}
z_* = - \int \frac{1}{z^2} \frac{1}{\sqrt{AB}} \rmd z \simeq \frac{1}{z} + A_1 \log z.
\end{align}
This indicates that the first term describes the outgoing mode but is non-normalizable and oscillates more rapidly as $z$ approaches $0$ in $(v,z)$-coordinate. Here, we want to impose the outgoing boundary condition. This situation is problematic. To resolve this we redefine $\Phi(z)$ as
\begin{align}
			\Phi(z) \sim e^{\frac{2 i \omega}{z}}z^{2 i A_1 \omega} \tilde{\Phi} (z) 
\end{align} 
and the new field $\tilde{\Phi} (z)$ becomes
\begin{align}
\tilde{\Phi} (z) = c_1 z - \frac{i c_2}{2 \omega} e^{- \frac{2 i \omega}{z}}z^{1- 2 i A_1 \omega}.
\end{align}
Then, the incoming mode becomes non-normalizable and is not captured by the numerical calculation due to the fast oscillation, while the outgoing mode becomes normalizable and behaves smoothly. Rewriting the perturbation equation (\ref{eq:PerbEq}) in terms of $\tilde{\Phi} (z)$, it yields
\begin{align}
			&\tilde{\Phi }''(z) + \bigg[ \frac{2 i \omega}{z^2}  \left(\frac{1}{\sqrt{A(z) B(z)}}+2 (A_1 z-1) \right)+ \frac{1}{2} \left(\frac{A'(z)}{A(z)}+\frac{B'(z)}{B(z)}\right)
			\bigg]\tilde{\Phi }'(z) \nonumber\\
			&+ \bigg[\frac{R_{\text{GB}} \tilde{f}_{\varphi^* \varphi}-\gamma  z^2}{z^4 B(z)}   -\frac{4 \omega^2 \left(A_1 z-1\right)}{z^4} \left(\frac{1}{\sqrt{A(z) B(z)}}+\left(A_1 z-1\right)\right) \nonumber\\
			&-\frac{i \omega}{z^3}  \left(2 A_1 z-4 -z \left(A_1 z-1\right) \left(\frac{A'(z)}{A(r)}+\frac{B'(z)}{B(r)}\right)+\frac{2}{\sqrt{A(z) B(z)}}\right)
			\bigg] \tilde{\Phi}(z)= 0. \label{eq:PerbEqRe}
\end{align}
By adopting the numerical method explained in Section 3.4, we will numerically calculate the quasinormal frequency $\omega$ in (\ref{eq:PerbEqRe}) and display their values in the following subsection. 

\subsection{Our Numerical Results}

\begin{figure}
    \centering
    \includegraphics[scale=0.36]{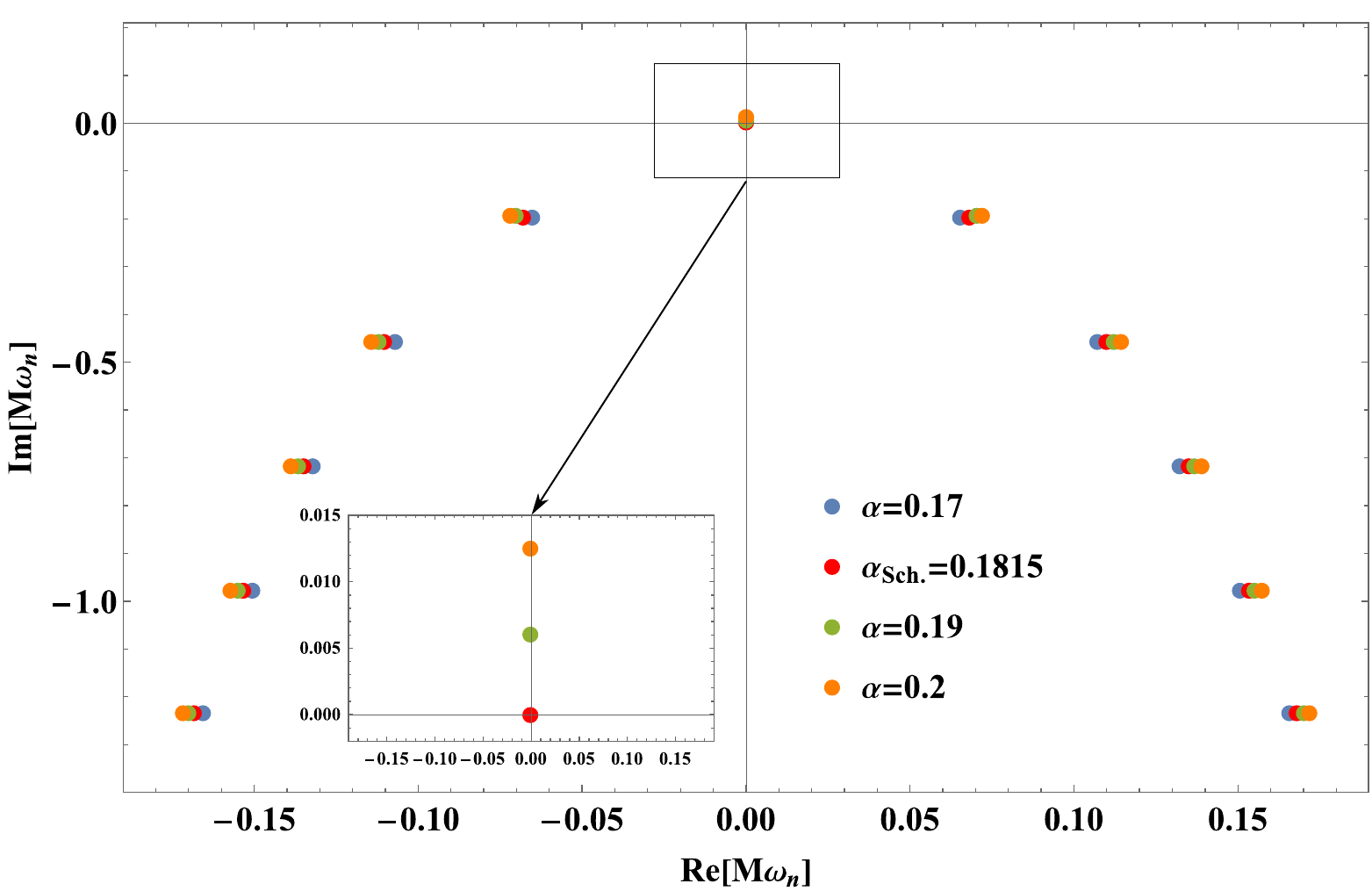} 
    \scriptsize{
    \begin{tabular}[b]{c|cc}\hline
        & Re[$M \omega_n$] & Im[$M \omega_n$] \\ \hline
      \multirow{4}{*}{$\alpha$=0.17} & $\pm$ 0.06528 & -0.19795 \\
       & $\pm$ 0.10718 & -0.45845 \\
       & $\pm$ 0.13196 & -0.71851 \\
       & $\pm$ 0.15043 & 0.97682 \\ \hline
       \multirow{5}{*}{$\alpha_{\textrm{Sch.}} \approx$ 0.1815} & 0 & 0.00002 \\
       & $\pm$ 0.06826 & -0.19618 \\ 
       & $\pm$ 0.11012 & -0.45776 \\
       & $\pm$ 0.13483 & -0.71820 \\
       & $\pm$ 0.15324 & -0.97669 \\ \hline
        \multirow{5}{*}{$\alpha$=0.19} & 0 & 0.00600 \\
       & $\pm$ 0.07013 & -0.19489 \\ 
       & $\pm$ 0.11208 & -0.45728 \\
       & $\pm$ 0.13677 & -0.71798 \\
       & $\pm$ 0.15513 & -0.97662 \\ \hline
          \multirow{5}{*}{$\alpha$=0.20} & 0 & 0.01254 \\
       & $\pm$ 0.07198 & -0.19340 \\ 
       & $\pm$ 0.11415 & -0.45675 \\
       & $\pm$ 0.13881 & -0.71775 \\
       & $\pm$ 0.15712 & -0.97654 \\ \hline
    \end{tabular}}
    \captionlistentry[table]{Sch}
    \captionsetup{labelformat=andtable}
    \caption{QNMs of Schwarzschild black holes for serval values $\alpha$} 
     \label{fig:alphaSch}
  \end{figure}

We first consider the Schwarzschild black hole, which is the solution with $\varphi=0$, and examine its stability according to the value of $\alpha$. By computing the quasinormal frequency, we found that a positive imaginary part of the frequency, Im$[\omega]$, emerges at a critical value, $\alpha_{\textrm{Sch.}} \approx 0.1815$ as shown in Figure \ref{fig:alphaSch}. This indicates that the corresponding system becomes unstable $\alpha \geq \alpha_{\textrm{Sch.}}$. Here, we want to argue that this instability triggers a transition from the Schwarzschild black hole to a hairy black hole, and the hairy black hole will be formed not in the symmetric phase but in the symmetry-broken phase. To support this argument, we show that the hairy black holes with $\alpha \geq \alpha_{\textrm{Sch.}}$ are unstable in the symmetric phase but stable in the symmetry-broken phase. This section focuses solely on the quasinormal frequencies of hairy black holes in the symmetric phase for various $\alpha$ values. The stability analysis of hairy black holes in the symmetry-broken phase will be addressed in Section 6.

\begin{figure}
	\begin{center}
	\includegraphics[scale=0.28]{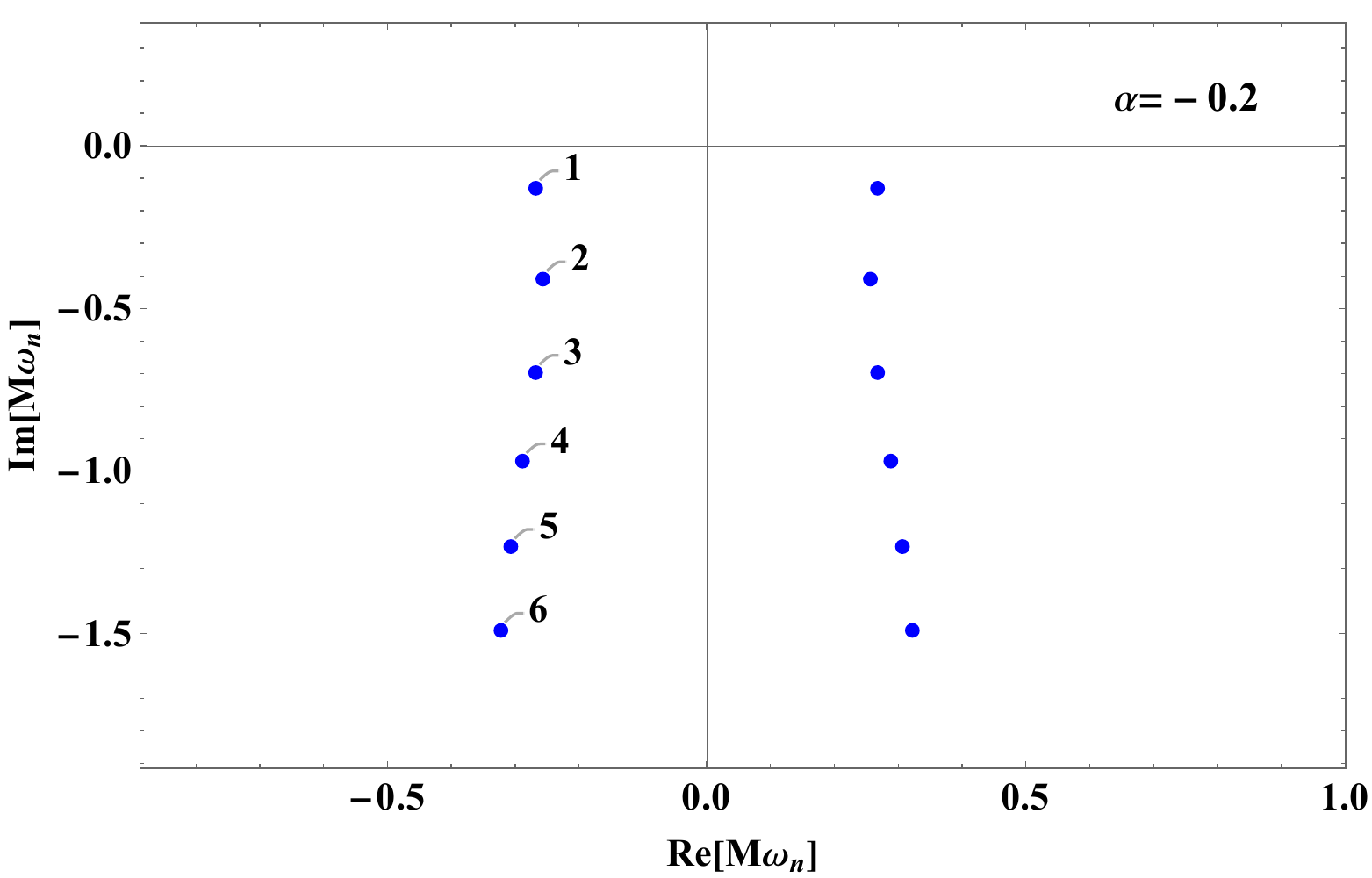} $\; \; \; \;$ \includegraphics[scale=0.28]{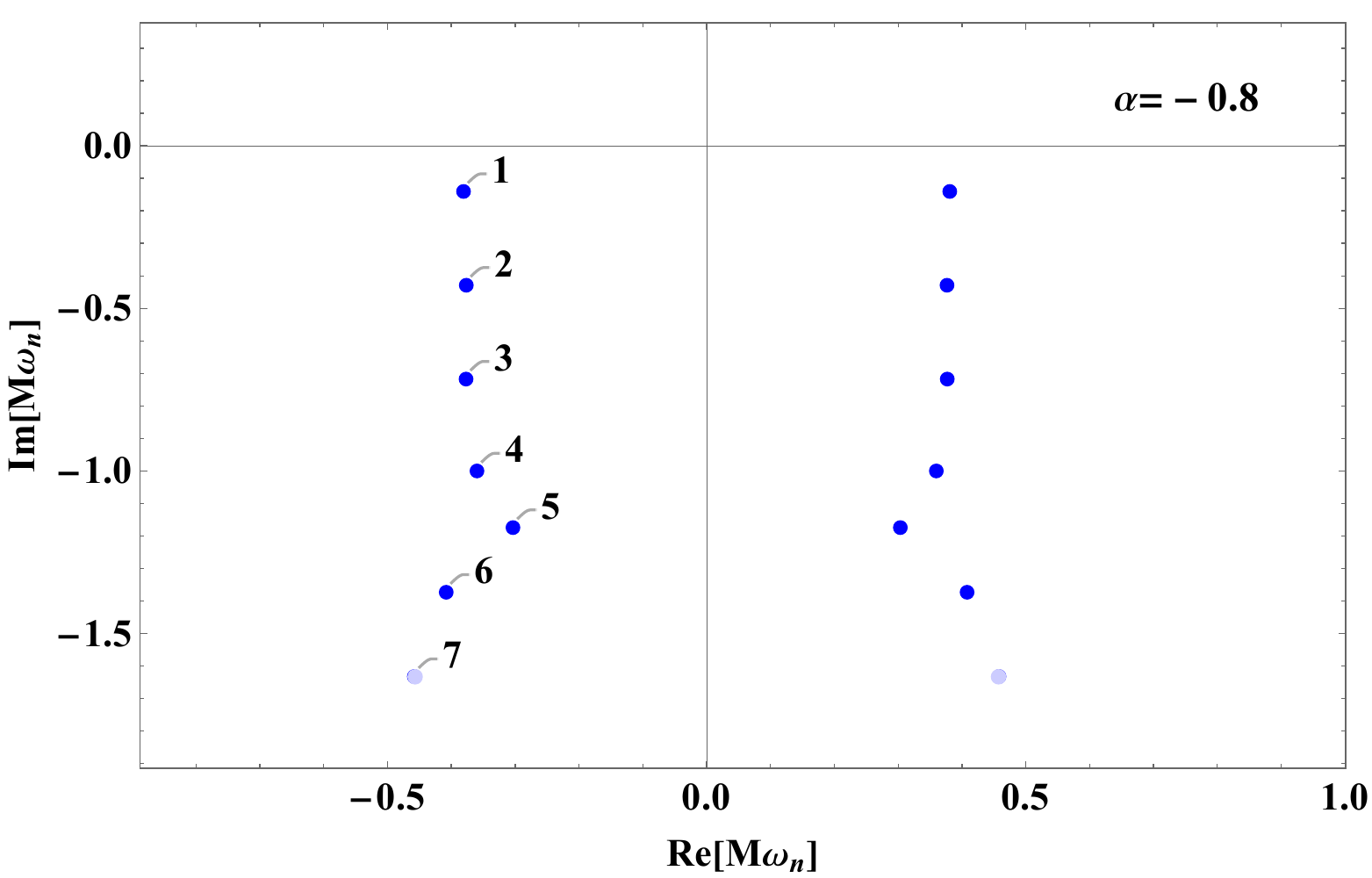}\\
	\includegraphics[scale=0.28]{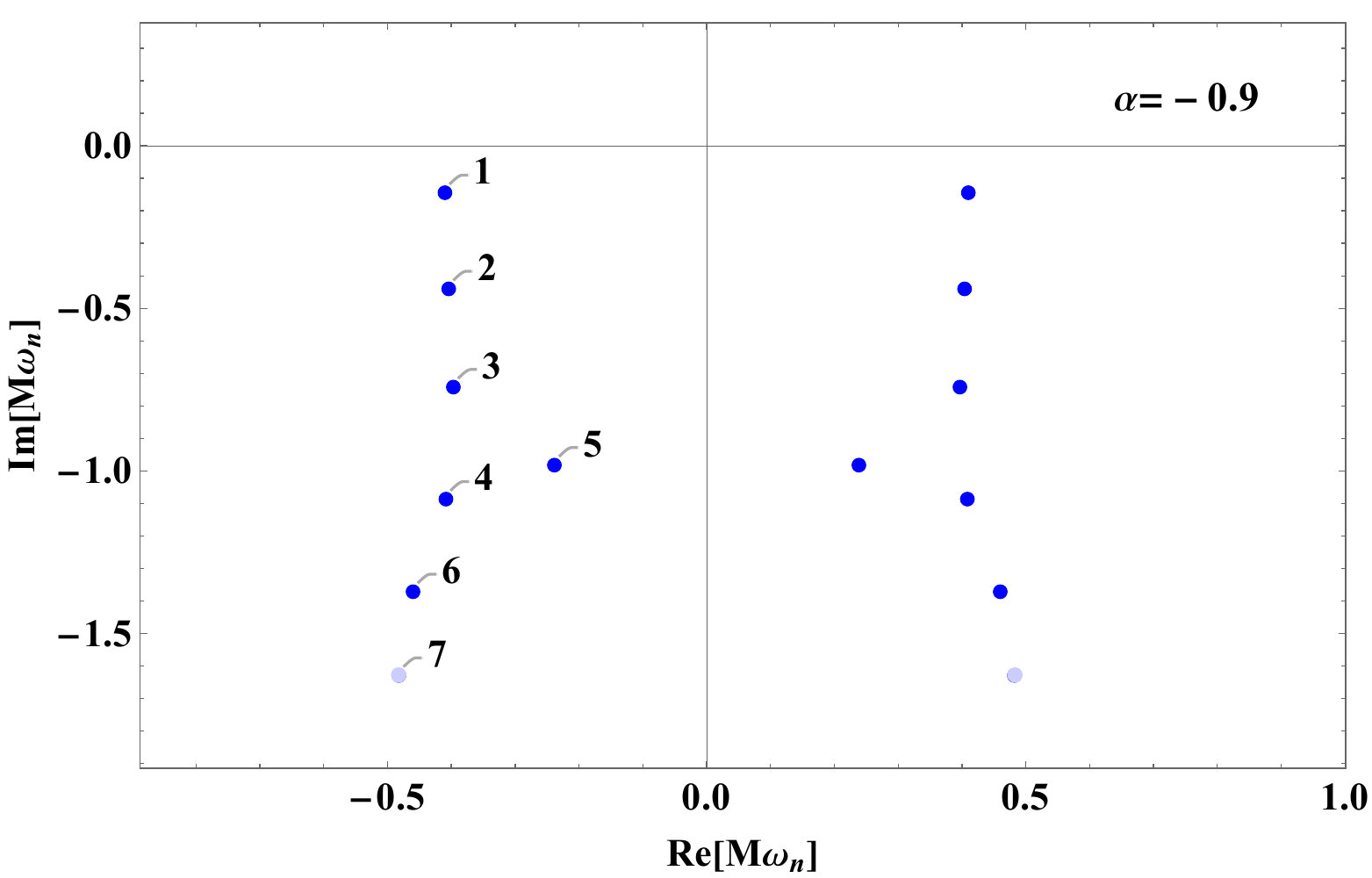} $\; \; \; \;$ \includegraphics[scale=0.28]{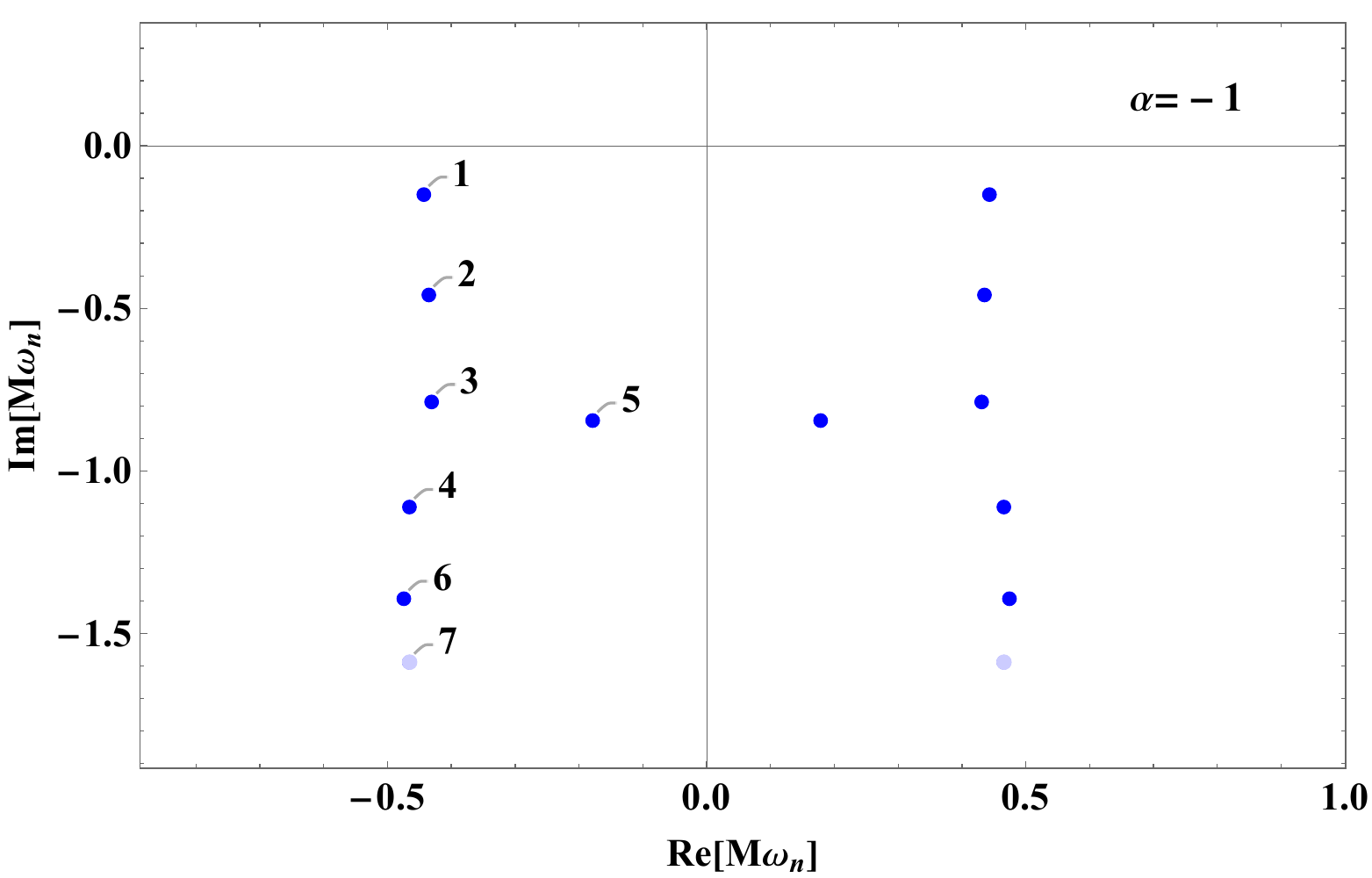}\\
	\includegraphics[scale=0.28]{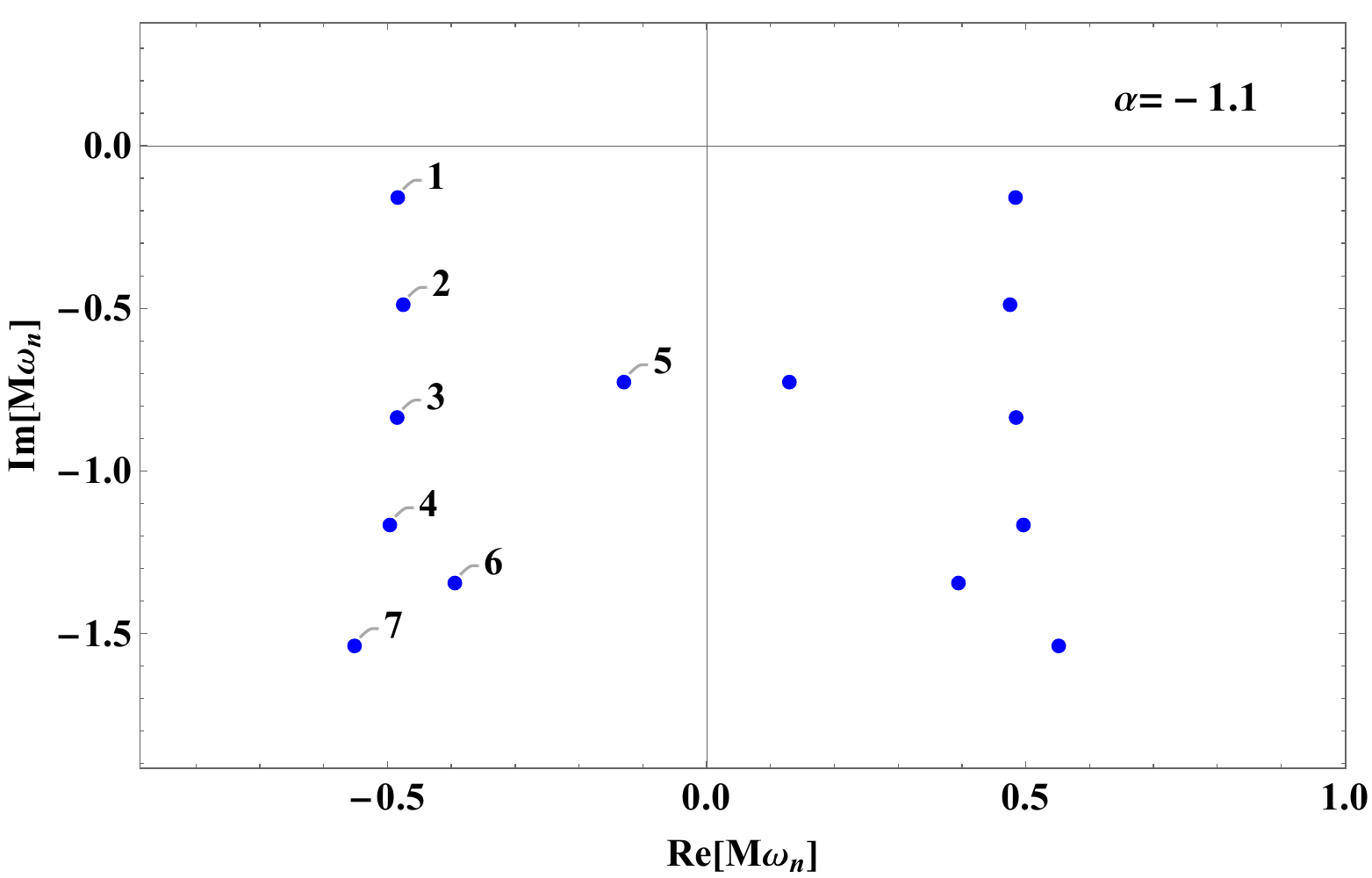} $\; \; \; \;$ \includegraphics[scale=0.28]{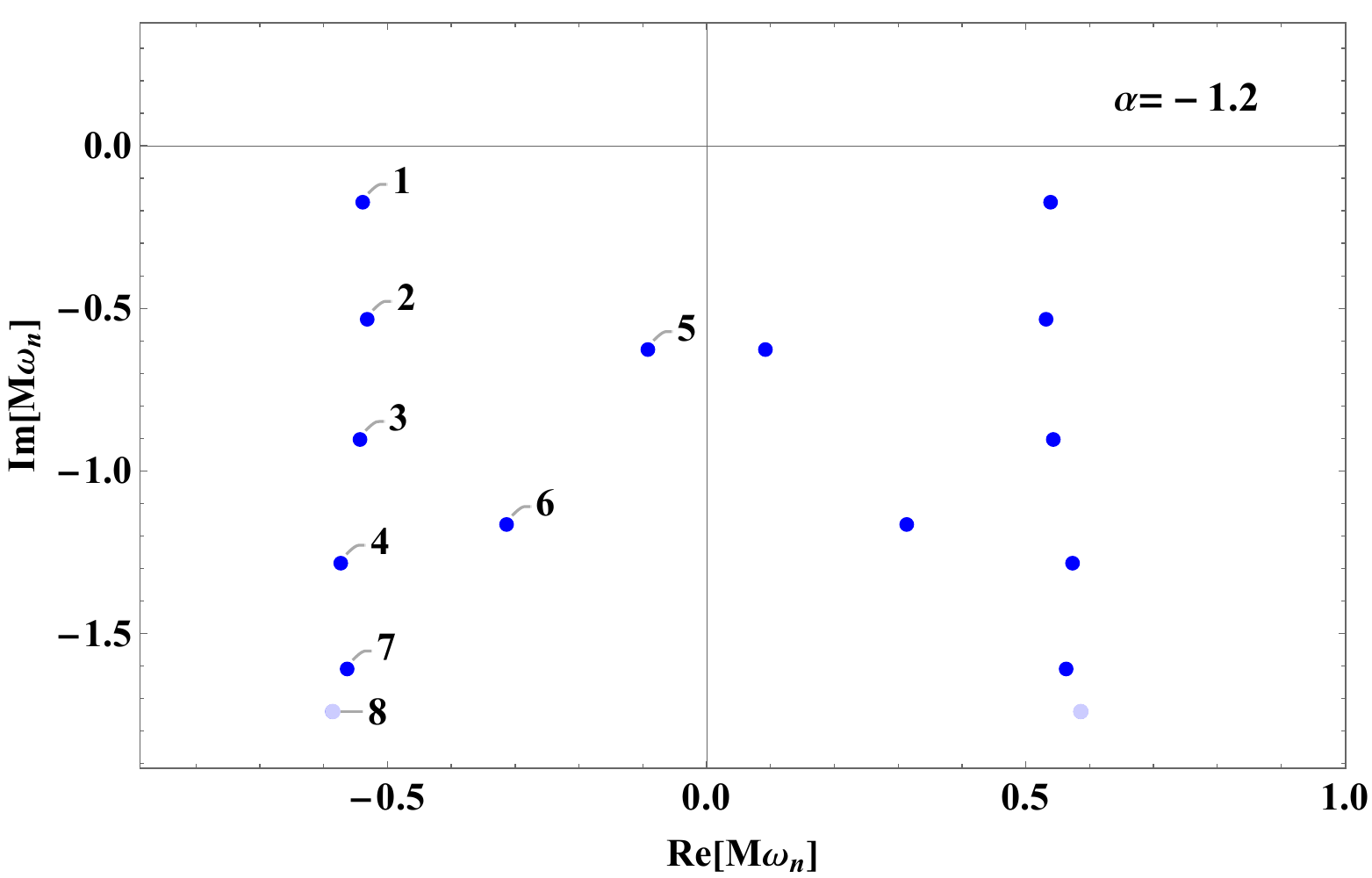}\\
	\includegraphics[scale=0.28]{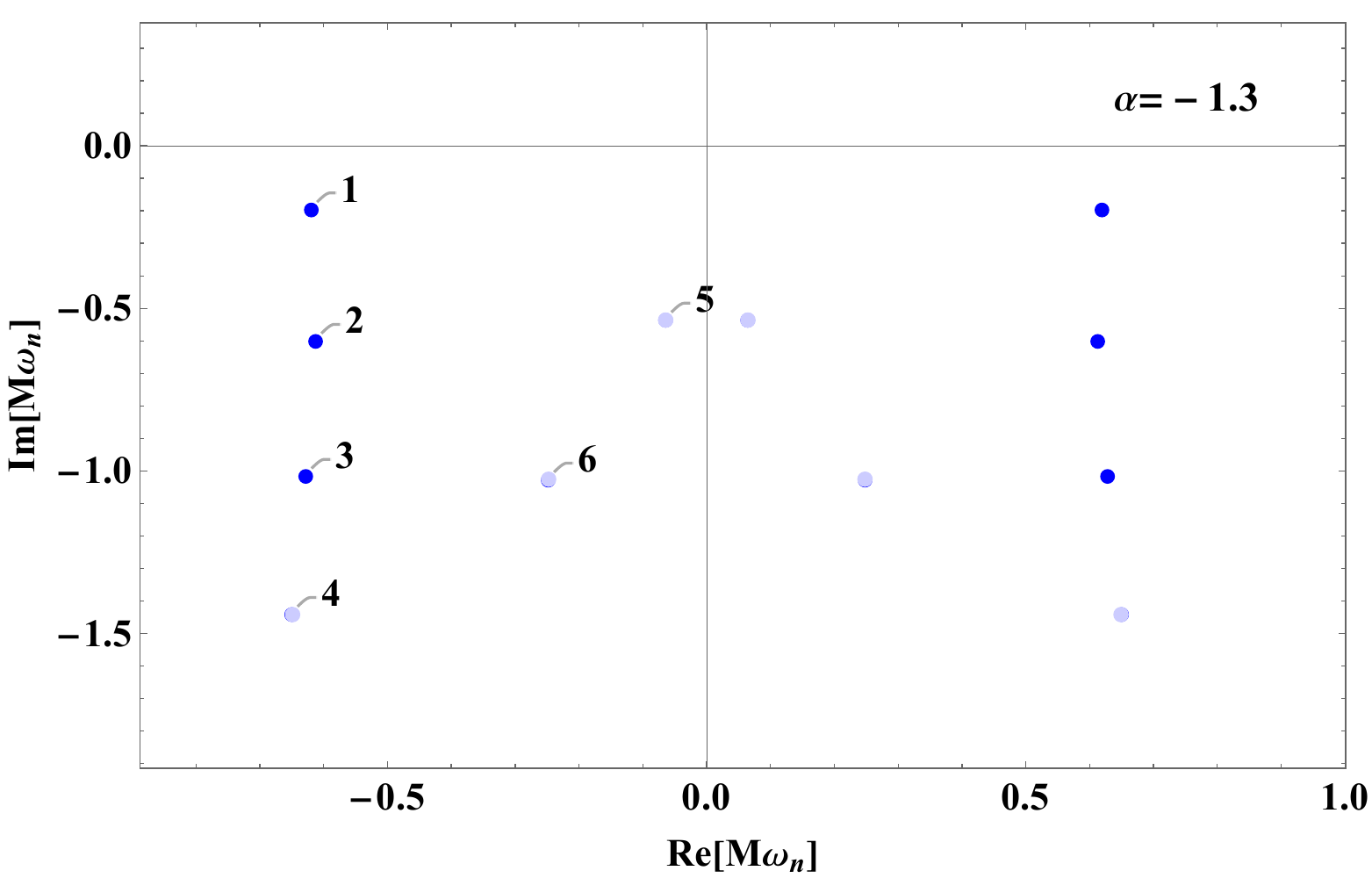} $\; \; \; \;$ \includegraphics[scale=0.28]{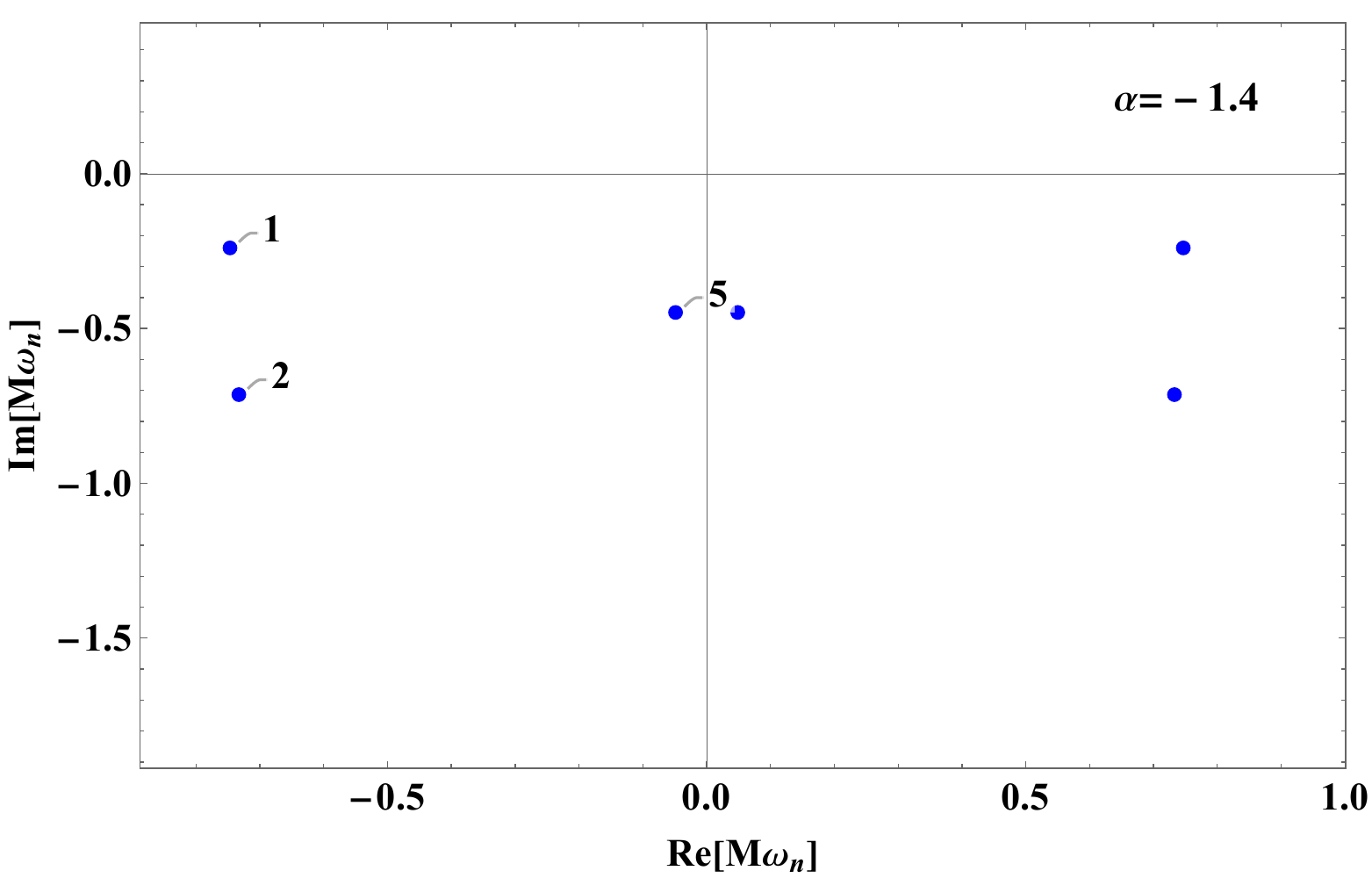}
	    \caption{QNMs with negative values of $\alpha$ for $\varphi_h = 0.01$. Blue dots are tolerable less than $10^{-3}$, whereas lighter blue dots have a tolerance of $10^{-2}$.}
    \label{fig:QNMvp001aN}
	\end{center}
\end{figure}

\begin{figure}
	\begin{center}
	\includegraphics[scale=0.28]{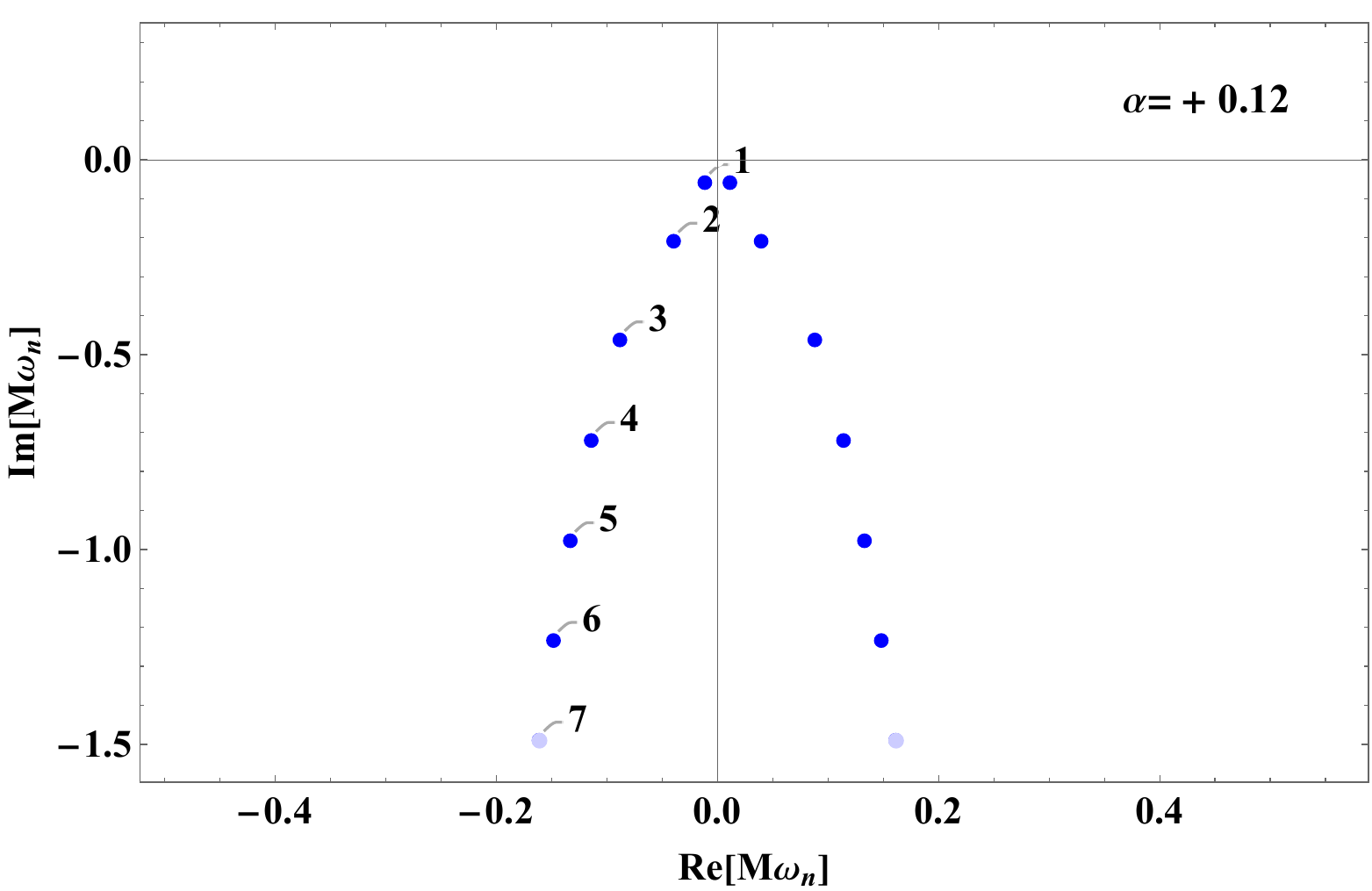} $\; \; \; \;$ \includegraphics[scale=0.28]{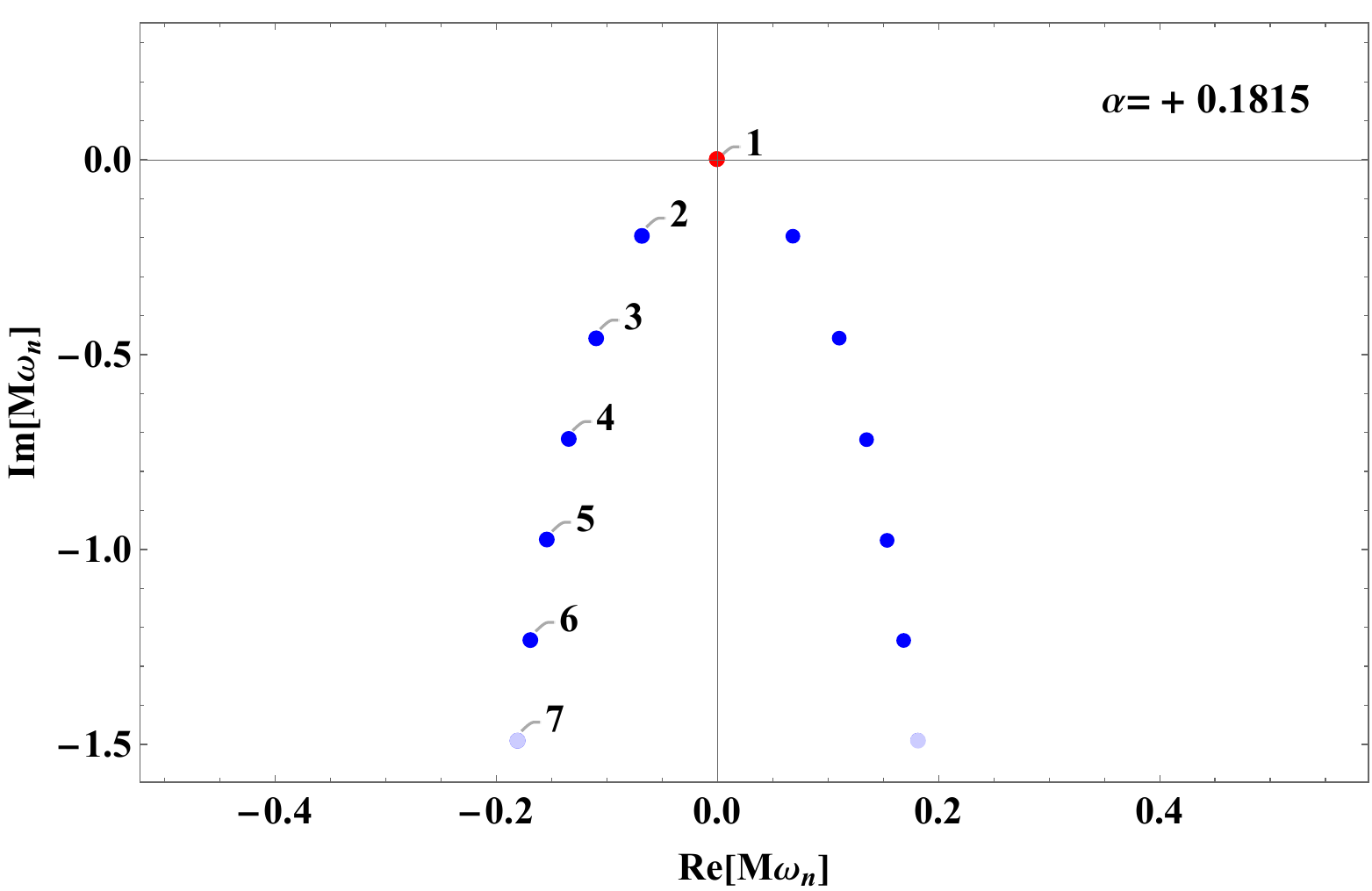}\\
	\includegraphics[scale=0.28]{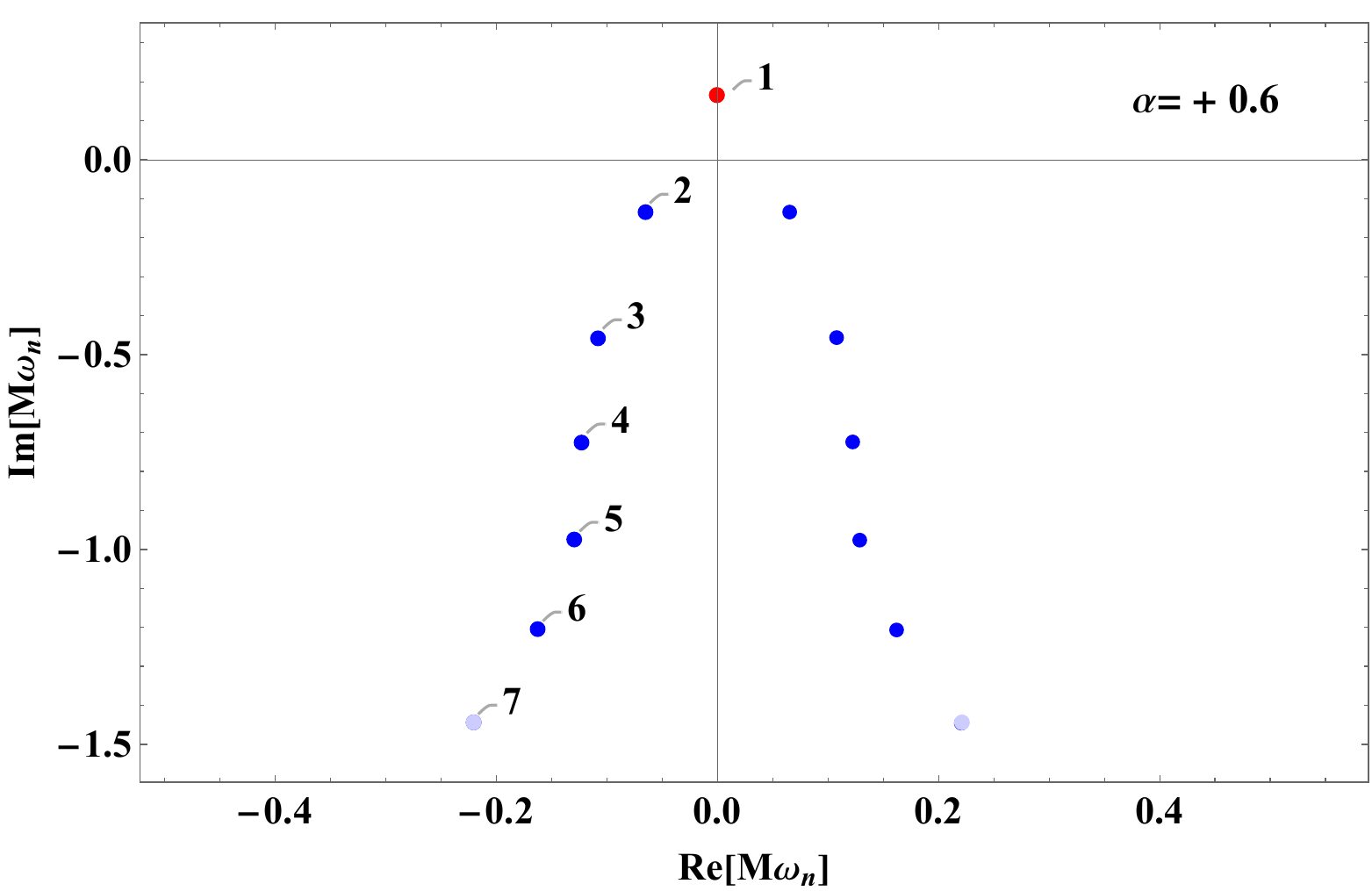} $\; \; \; \;$ \includegraphics[scale=0.28]{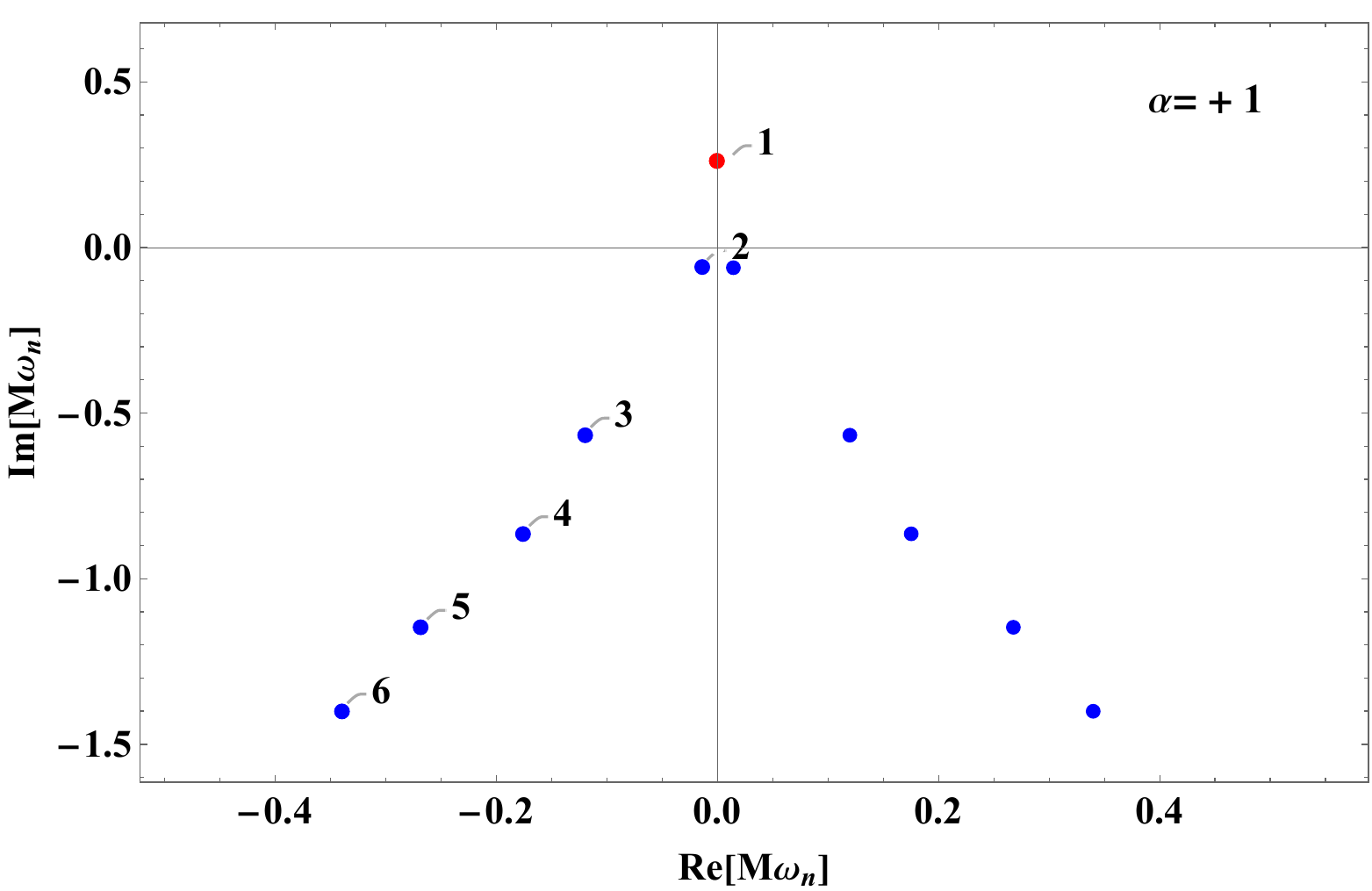}\\
	\includegraphics[scale=0.28]{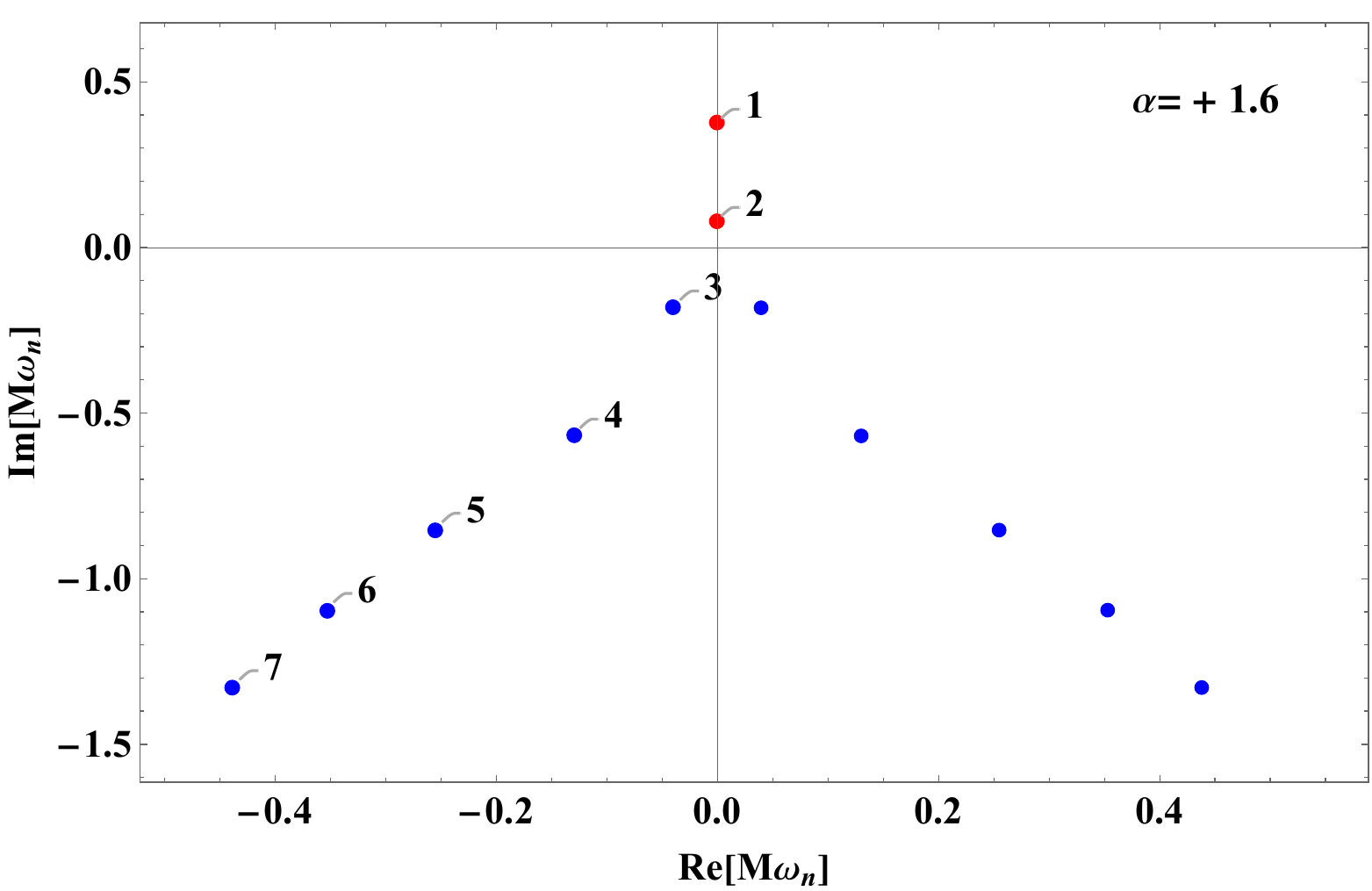} $\; \; \; \;$ \includegraphics[scale=0.28]{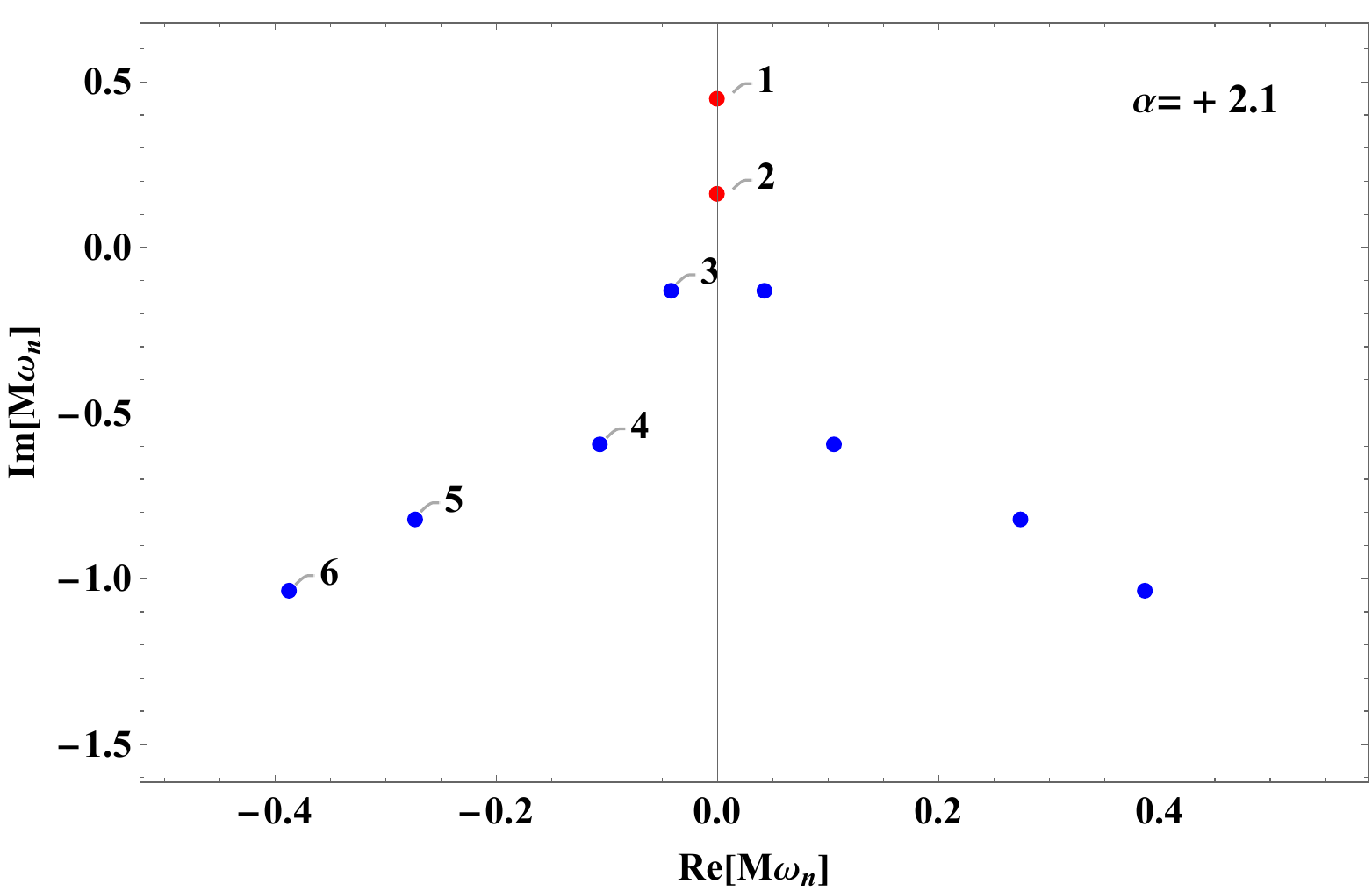}\\
	\includegraphics[scale=0.28]{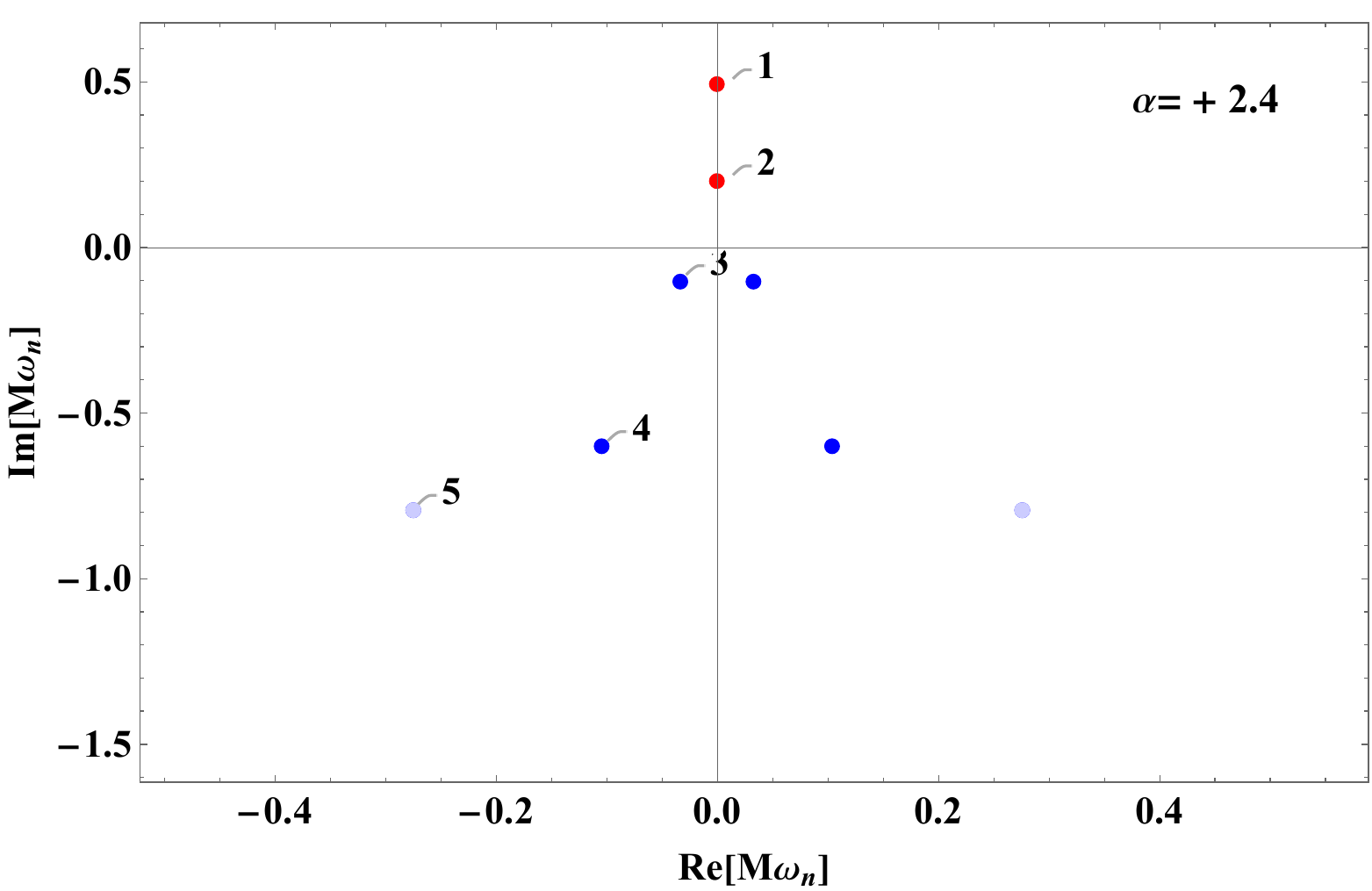} $\; \; \; \;$ \includegraphics[scale=0.28]{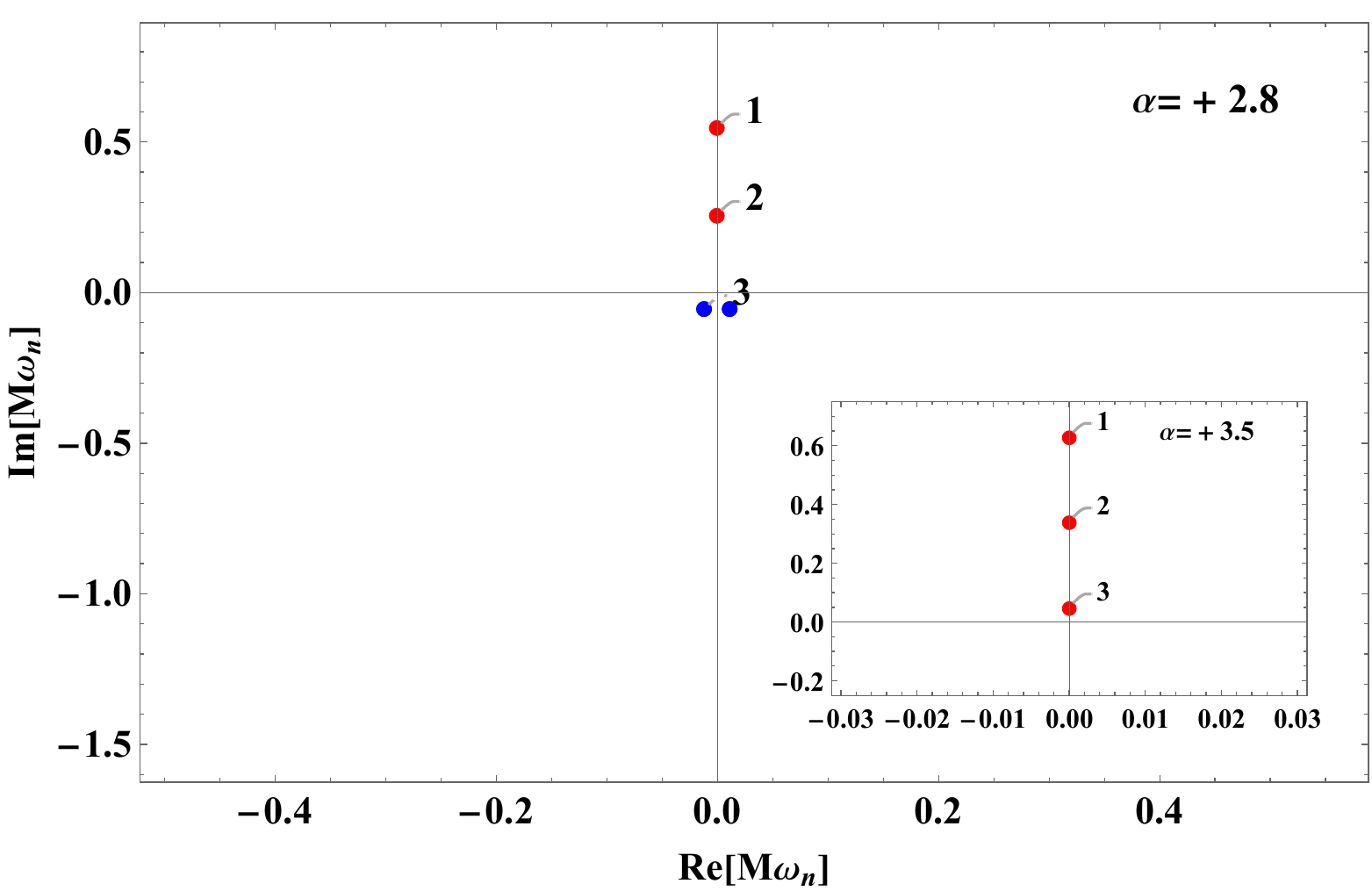}
	    \caption{QNMs with positive values of $\alpha$ for $\varphi_h = 0.01$. Blue dots are tolerable less than $10^{-3}$, whereas lighter blue dots have a tolerance of $10^{-2}$.}
    \label{fig:QNMvp001aP}
	\end{center}
\end{figure}

To compute the eigenvalues, we performed calculations with grids of size 100 and 140, resulting in 100 and 140 eigenvalues, respectively. To identify reliable eigenvalues, we considered only those for which both the real and imaginary parts agreed at least to the second decimal places (i.e. tolerance of $10^{-2}$) between the two grid sizes. Finally, we plotted the corresponding eigenfunctions associated with these converged eigenvalues to verify that they satisfy the boundary conditions. Figure~\ref{fig:QNMvp001aN} shows the quasinormal frequencies for $\alpha = -0.2, -0.8, -0.9, -1, -1.1, -1.2, -1.3$ and $-1.4$, generated on the background solution with $\varphi_h = 0.01$ for a fixed value of $\lambda = 0.1$. Figure~\ref{fig:QNMvp001aP} demonstrates the quasinormal frequencies for $\alpha = 0.12, 0.1815, 0.6, 1, 1.6, 2.1, 2.4$ and $2.8$ on the same background. The detailed numerical values for the QNMs presented in Figure~\ref{fig:QNMvp001aN} and~\ref{fig:QNMvp001aP} can be found in Appendix A.

Our results show that the imaginary parts of the QNM frequencies are all negative for the negative values of $\alpha$. In Figure~\ref{fig:QNMvp001aN}, the QNMs exhibit a well arranged line of spectrum at $\alpha = -0.2$. As $\alpha$ decreases towards more negative values, a progressive shift is observed in the higher-order QNMs. The fifth QNM exhibits this behavior first, moving inwards and upwards relative to the main QNMs' array. Subsequently, at larger negative $\alpha$ values, the sixth QNM starts displaying the same behavior. This sequence of inward and upward shifts with increasing negative $\alpha$ forms the additional array of the QNM spectrum inside the outer array of the QNMs. However, pushing $\alpha$ to even more negative values leads to a depletion of higher-frequency QNMs, resulting in a sparser spectrum at $\alpha = -1.4$. Figure~\ref{fig:SM001sub} reveals that the metric exhibits increasing anisotropy ($A/B \neq 1$) with more negative values of $\alpha$. This deviation from isotropy ($A/B=1$) is suspected to be a contributing factor to the emergence of the exotic (cases of $\alpha=-0.8$ to $-1.2$) and diminishing (cases of $\alpha = -1.3$ and $-1.4$) QNM spectrum observed at large negative $\alpha$ values.

\begin{figure}
    \centering
    \includegraphics[scale=0.3]{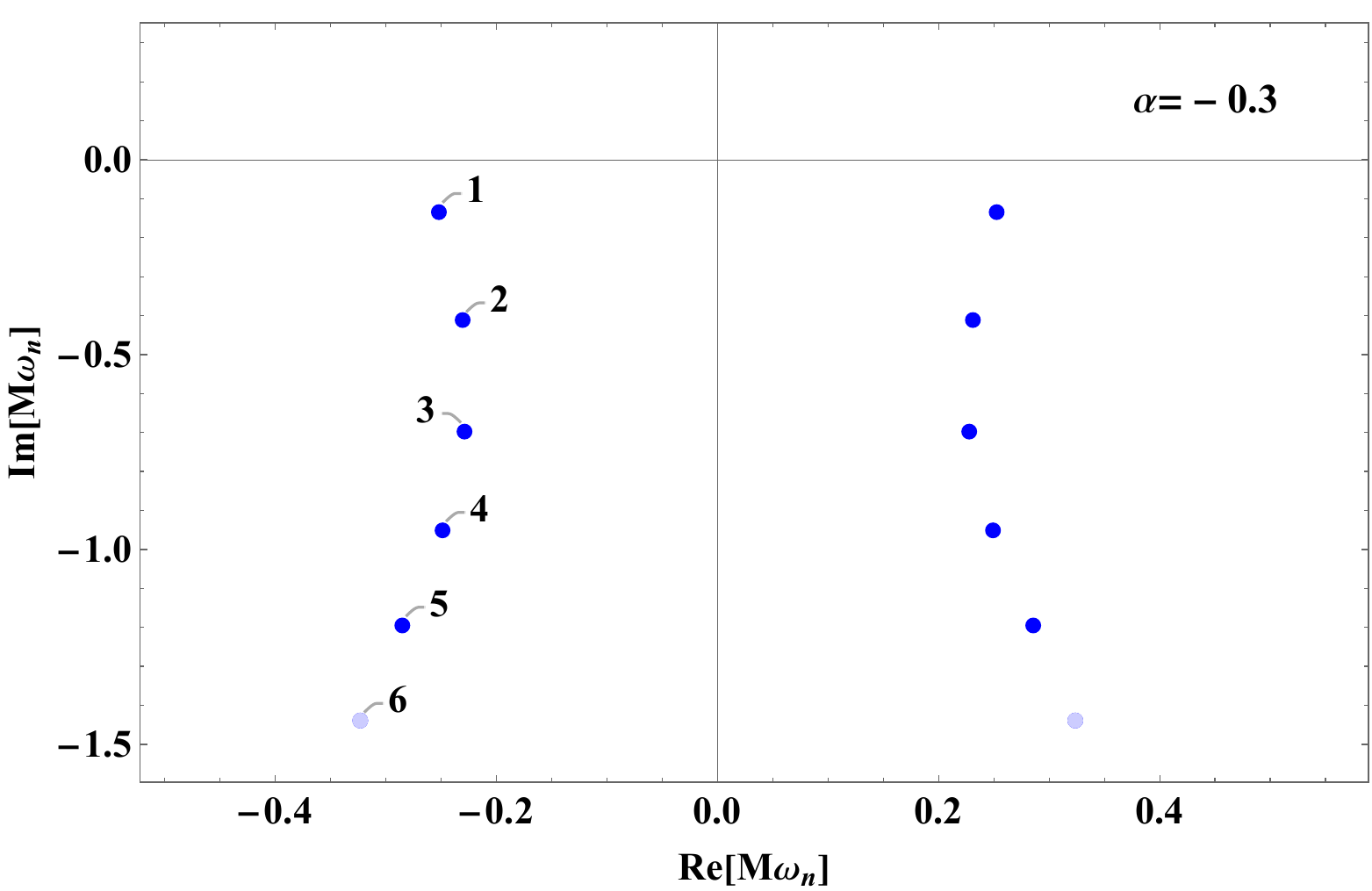}
    \qquad
    \begin{tabular}[b]{c cc}\hline
      & Re[$M\omega_n$] & Im[$M\omega_n$] \\ \hline
           1& $\pm 0.25244$ & $-0.13243$ \\
     2& $\pm 0.23104$ & $-0.41327$ \\
     3& $\pm 0.22792$ & $-0.69714$ \\
     4& $\pm 0.24934$ & $-0.95078$ \\
     5&$\pm 0.28562$ & $-1.19428$ \\
     6&$\pm 0.32\textcolor{gray}{339}$ & $-1.44\textcolor{gray}{088}$ \\ \hline
           \vspace{0.2cm}
    \end{tabular}
                      \vspace{0.1cm} \\
\includegraphics[scale=0.3]{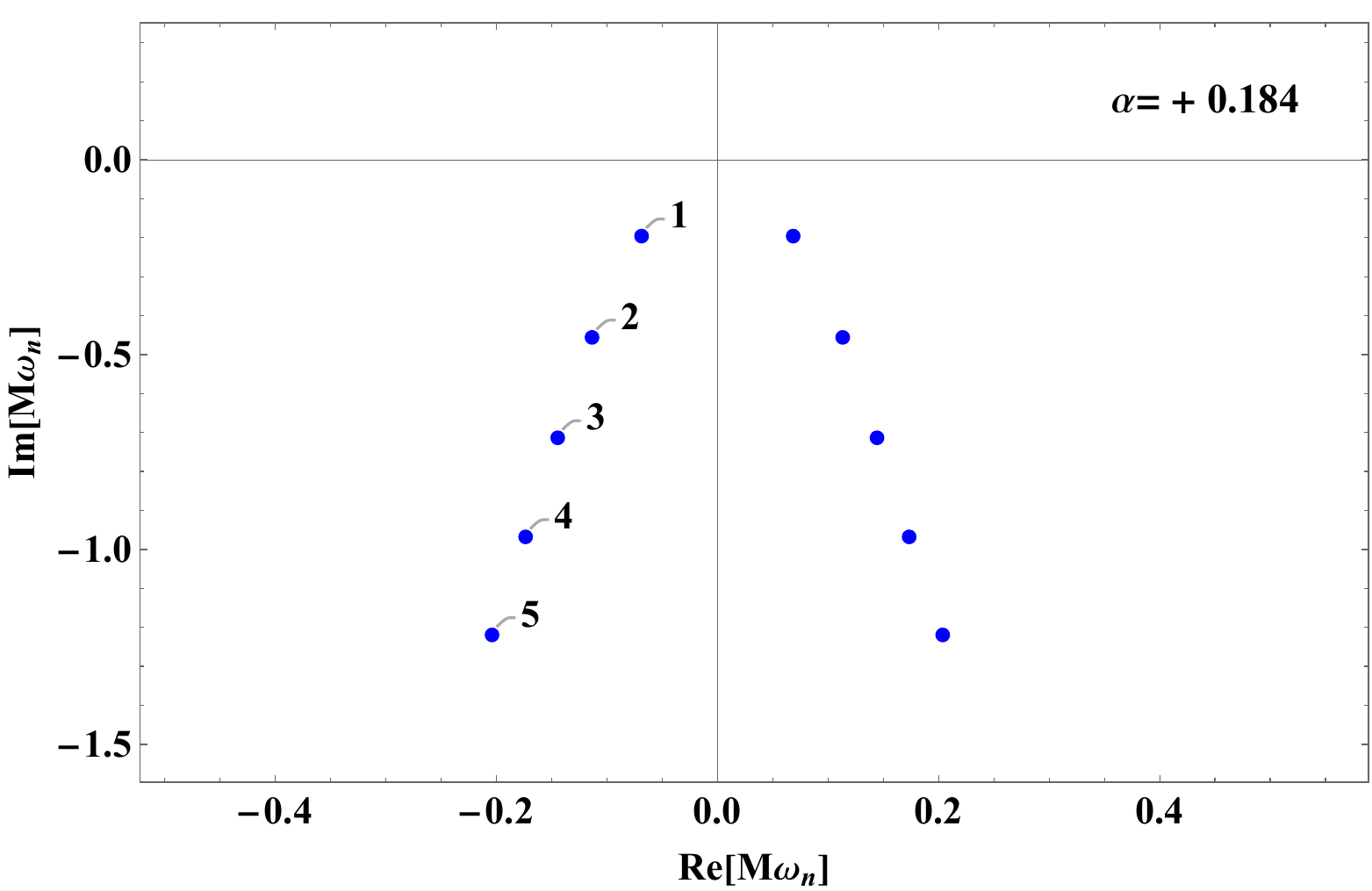}
    \qquad
    \begin{tabular}[b]{c cc}\hline
      & Re[$M\omega_n$] & Im[$M\omega_n$] \\ \hline
     1& $\pm 0.06855$ & $-0.19589$ \\
     2& $\pm 0.11334$ & $-0.45581$ \\
     3& $\pm 0.1443\textcolor{gray}{5}$ & $-0.7134\textcolor{gray}{4}$ \\
     4& $\pm 0.1734\textcolor{gray}{1}$ & $-0.9677\textcolor{gray}{1}$ \\
     5&$\pm 0.203\textcolor{gray}{76}$ & $-1.219\textcolor{gray}{34}$ \\ \hline
           \vspace{0.5cm}
    \end{tabular}
                      \vspace{0.1cm} \\
\includegraphics[scale=0.3]{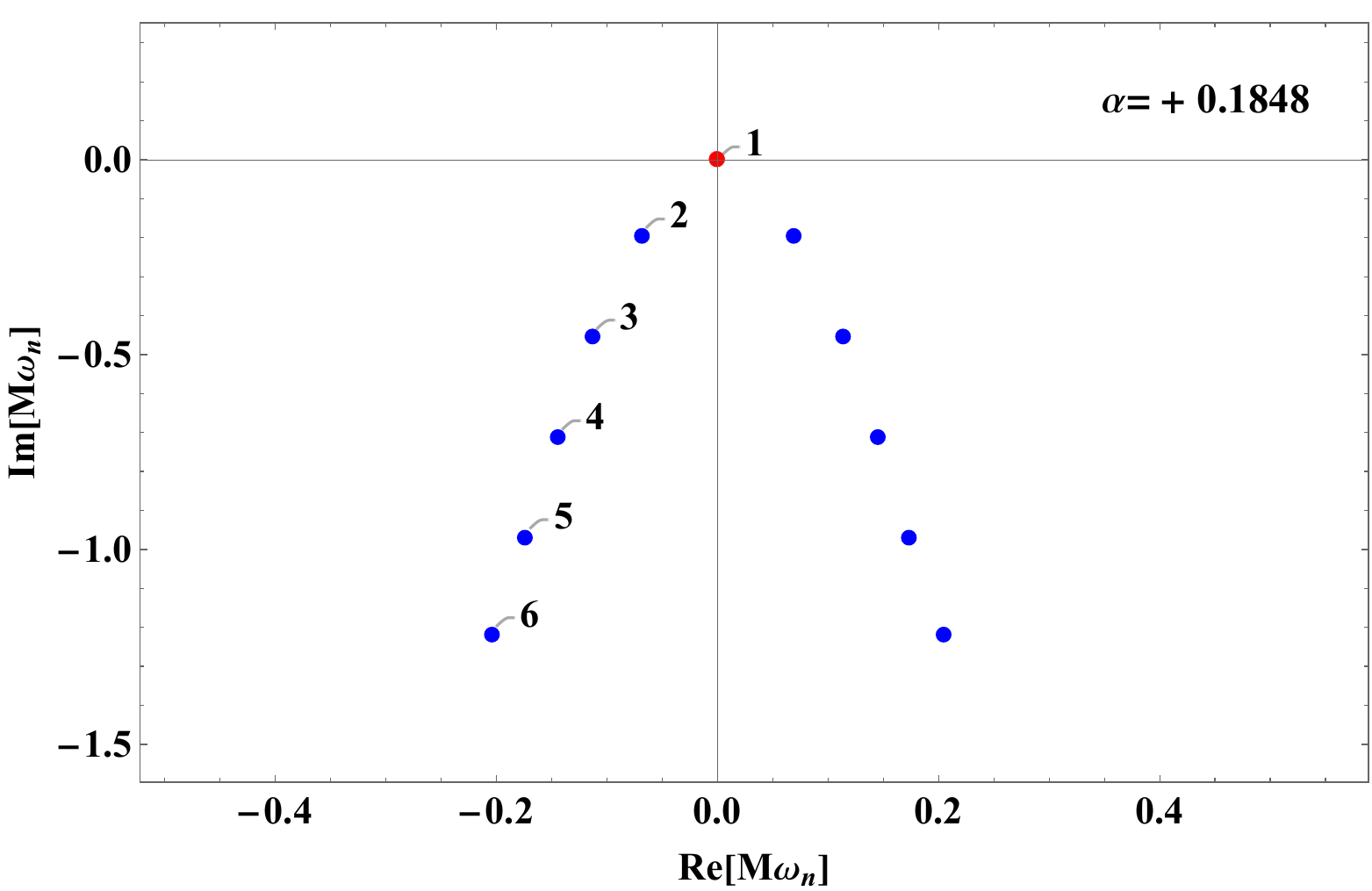}
    \qquad
    \begin{tabular}[b]{c cc}\hline
      & Re[$M\omega_n$] & Im[$M\omega_n$] \\ \hline
      1& $0 $ & $0.00015$ \\
      2& $\pm 0.06876$ & $-0.19576$ \\
      3& $\pm 0.11359$ & $-0.45573$ \\
      4& $\pm 0.1446\textcolor{gray}{7}$ & $-0.7133\textcolor{gray}{4}$ \\
      5& $\pm 0.173\textcolor{gray}{85}$ & $-0.967\textcolor{gray}{57}$ \\
      6& $\pm 0.204\textcolor{gray}{36}$ & $-1.219\textcolor{gray}{14}$ \\ \hline
      \vspace{0.4cm}
    \end{tabular}
                      \vspace{0.1cm} \\
\includegraphics[scale=0.3]{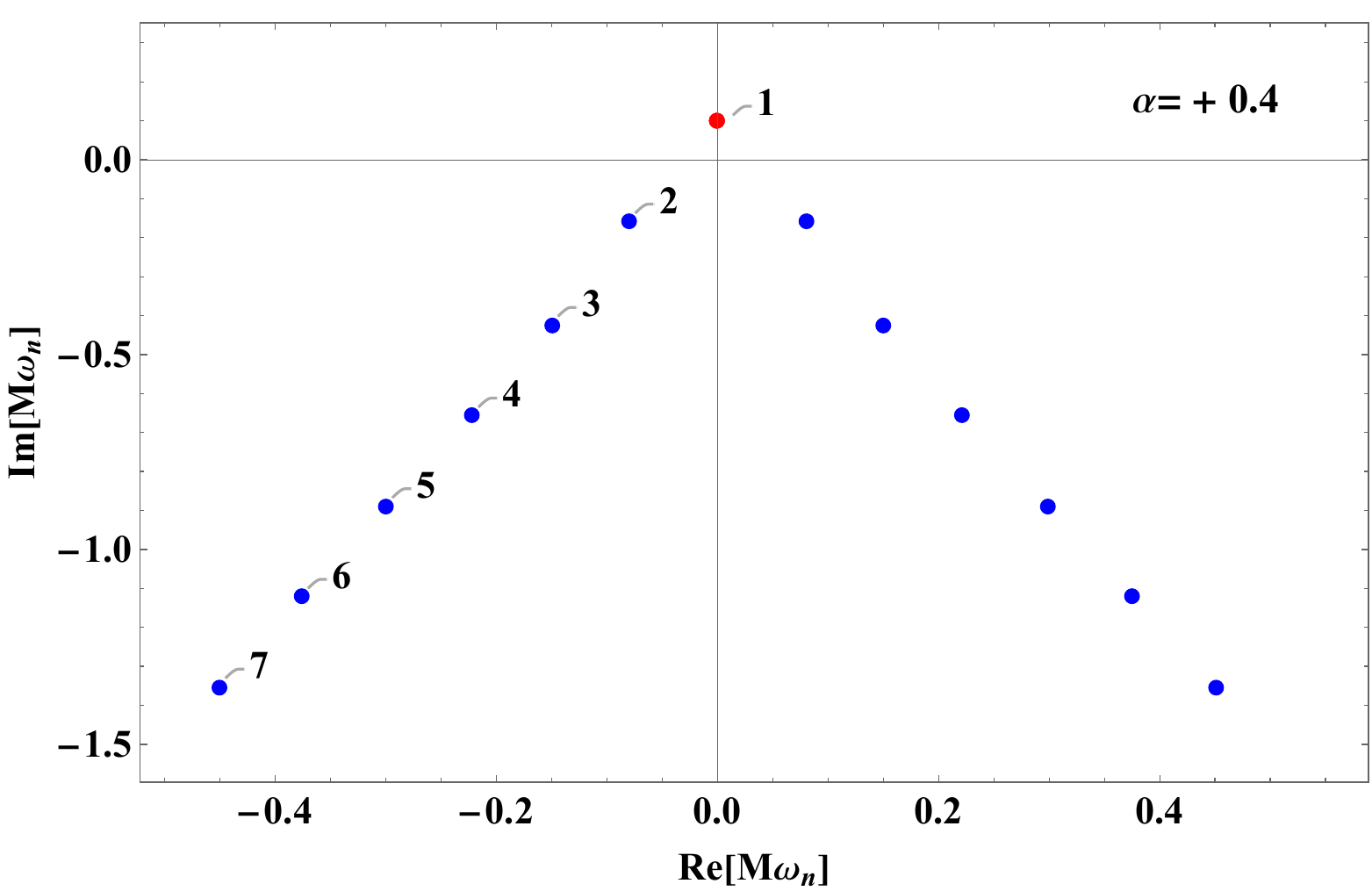}
    \qquad
    \begin{tabular}[b]{c cc}\hline
      & Re[$M\omega_n$] & Im[$M\omega_n$] \\ \hline
      1& $0$ & $0.10237$ \\
      2& $\pm 0.07985$ & $-0.15911$ \\
      3& $\pm 0.14962$ & $-0.42391$ \\
      4& $\pm 0.22154$ & $-0.65686$ \\
      5& $\pm 0.29939$ & $-0.88937$ \\
      6& $\pm 0.37565$ & $-1.12184$ \\
      7& $\pm 0.450\textcolor{gray}{30}$ & $-1.352\textcolor{gray}{39}$ \\ \hline
      \vspace{0.1cm}
    \end{tabular}
    \captionlistentry[table]{A table beside a figure}
    \captionsetup{labelformat=andtable}
    \caption{QNMs with various values of $\alpha$ for $\varphi_h = 0.1$. Blue dots are tolerable less than $10^{-3}$, whereas lighter blue dots have a tolerance of $10^{-2}$.}
    \label{fig:QNMvph01}
  \end{figure}

For the positive values of $\alpha$, we found that the imaginary parts of the QNM frequencies become positive at $\alpha = 0.1815$ (denoted as $\alpha_{\textrm{crit.}}$). This positive imaginary part indicates that the hairy black hole becomes dynamically unstable in the regime $\alpha \geq \alpha_{\textrm{crit.}}$. We demonstrated the corresponding eigenfunctions to ensure these eigenvalues and examined their behaviors in Figure~\ref{fig:eigenfunctions} in Appendix B. To investigate the angular dependence of the QNM spectrum in this case, we further checked the QNM behavior for different values of $l$. The results are presented in Figure~\ref{fig:QNMwithL} in Appendix C. Figure~\ref{fig:QNMvp001aP} illustrates that as the value of $\alpha$ increases, lower-frequency QNMs (with smaller imaginary parts) transition to having positive imaginary parts. As the value of $\alpha$ increases, our current numerical approach struggles to capture high-frequency QNMs with negative imaginary value. These high-frequency QNMs might become sparser, meaning they are fewer in number and more challenging to detect with our current methods. Also the nature of these modes could be affected by the metric's instability, causing them to lose their characteristic of oscillating and decaying over time and potentially transitioning into unstable modes themselves. For $\alpha = 3.5$, the QNMs exhibit purely positive imaginary frequencies. This implies that the oscillations of the perturbed scalar fields vanish and are only damped or growing at this value. We also computed the QNMs frequencies for $\alpha=-0.3, 0.184, 0.1848$ and $0.4$ for $\varphi_h = 0.1$ and $\lambda = 0.1$ and plotted them in Figure~\ref{fig:QNMvph01}. Table 2 presents the numerically computed QNM frequencies. Black colored numbers indicate a high degree of accuracy, with a numerical error less than $10^{-5}$ between the results obtained from grids of size $100$ and $140$. Conversely, gray colored numbers signify potential numerical inaccuracy in those digits.  In this case, we found that the critical value of $\alpha$ appears at $\alpha_{\textrm{crit.}}=0.1848$. Since these critical values of $\alpha$ are involved in dynamical stability, their values differ from those computed in \cite{Latosh:2023cxm}, which are sufficient conditions.

%%%%%%%%%%%%%%%%%%%%%%%%%%%%%%%%%%%%%%%%%%%%%%%%
\section{Quasinormal Modes for Hairy Black Holes in the Symmetry-Broken Phase}
%%%%%%%%%%%%%%%%%%%%%%%%%%%%%%%%%%%%%%%%%%%%%%%%

This section demonstrates the numerically generated hairy black holes' solution in the symmetry-broken phase and shows their QNMs. 

\subsection{Background solution}

When $\alpha$ is positive, the interaction potential $V$ forms degenerate vacuums, and the stable minima are described by
\begin{align}
  \langle \varphi \rangle = v_{\textrm{vac}} \, e^{i \beta},  \qquad v_{\textrm{vac}}= \sqrt{\frac{\alpha}{2\, \lambda}} ,
\end{align}
where the vacuum states are labeled by the parameter $\beta$. These ground states do not respect the symmetry of the Lagrangian, which indicates that symmetry is spontaneously broken in the vacuum. We expand a field around a ground state $v$ by reparameterizing it as follows
\begin{align}
 \varphi(z) = \bigg(v_{\textrm{vac}} + \frac{\sigma(z)}{\sqrt{2}} \bigg)e^{i \theta(z)} .
\end{align}
Here, the new variables $\sigma(z)$ and $\theta(z)$ are physical fields because they are excitations above the vacuum. Then we rewrite the Lagrangian in terms of $\sigma(z)$ and $\theta(z)$
\begin{align}
  \mathcal{L}_{\varphi} = - \frac{1}{2}\nabla_{\alpha} \sigma(z) \nabla^{\alpha} \sigma(z) - \bigg(v_{\textrm{vac}} + \frac{\sigma(z)}{\sqrt{2}} \bigg)^2 \nabla_{\alpha} \theta(z) \nabla^{\alpha} \theta(z) + f(\sigma) \, \mathcal{G}
\end{align}
where
\begin{align}
  f(\sigma) = -\alpha \, \sigma (z)^2- \sqrt{\alpha\,  \lambda} \; \sigma (z)^3-\frac{\lambda}{4}\, \sigma (z)^4.
\end{align}
Einstein equations and scalar field equations read
\begin{align}
  &\frac{1}{\kappa^2} \bigg(R_{\mu \nu} - \frac{1}{2} R g_{\mu \nu} \bigg) =  \bigg[-\frac{1}{2} \nabla_{\gamma} \sigma \nabla^{\gamma} \sigma - \bigg(v_{\textrm{vac}} + \frac{\sigma}{\sqrt{2}} \bigg)^2 \nabla_{\gamma} \theta \nabla^{\gamma} \theta \bigg]g_{\mu \nu} \nonumber\\
  &\qquad \qquad + \nabla_{\mu} \sigma \nabla_{\nu} \sigma + 2 \bigg(v_{\textrm{vac}} + \frac{\sigma}{\sqrt{2}} \bigg)^2 \nabla_{\mu} \theta \, \nabla_{\nu} \theta - (g_{\rho \mu} g_{\lambda \nu} + g_{\lambda \mu} g_{\rho \nu}) \eta^{\kappa \lambda \alpha \beta} \tilde{R}^{\rho \gamma}{}_{\alpha \beta} \nabla_{\gamma} \nabla_{\kappa} f(\sigma), \\
  &\nabla^{\alpha} \nabla_{\alpha} \sigma - \sqrt{2} \bigg(v_{\textrm{vac}} + \frac{\sigma}{\sqrt{2}}\bigg) \nabla^{\alpha} \theta\, \nabla_{\alpha} \theta + f_{\sigma}\, \mathcal{G} = 0, \qquad \bigg(v_{\textrm{vac}} + \frac{\sigma}{\sqrt{2}} \bigg)  \nabla^{\alpha} \nabla_{\alpha} \theta + \sqrt{2}\, \nabla^{\alpha} \sigma \nabla_{\alpha} \theta = 0 . \label{eq:SBsigmaEQ}
\end{align}
Consequently, they take the following explicit form:
\begin{align}
&\frac{1}{z^2} \bigg(1-\frac{1}{B}-\frac{zB'}{B} \bigg)+\kappa ^2\left(v+\frac{\sigma}{\sqrt{2}}\right)^2 \theta'^2+\frac{1}{2} \kappa ^2 (\sigma')^2\nonumber\\
&\qquad +4 \kappa ^2 \left[\frac{z^2 (1-3 B) B' \sigma' f_{\sigma}}{B}-2 z (B-1) \bigg(z  f_{\sigma \sigma} (\sigma')^2 +f_{\sigma} \left(z \sigma''+2 \sigma' \right)\bigg)\right] = 0, \\
&\frac{1}{z} \bigg(\frac{B'}{B}-\frac{A'}{A} \bigg)-2 \kappa ^2 \left(v+\frac{\sigma}{\sqrt{2}}\right)^2 \theta'^2-\kappa ^2 (\sigma')^2 \nonumber\\
&\qquad+ \kappa ^2 \left[4 z^2 (3 B-1) \left(\frac{B'}{B}-\frac{A'}{A}\right) \sigma' f_{\sigma}+8 z (B-1) \bigg(z  f_{\sigma \sigma} (\sigma')^2+f_{\sigma} \left(z \sigma''+2 \sigma'\right)\bigg)\right] = 0,\\
&\frac{A''}{2 A}+\frac{1}{4 z} \left(\frac{z A'}{A}-2\right) \left(\frac{B'}{B}-\frac{A'}{A}\right)+\kappa ^2 \left(v+\frac{\sigma}{\sqrt{2}}\right)^2 \theta'^2+\frac{1}{2} \kappa ^2 (\sigma')^2 \nonumber\\
&\qquad+2 z^2 B \kappa ^2 \left[ \frac{2 z A' }{A} \left(\sigma'^2 f_{\sigma \sigma}+\sigma'' f_{\sigma} \right)+\sigma' f_{\sigma} \left(\frac{z A'}{A} \left(\frac{3 B'}{B}-\frac{A'}{A}\right)+ 2 \left(\frac{z A''}{A}+\frac{4 A'}{A}\right)\right)\right] = 0,
\end{align}
and
\begin{align}
  &\sigma '' + \bigg(\frac{A'}{2 A}+\frac{B'}{2 B} \bigg) \sigma' - \sqrt{2} \bigg(v_{\textrm{vac}}+ \frac{\sigma}{\sqrt{2}} \bigg)\left(\theta '\right)^2  \nonumber \\
  & \qquad \qquad+ 2 z f_{\sigma}\left[\frac{(3 B-1) A' B'}{A B}-(B-1) \left(\frac{z A'^2}{A^2}-\frac{4 A'}{A} -\frac{2 z A''}{A} \right)\right] = 0, \\
  &\bigg(v_{\textrm{vac}}+\frac{\sigma}{\sqrt{2}} \bigg) \theta '' + \bigg[\sqrt{2} \sigma '  + \bigg(\frac{A'}{2 A}+\frac{B'}{2 B} \bigg) \bigg(v_{\textrm{vac}} + \frac{\sigma}{\sqrt{2}} \bigg) \bigg] \theta ' = 0.
\end{align}
Field $\theta(z)$ is decoupled from the system, and the solution for $\theta'(z)$ reads
\begin{align}
  \theta'(z) =  \frac{c_2}{4\sqrt{A(z)B(z)}} \bigg(v_{\textrm{vac}}+ \frac{\sigma(z)}{\sqrt{2}}\bigg)^{-2} . \label{eq:thetaEQ}
\end{align}
For the scalar field flux to be finite, $c_2$ should be zero \cite{Latosh:2023cxm}.

\begin{figure}[t]
\begin{center}
	\includegraphics[scale=0.29]{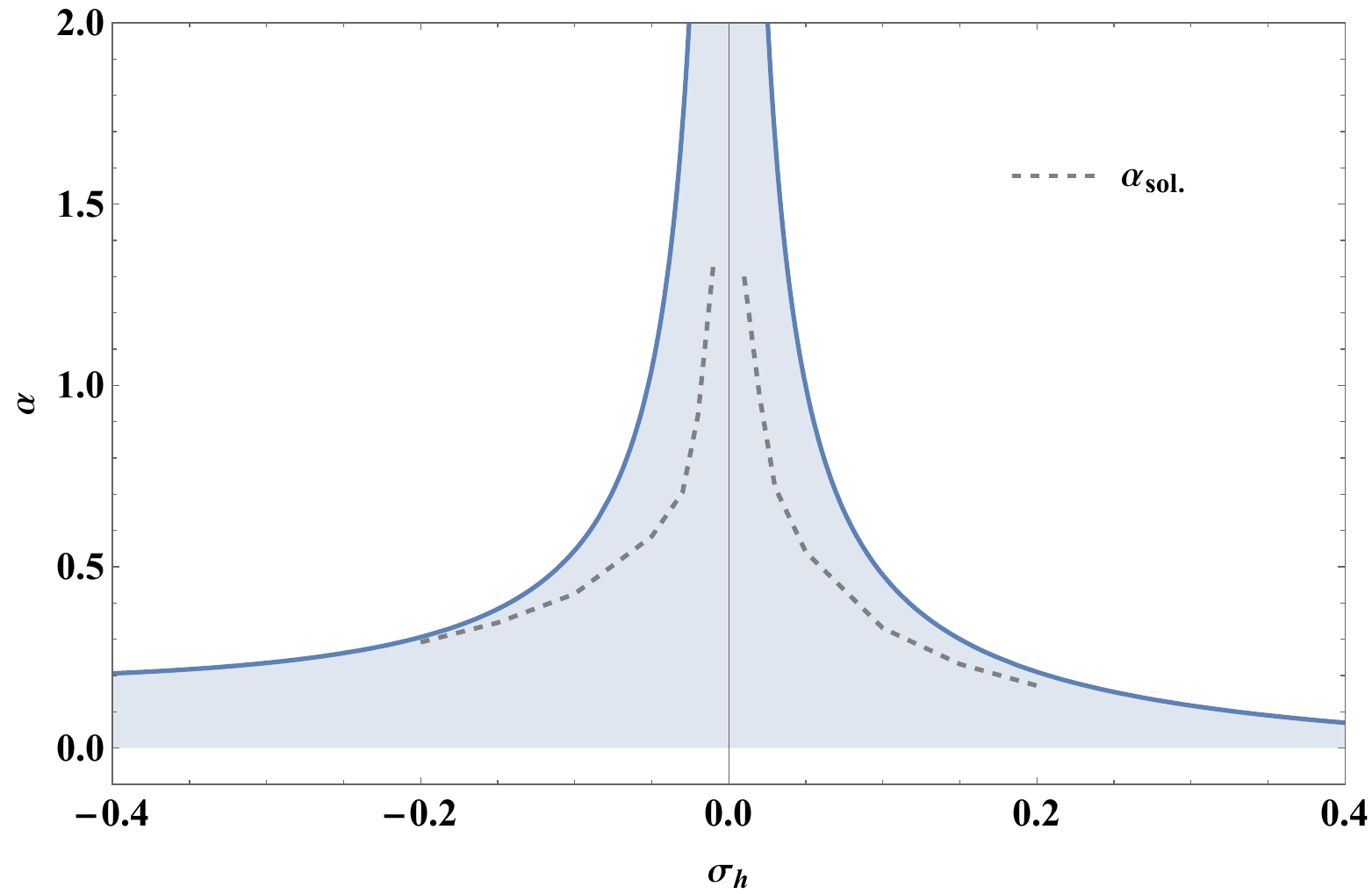}
	\caption{Phase space for $\alpha$ and $\sigma_h$ with fixed values of $\lambda$ in the symmetry-broken phase}
	\label{fig:SMBphase}
\end{center}
\end{figure}

\begin{figure}[h!]
\subfloat[The metric solutions of $A$ and $B$ with $\alpha=0.2$]{\includegraphics[scale=0.27]{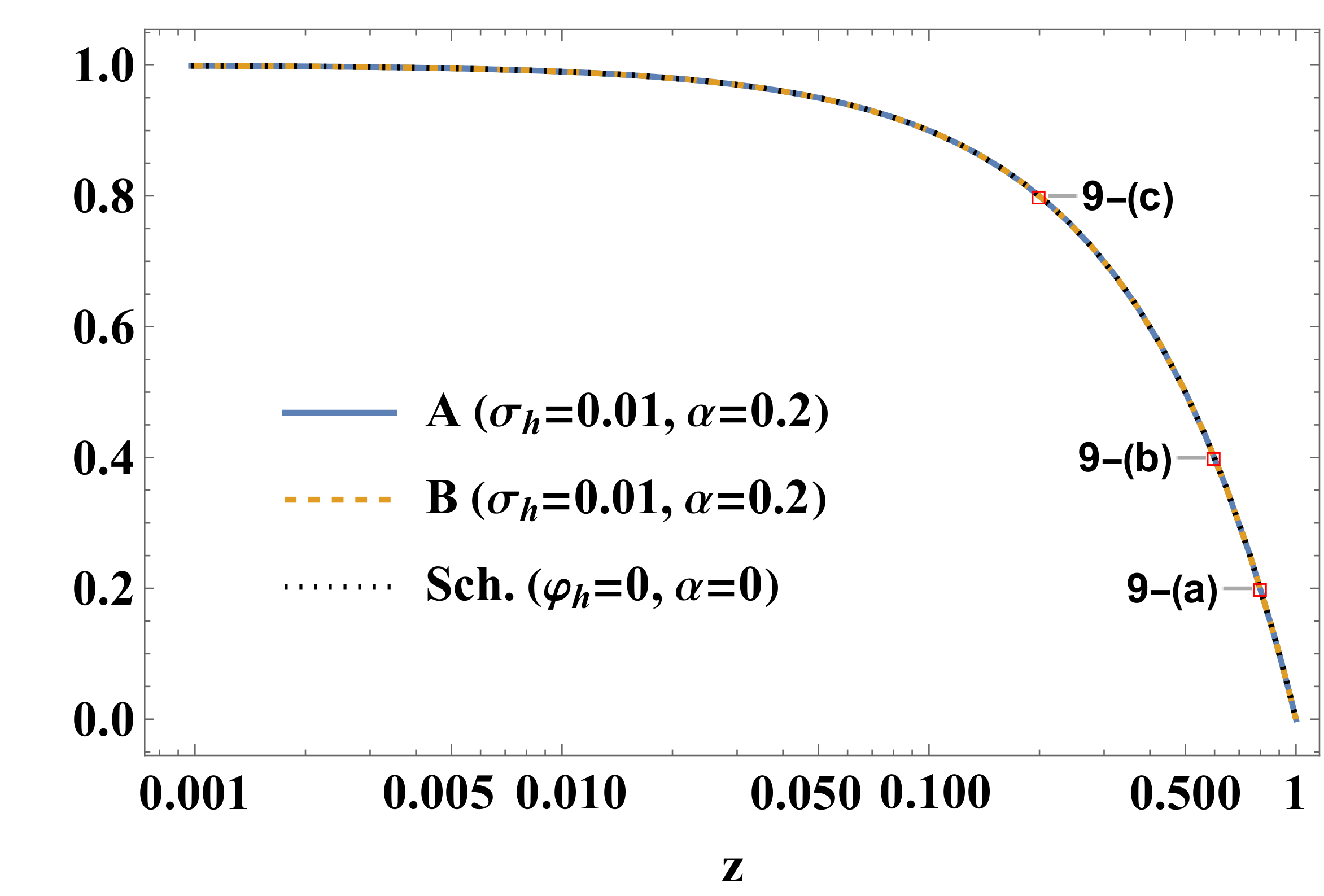}}
\subfloat[ii][The metric solutions of $A$ and $B$ with $\alpha=0.6$]{\includegraphics[scale=0.27]{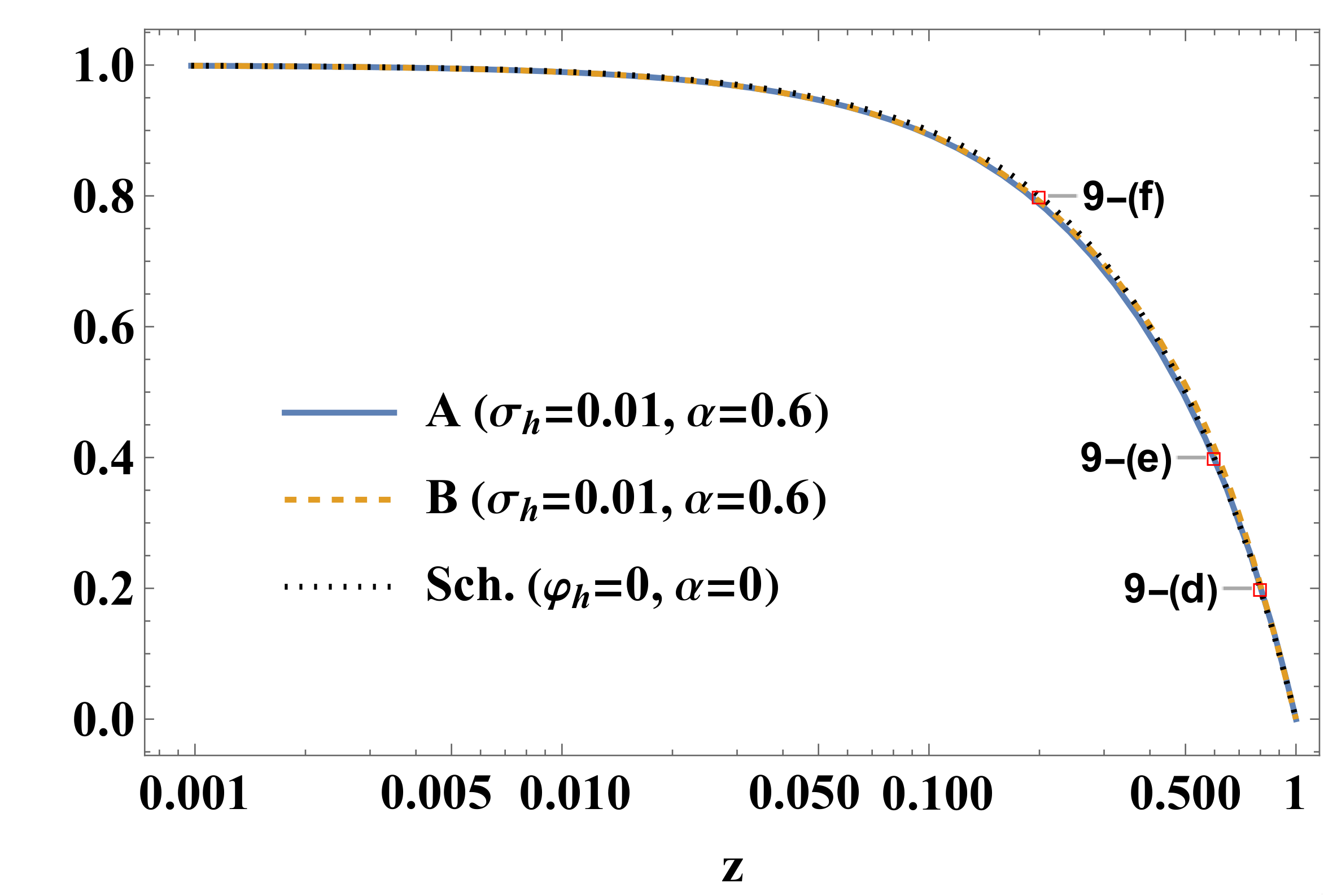}}
\subfloat[The scalar field solutions $\varphi$ for various values of $\alpha$]{\includegraphics[scale=0.27]{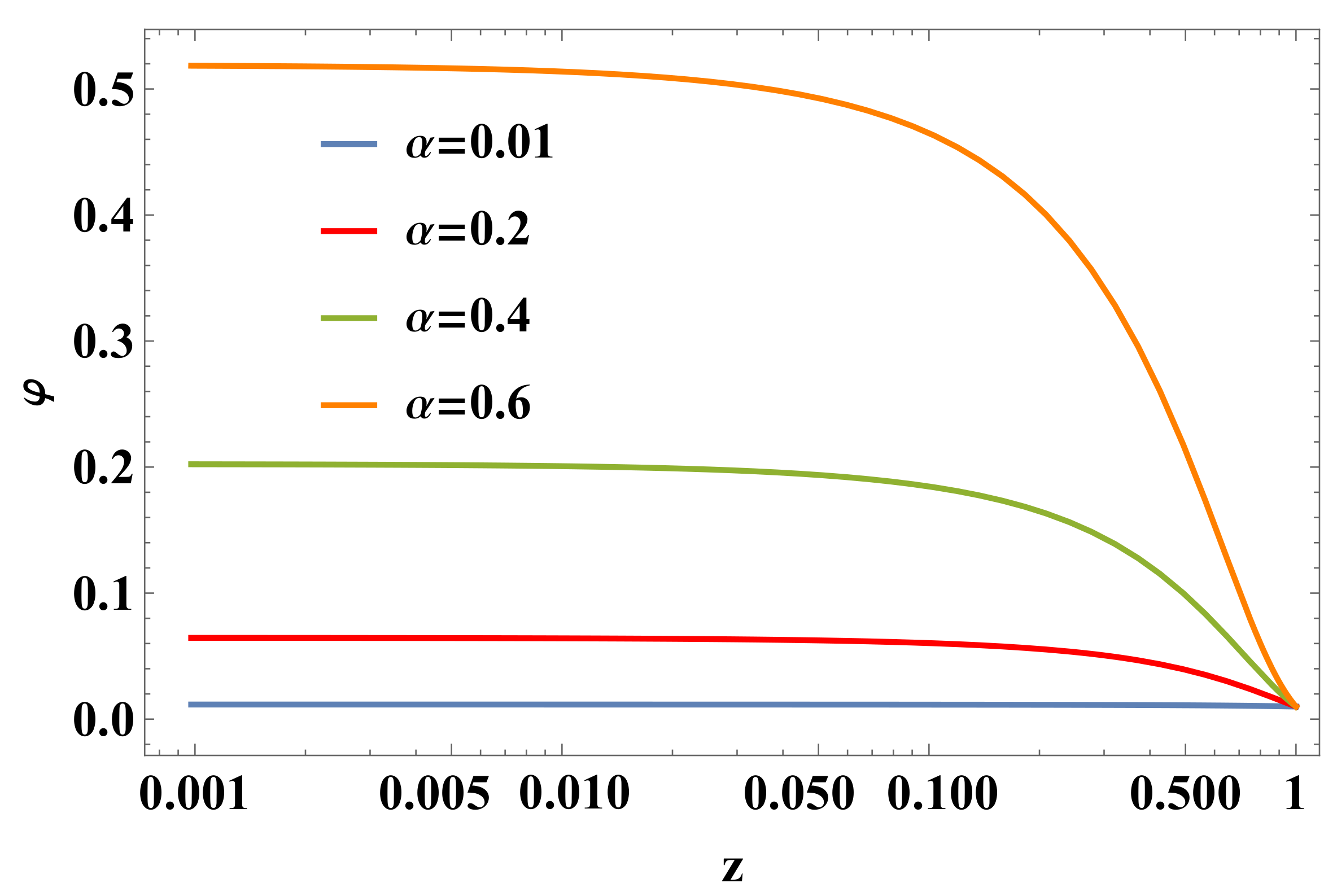}}
\caption{Hairy black hole solutions with $\sigma_h = 0.01$ for various values of $\alpha$. The black dotted line denotes the solutions of Schwarzschild black hole. The small regimes with the callout index (a)$\sim$(f) are plotted in Figure~\ref{fig:SB01}.}
	\label{fig:SB001}

\vspace{1cm}

\subfloat[Case of $\alpha=0.2$]{\includegraphics[scale=0.27]{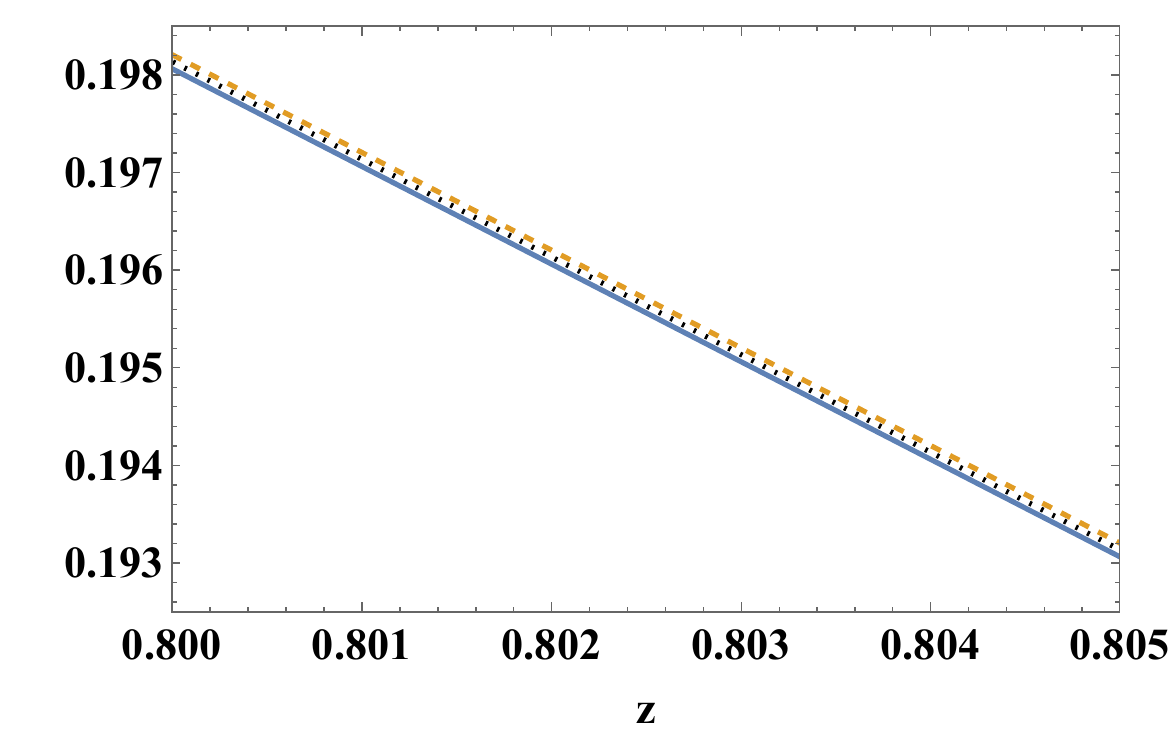}}
\subfloat[Case of $\alpha=0.2$]{\includegraphics[scale=0.27]{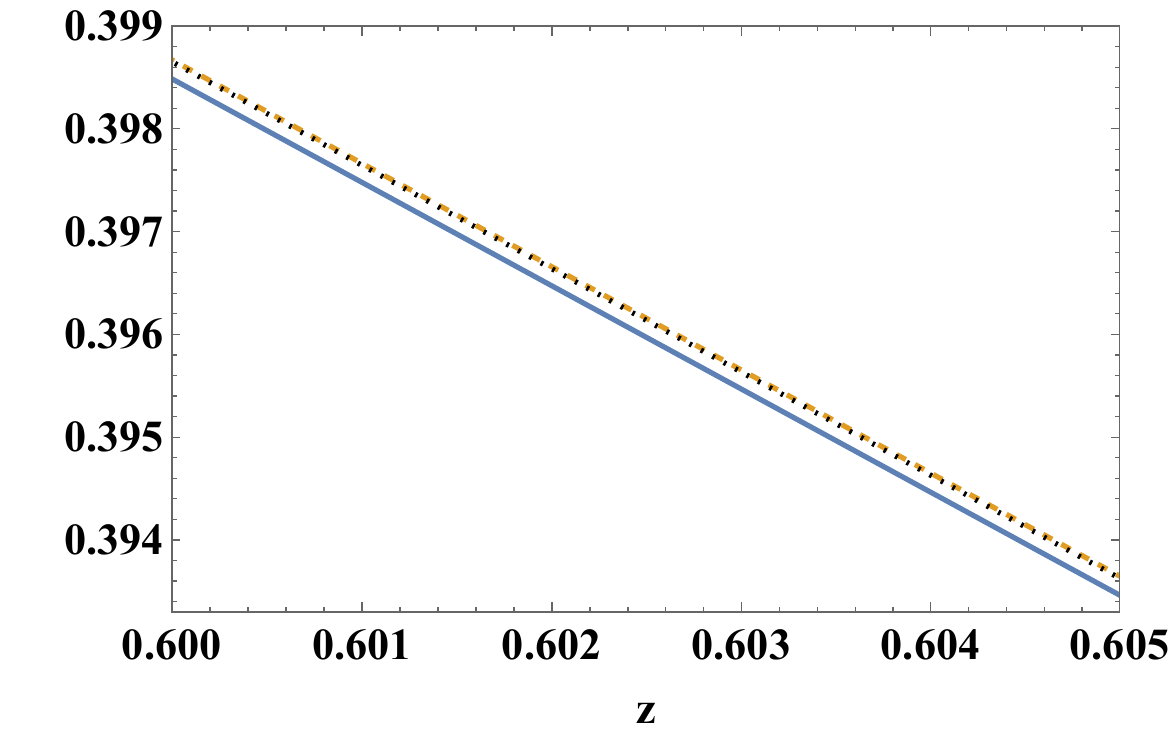}}
\subfloat[Case of $\alpha=0.2$]{\includegraphics[scale=0.27]{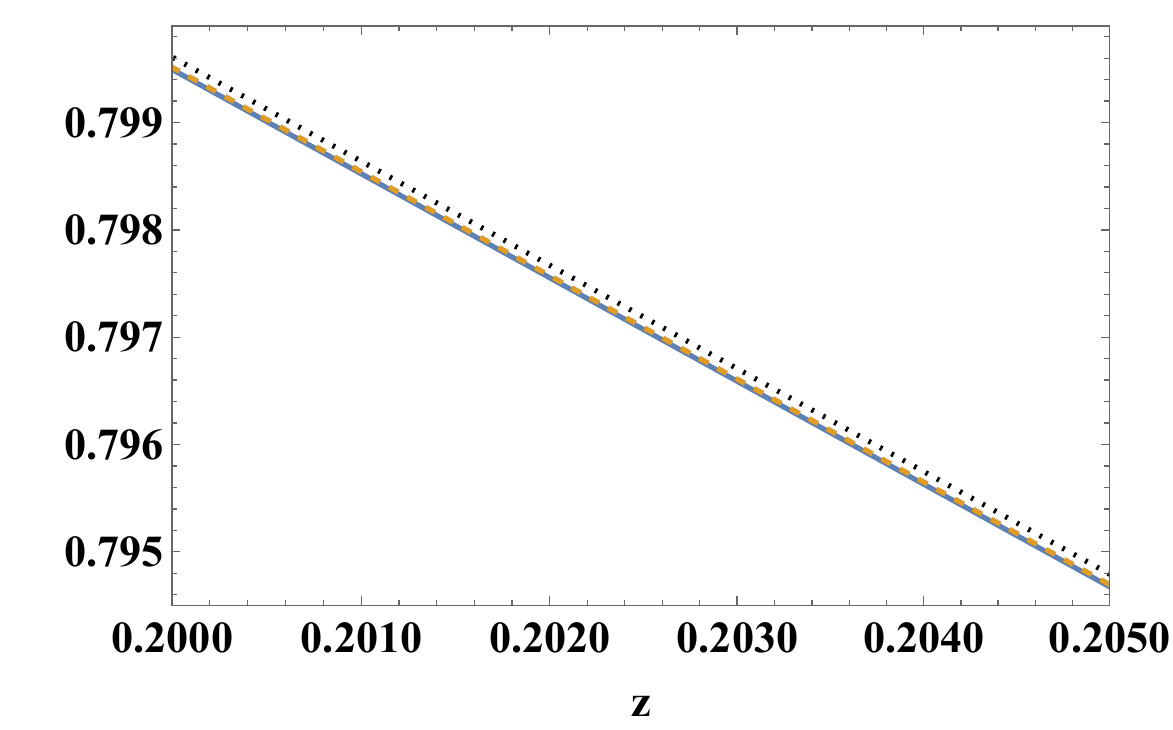}}

\subfloat[Case of $\alpha=0.6$]{\includegraphics[scale=0.27]{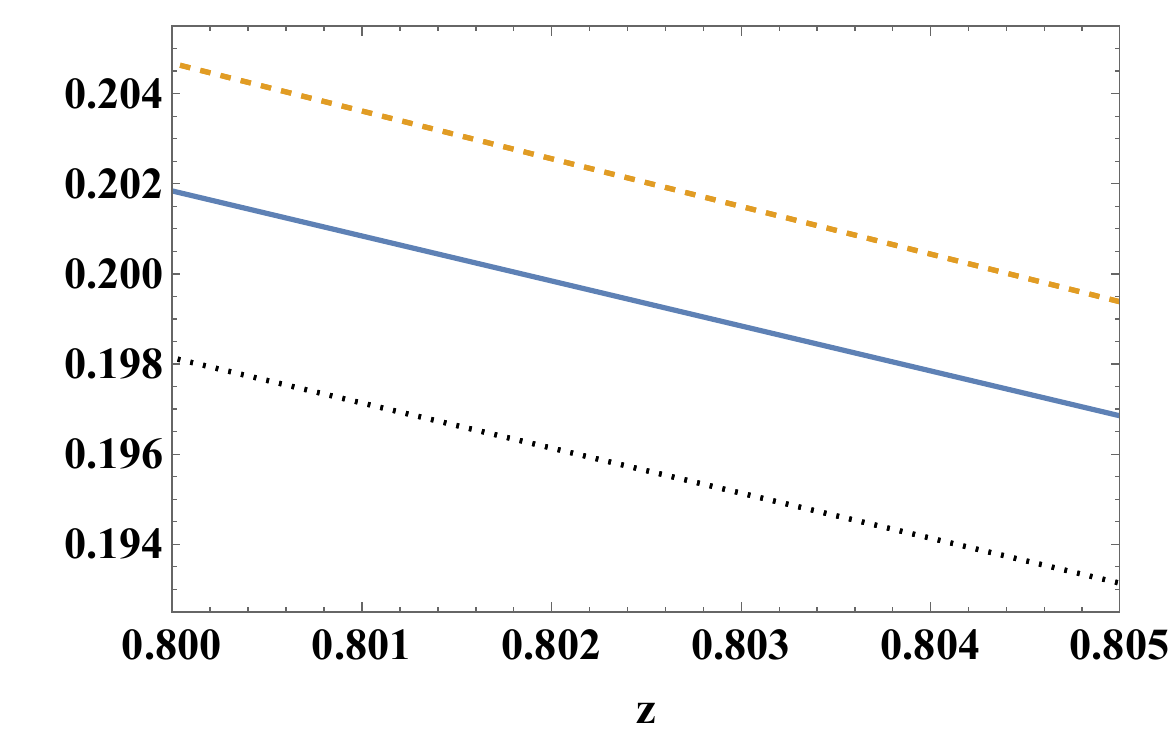}}
\subfloat[Case of $\alpha=0.6$]{\includegraphics[scale=0.27]{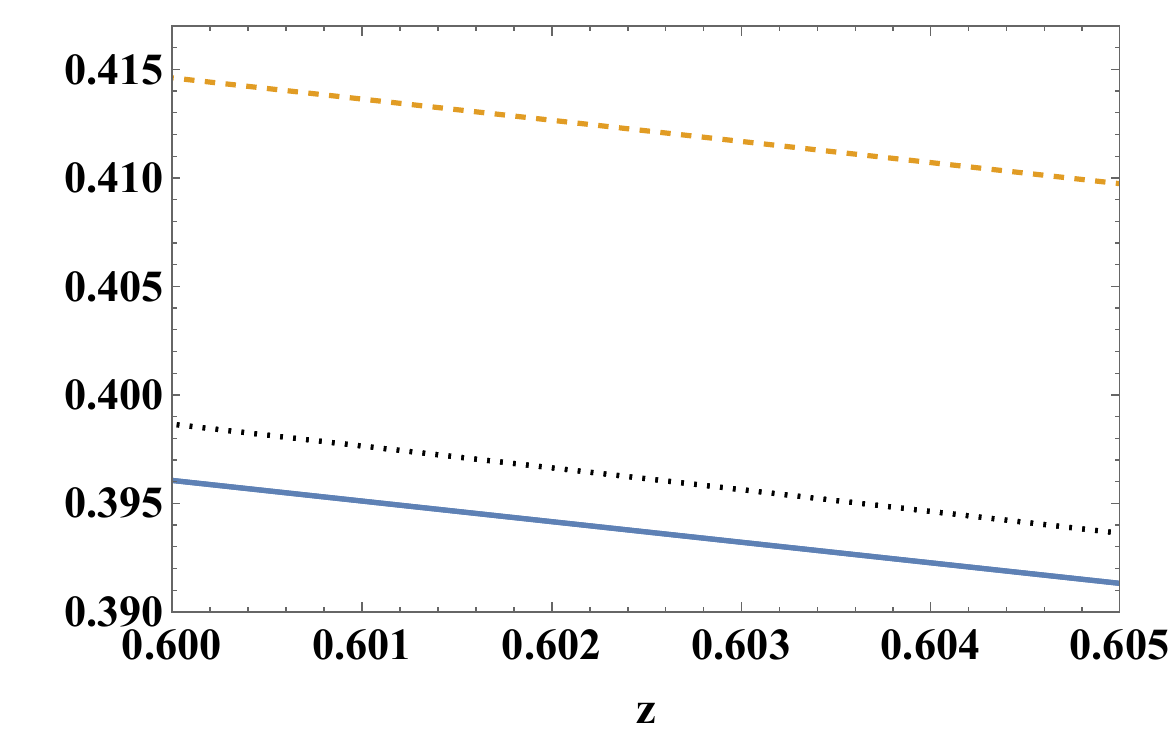}}
\subfloat[Case of $\alpha=0.6$]{\includegraphics[scale=0.27]{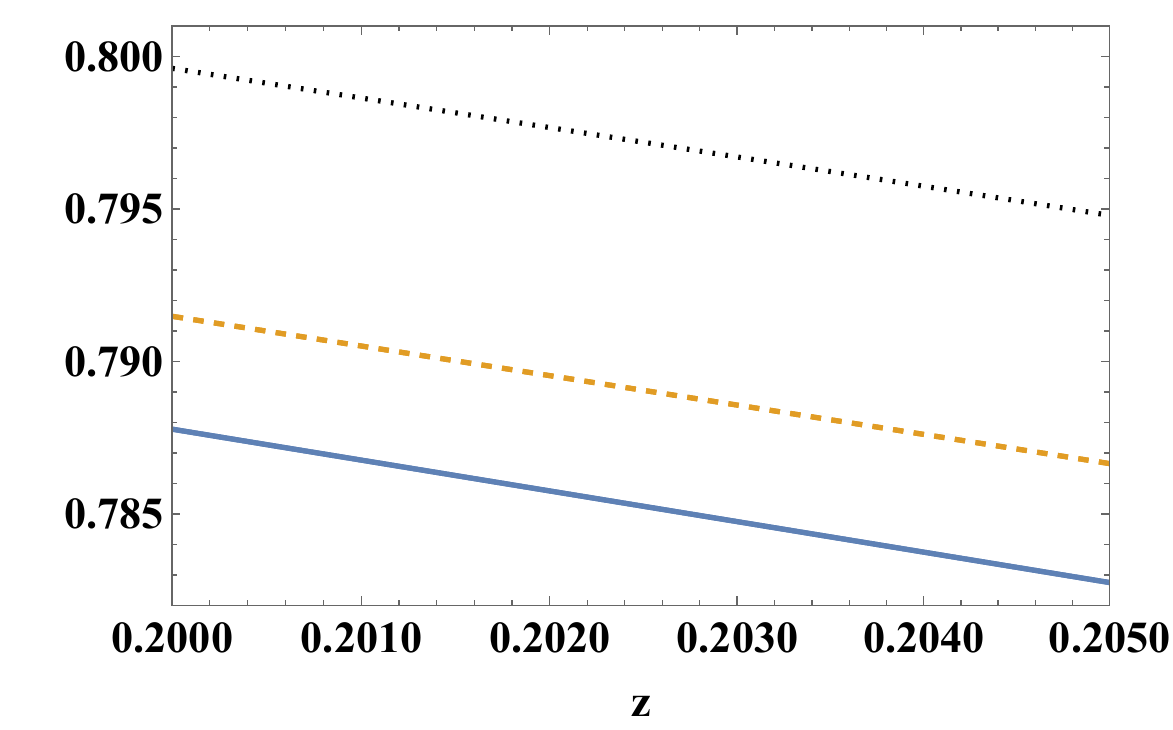}}
\caption{The comparison of metric solutions for hairy black holes and the Schwarzschild black hole. When $\alpha$ increases, hairy black holes exhibit a slight departure from the Schwarzschild black holes and show the anisotropic property of the metric ($A/B \neq 1$). This deviation, though subtle, leads to noticeably distinct behavior in their QNMs for large $\alpha$ values.}
	\label{fig:SB01}
\end{figure}

As we consider the regular black hole solutions, we impose the boundary condition near the horizon as follows
\begin{align}
&A(z) \sim A_h \epsilon + \mathcal{O}(\epsilon^2) , \; \; \; B(z) \sim B_h \epsilon  + \mathcal{O}(\epsilon^2), \; \; \; \sigma(z) \sim \sigma_{h} + {\sigma_{h,1}} \epsilon  + \mathcal{O}(\epsilon^2) , \label{eq:NHexpSB}  
\end{align}
where the near horizon expansion of the metric and the scalar field $\sigma(z)$ take the following forms from the equations of motion
\begin{align}
&B_h = \frac{1 - \sqrt{1-96 z_h^4 (f_{\sigma_h})^2}}{48 z_h^5 (f_{\sigma_h})^2} , \\
&\sigma_{h,1} =\frac{- 1 + \sqrt{1-96 z_h^4 (f_{\sigma_h})^2}}{4 z_h^3 f_{\sigma_h}}.
\end{align}
The two independent variables $A_h$ and $\sigma_h$ and the coupling constants $\alpha$ and $\lambda$ fully describe the boundary conditions. The regularity condition requires
\begin{align}
(f_{\sigma _h})^2 < \frac{1}{96 z_h^4} , \label{eq:SMrgcnd}
\end{align}
which limits the values of $\alpha, \lambda$ and $\sigma_h$. Figure~\ref{fig:SMBphase} illustrates the valid parameter space for hairy black hole solutions at fixed values of $\lambda$. The shaded regions bounded by the solid lines represent valid values of $\alpha$ and $\sigma_h$, excluding points that fall exactly on the solid lines.
Within the parameter values in the phase space, we numerically generate hairy black hole solutions in the symmetry-broken phase with the same precision and accuracy used in the symmetric phase. For the choice of $\alpha=0.01$ with the fixed value of $\lambda = 0.1$, we plot the solutions in Figure~{\ref{fig:SB001}} and Figure~\ref{fig:SB01}. By increasing the value of $\alpha$, we observed that the numerical solutions become increasingly difficult to find for values of $\alpha$ exceeding $\alpha_{\textrm{sol}}$, as discussed in \cite{Latosh:2023cxm}.

\subsection{Scalar Field Perturbation}

We consider a linearized perturbation of the real scalar field as follows
\begin{align}
\sigma \rightarrow \sigma(z) + \delta \sigma(v,z,\theta,\phi).
\end{align}
Applying $c_2=0$ in (\ref{eq:thetaEQ}) to the scalar field equations (\ref{eq:SBsigmaEQ}), the linearized part becomes 
\begin{align}
\nabla_{\alpha} \nabla^{\alpha} \delta \sigma + f_{\sigma \sigma} \mathcal{G} \delta \sigma = 0.
\end{align}
Using the following ansatz 
\begin{align}
\delta \sigma (v,z,\theta,\phi) = \sum_{l,m} \Sigma (z) Y_{lm}(\theta, \phi ) e^{-i \omega v}
\end{align}
and (\ref{eq:sphericalY}), the linearized equation is written as
\begin{align}
&z^2 A(z) B(z) \Sigma''(z) + \frac{1}{2}  \bigg( z^2 B(z) A'(z)+z^2 A(z) B'(z)+4 i \omega  B(z) \sqrt{\frac{A(z)}{B(z)}} \bigg) \Sigma'(z) \\
&+ \frac{1}{z^2} \bigg( A(z) \left( \mathcal{G} f_{\sigma \sigma}(\sigma)-\gamma  z^2\right)-2 i \omega  z B(z) \sqrt{\frac{A(z)}{B(z)}} \bigg) \Sigma(z) = 0 , \label{eq:PerbEqSB}
\end{align}
where
\begin{align}
f_{\sigma \sigma} (\sigma) = -2 \alpha -6 \sqrt{\alpha  \lambda}~ \sigma -3 \lambda  \sigma ^2. 
\end{align}
The equation structure of (\ref{eq:PerbEqSB}) remains unchanged from (\ref{eq:PerbEq}). Thus we utilize (\ref{eq:PerbEqRe}), substituting $\tilde{f}_{\varphi^* \varphi}$ with $f_{\sigma \sigma}$, to test the instability of the QNMs.

\subsection{Our Numerical Results}

\begin{figure}
    \centering
    \includegraphics[scale=0.3]{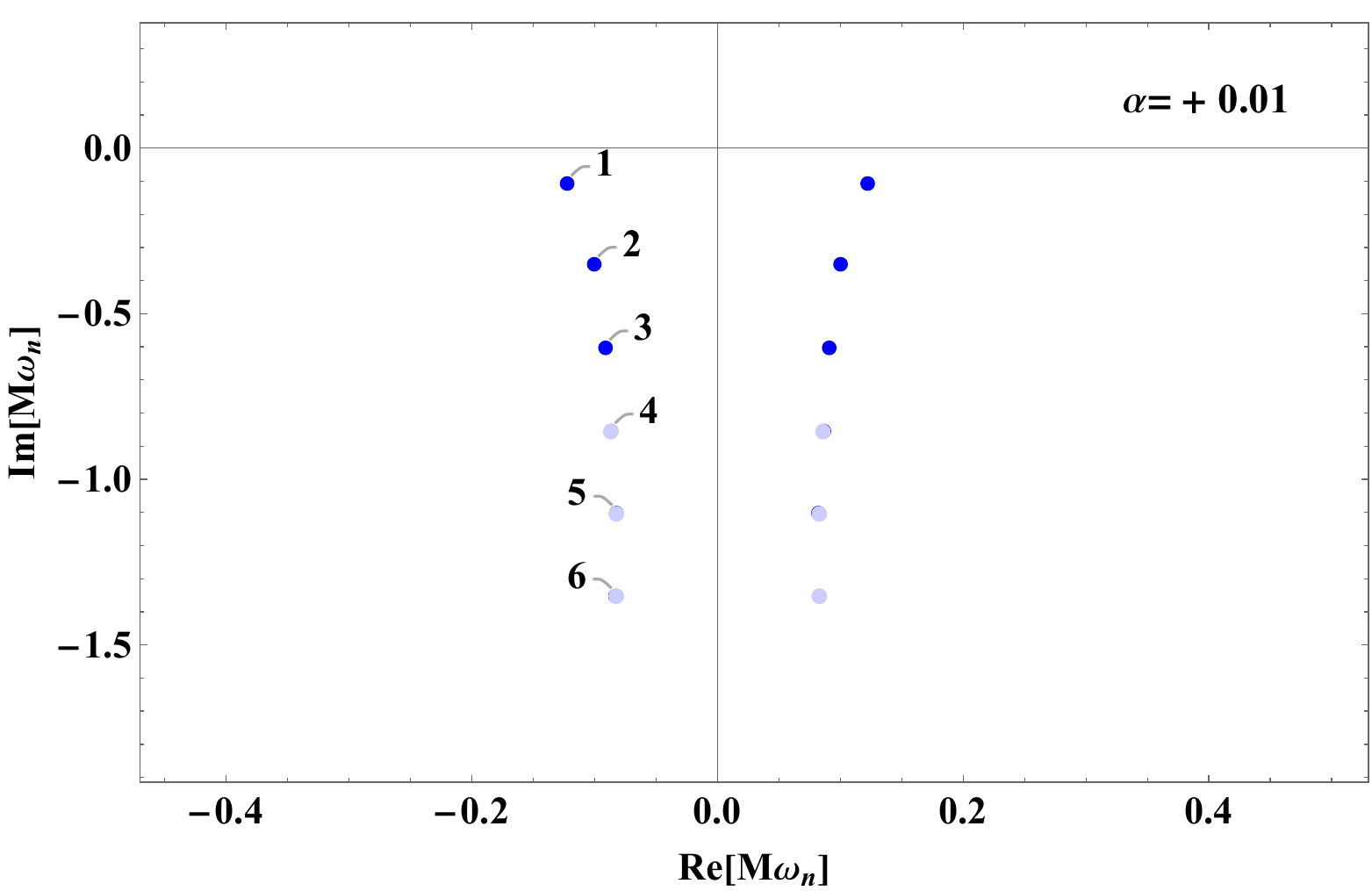}
    \qquad
    \begin{tabular}[b]{c cc}\hline
      & Re[$M\omega_n$] & Im[$M\omega_n$] \\ \hline
     1& $\pm 0.12229$ & $-0.10701$ \\
     2& $\pm 0.10025$ & $-0.35077$ \\
     3& $\pm 0.091\textcolor{gray}{04}$ & $-0.603\textcolor{gray}{12}$ \\
     4& $\pm 0.08\textcolor{gray}{662}$ & $-0.85\textcolor{gray}{448}$ \\
     5&$\pm 0.08\textcolor{gray}{226}$ & $-1.10\textcolor{gray}{205}$ \\
     6&$\pm 0.08\textcolor{gray}{295}$ & $-1.35\textcolor{gray}{246}$ \\ \hline
           \vspace{0.2cm}
    \end{tabular}
                  \vspace{0.1cm} \\    
\includegraphics[scale=0.3]{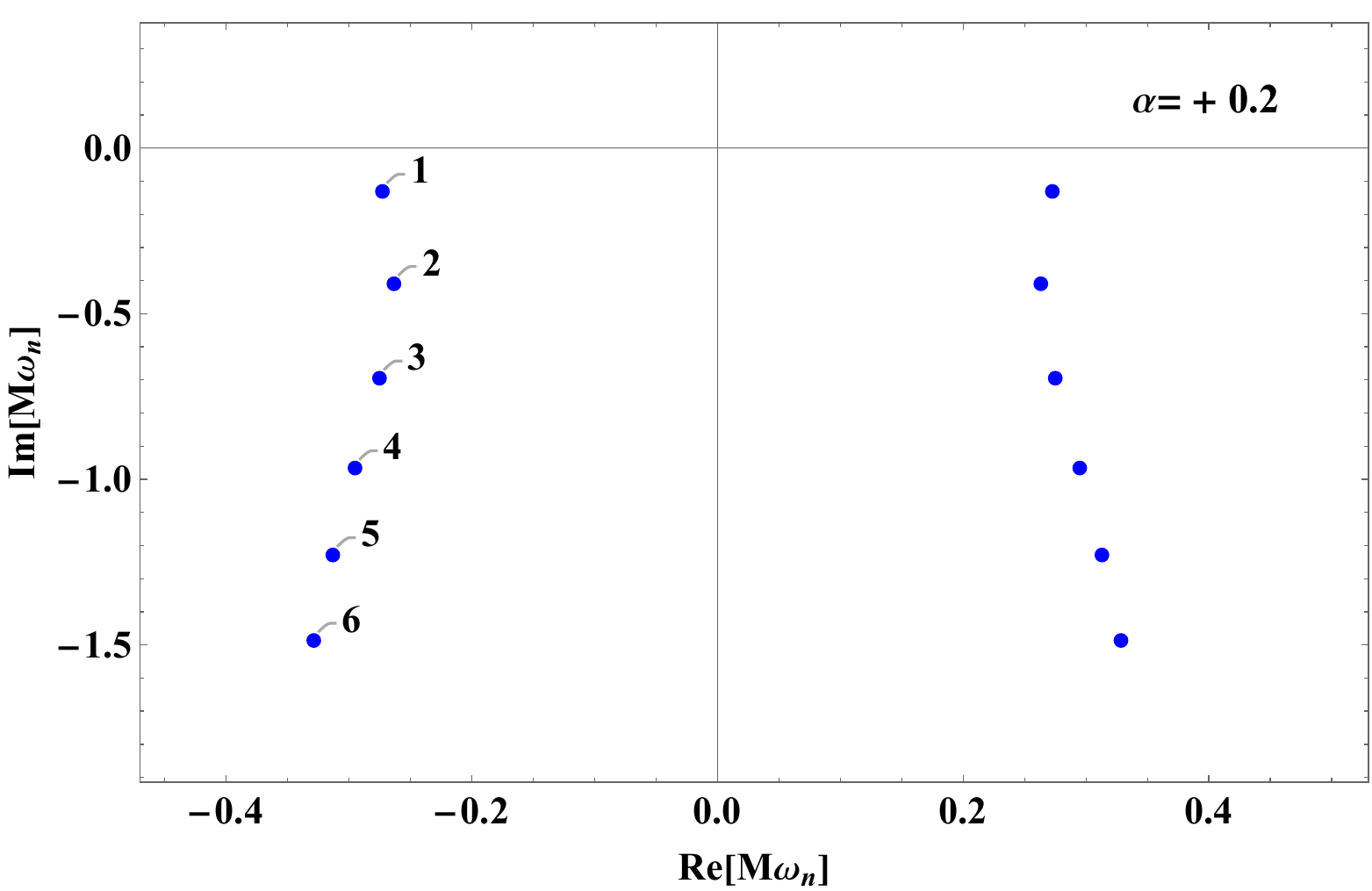}
    \qquad
    \begin{tabular}[b]{c cc}\hline
      & Re[$M\omega_n$] & Im[$M\omega_n$] \\ \hline
     1& $\pm 0.27261$ & $-0.13084$ \\
     2& $\pm 0.26325$ & $-0.40969$ \\
     3& $\pm 0.27500$ & $-0.69474$ \\
     4& $\pm 0.29492$ & $-0.96587$ \\
     5&$\pm 0.31297$ & $-1.22847$ \\
     6&$\pm 0.3284\textcolor{gray}{8}$ & $-1.486\textcolor{gray}{5}$ \\ \hline
           \vspace{0.2cm}
    \end{tabular}
                      \vspace{0.1cm} \\
\includegraphics[scale=0.3]{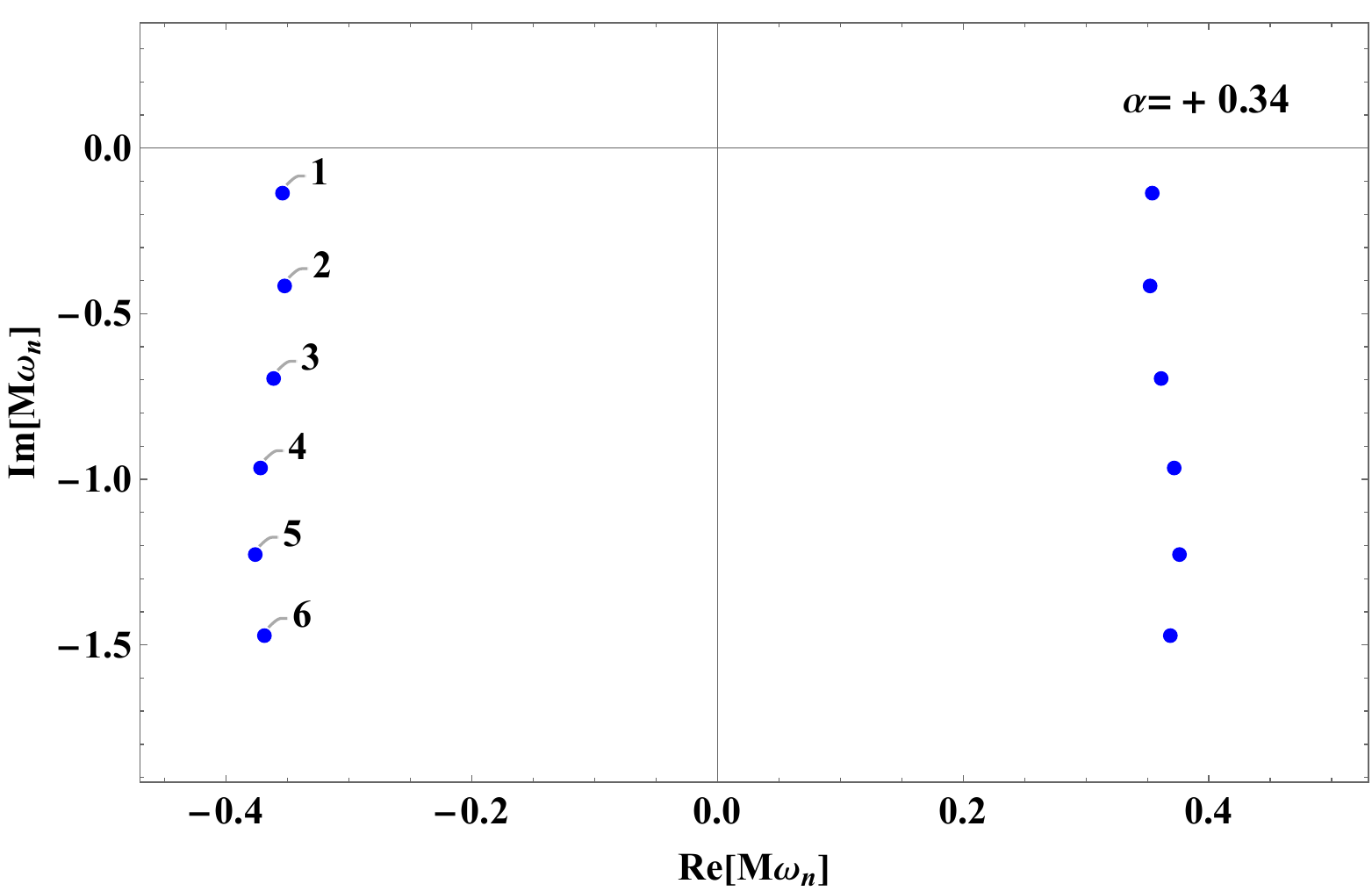}
    \qquad
    \begin{tabular}[b]{c cc}\hline
      & Re[$M\omega_n$] & Im[$M\omega_n$] \\ \hline
      1& $\pm 0.35393$ & $-0.13610$ \\
      2& $\pm 0.35217$ & $-0.41623$ \\
      3& $\pm 0.36116$ & $-0.69570$ \\
      4& $\pm 0.37179$ & $-0.96582$ \\
      5& $\pm 0.37610$ & $-1.22716$ \\
      6& $\pm 0.3686\textcolor{gray}{6}$ & $-1.4719\textcolor{gray}{3}$ \\ \hline
      \vspace{0.2cm}
    \end{tabular}
                      \vspace{0.1cm} \\
\includegraphics[scale=0.3]{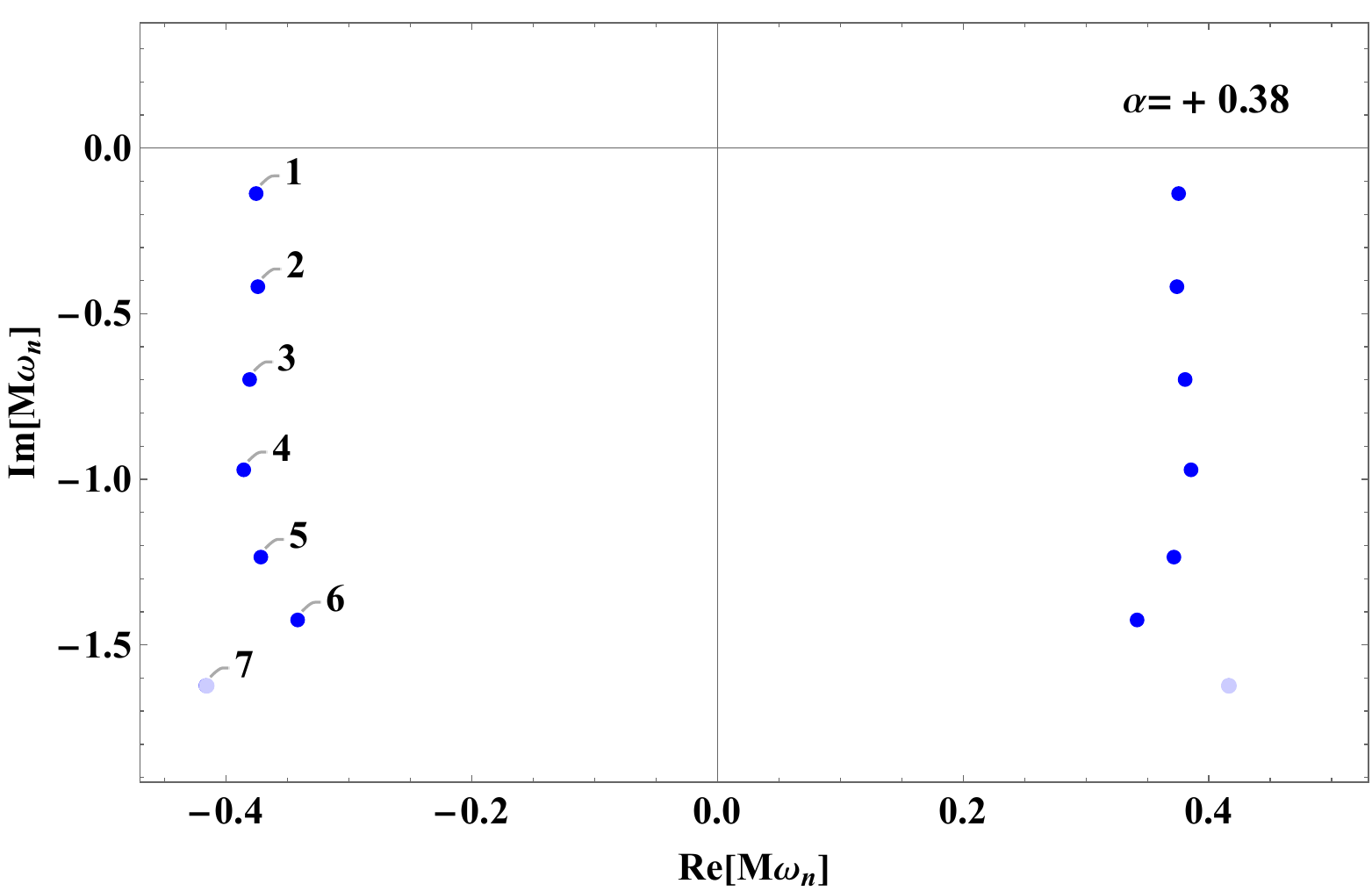}
    \qquad
    \begin{tabular}[b]{c cc}\hline
      & Re[$M\omega_n$] & Im[$M\omega_n$] \\ \hline
      1& $\pm 0.37538$ & $-0.13731$ \\
      2& $\pm 0.37399$ & $-0.41842$ \\
      3& $\pm 0.38069$ & $-0.69878$ \\
      4& $\pm 0.38549$ & $-0.97137$ \\
      5& $\pm 0.37156$ & $-1.23482$ \\
      6& $\pm 0.341\textcolor{gray}{59}$ & $-1.424\textcolor{gray}{65}$ \\
      7& $\pm 0.41\textcolor{gray}{624}$ & $-1.62\textcolor{gray}{298}$ \\ \hline
      \vspace{0.1cm}
    \end{tabular}
    \captionlistentry[table]{A table beside a figure}
    \captionsetup{labelformat=andtable}
    \caption{QNMs with various values of $\alpha$ for $\sigma_h = 0.01$. Blue dots are tolerable less than $10^{-3}$, whereas lighter blue dots have a tolerance of $10^{-2}$.}
    \label{fig:QNMsigmah001a}
  \end{figure}

\begin{figure}
    \centering
    \includegraphics[scale=0.3]{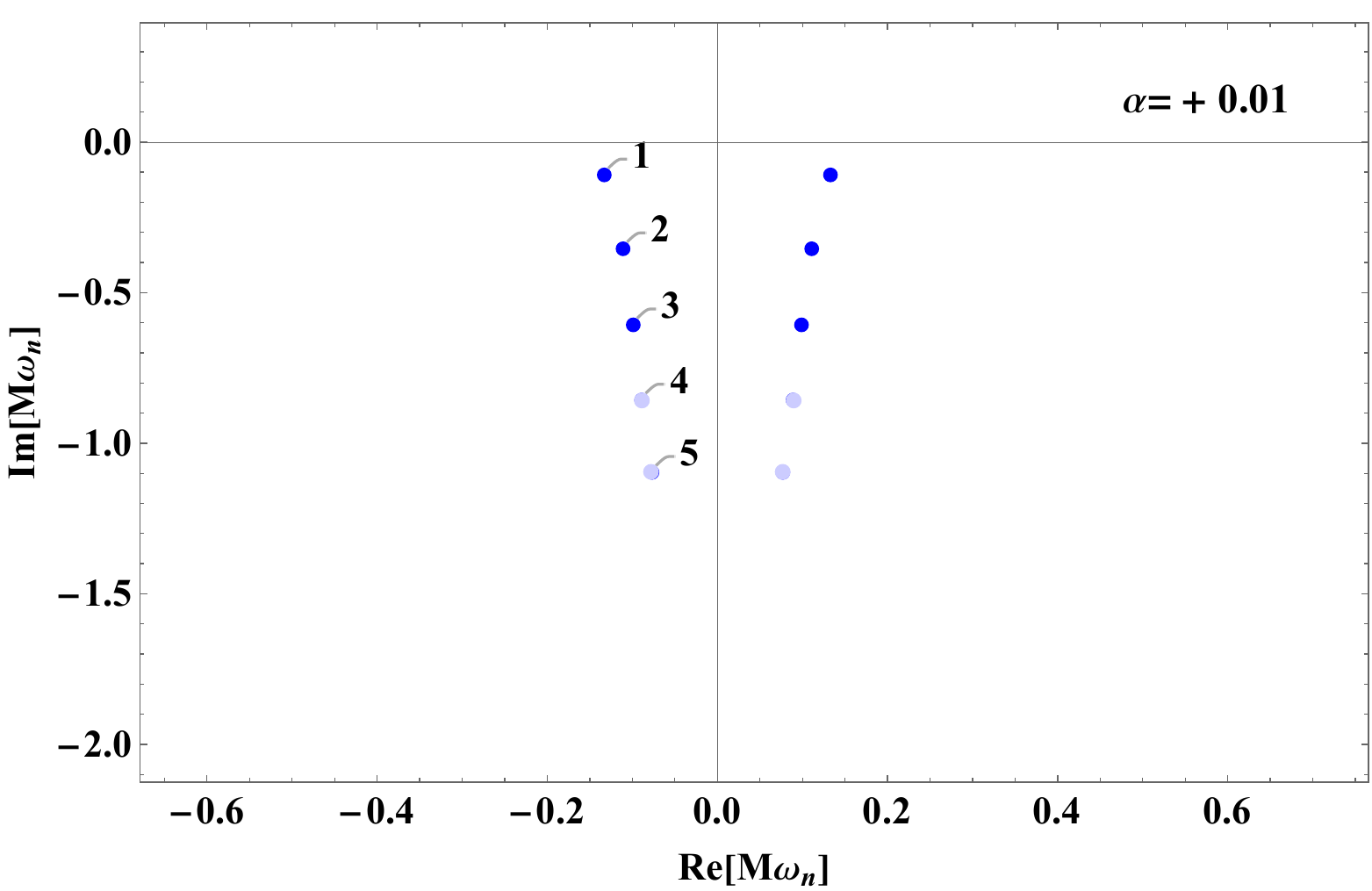}
    \qquad
    \begin{tabular}[b]{c cc}\hline
       & Re[$M\omega_n$] & Im[$M\omega_n$] \\ \hline
      1 & $\pm 0.13296$ & $-0.10915$ \\
      2 & $\pm 0.11089$ & $-0.35434$ \\
      3 & $\pm 0.098\textcolor{gray}{88}$ & $-0.607\textcolor{gray}{10}$ \\
      4 & $\pm 0.08\textcolor{gray}{903}$ & $-0.85\textcolor{gray}{684}$ \\
      5 & $\pm 0.07\textcolor{gray}{694}$ & $-1.09\textcolor{gray}{692}$ \\ \hline
      \vspace{0.4cm}
    \end{tabular}
             \vspace{0.1cm} \\
    \includegraphics[scale=0.3]{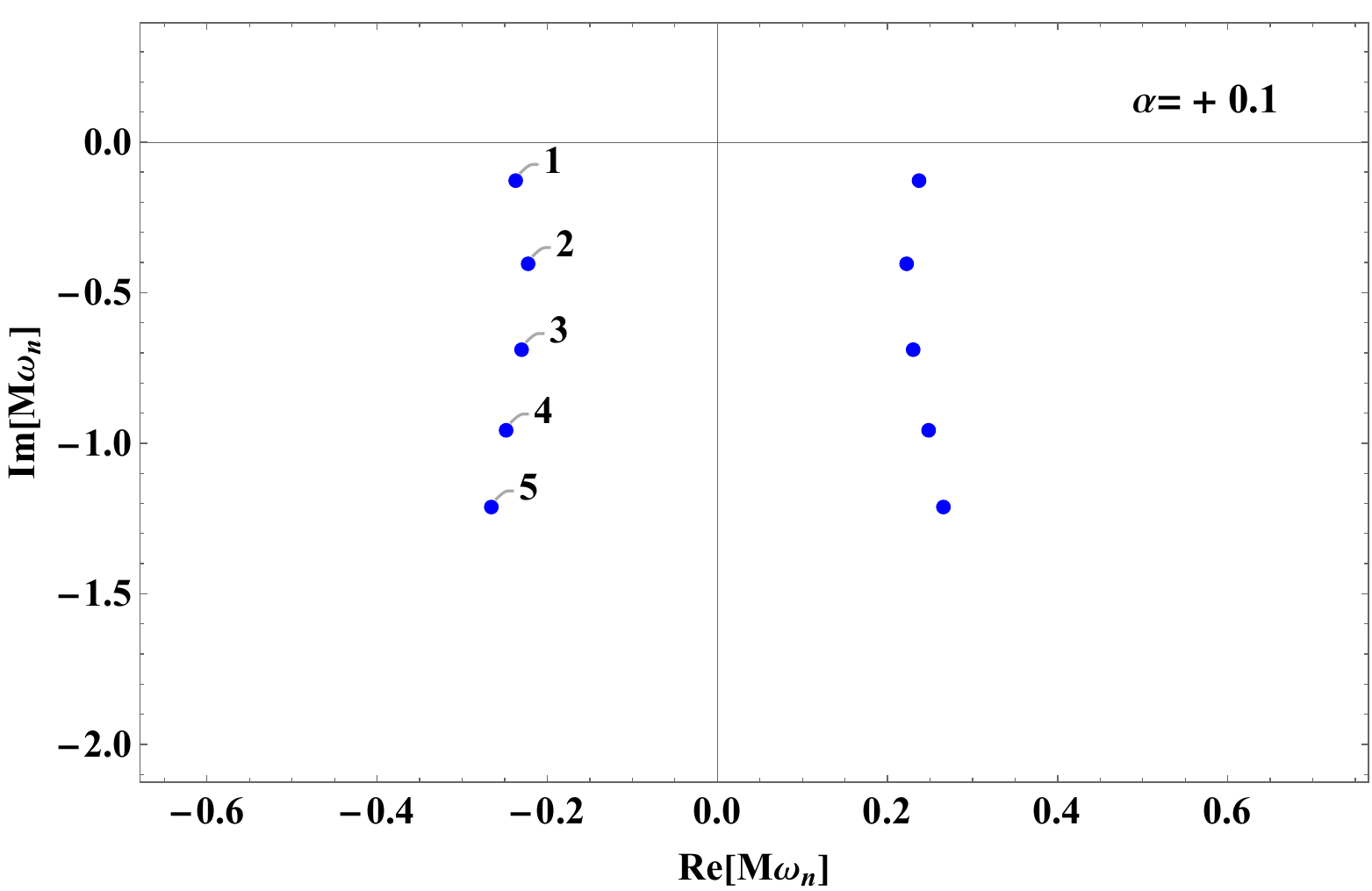}
    \qquad
    \begin{tabular}[b]{c cc}\hline
       & Re[$M\omega_n$] & Im[$M\omega_n$] \\ \hline
      1 & $\pm 0.23707$ & $-0.12804$ \\
      2 & $\pm 0.22250$ & $-0.40429$ \\
      3 & $\pm 0.23017$ & $-0.68951$ \\
      4 & $\pm 0.24834$ & $-0.95723$ \\
      5 & $\pm 0.2657\textcolor{gray}{3}$ & $-1.2120\textcolor{gray}{4}$ \\ \hline
      \vspace{0.4cm}
    \end{tabular}
                  \vspace{0.1cm} \\
    \includegraphics[scale=0.3]{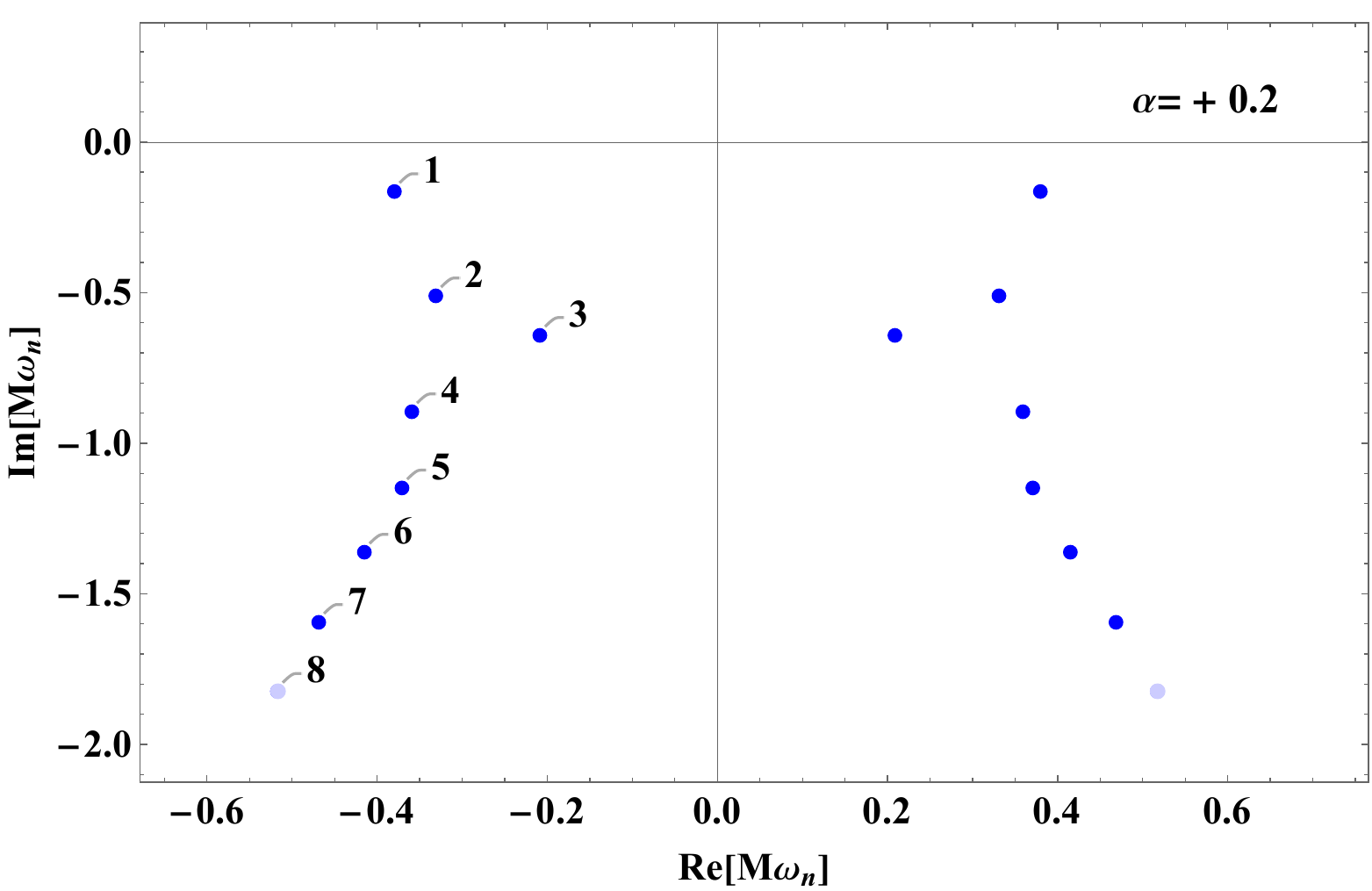}
    \qquad
    \begin{tabular}[b]{c cc}\hline
       & Re[$M\omega_n$] & Im[$M\omega_n$] \\ \hline
      1 & $\pm 0.37969$ & $-0.16432$ \\
      2 & $\pm 0.33104$ & $-0.51077$ \\
      3 & $\pm 0.20870$ & $-0.64208$ \\
      4 & $\pm 0.35918$ & $-0.89566$ \\
      5 & $\pm 0.37078$ & $-1.14874$ \\
      6 & $\pm 0.41497$ & $-1.36210$ \\
      7 & $\pm 0.4686\textcolor{gray}{2}$ & $-1.5949\textcolor{gray}{5}$ \\
      8 & $\pm 0.51\textcolor{gray}{717}$ & $-1.82\textcolor{gray}{424}$ \\ \hline
      \vspace{0.1cm}
    \end{tabular}
          \vspace{0.2cm} \\
    \includegraphics[scale=0.3]{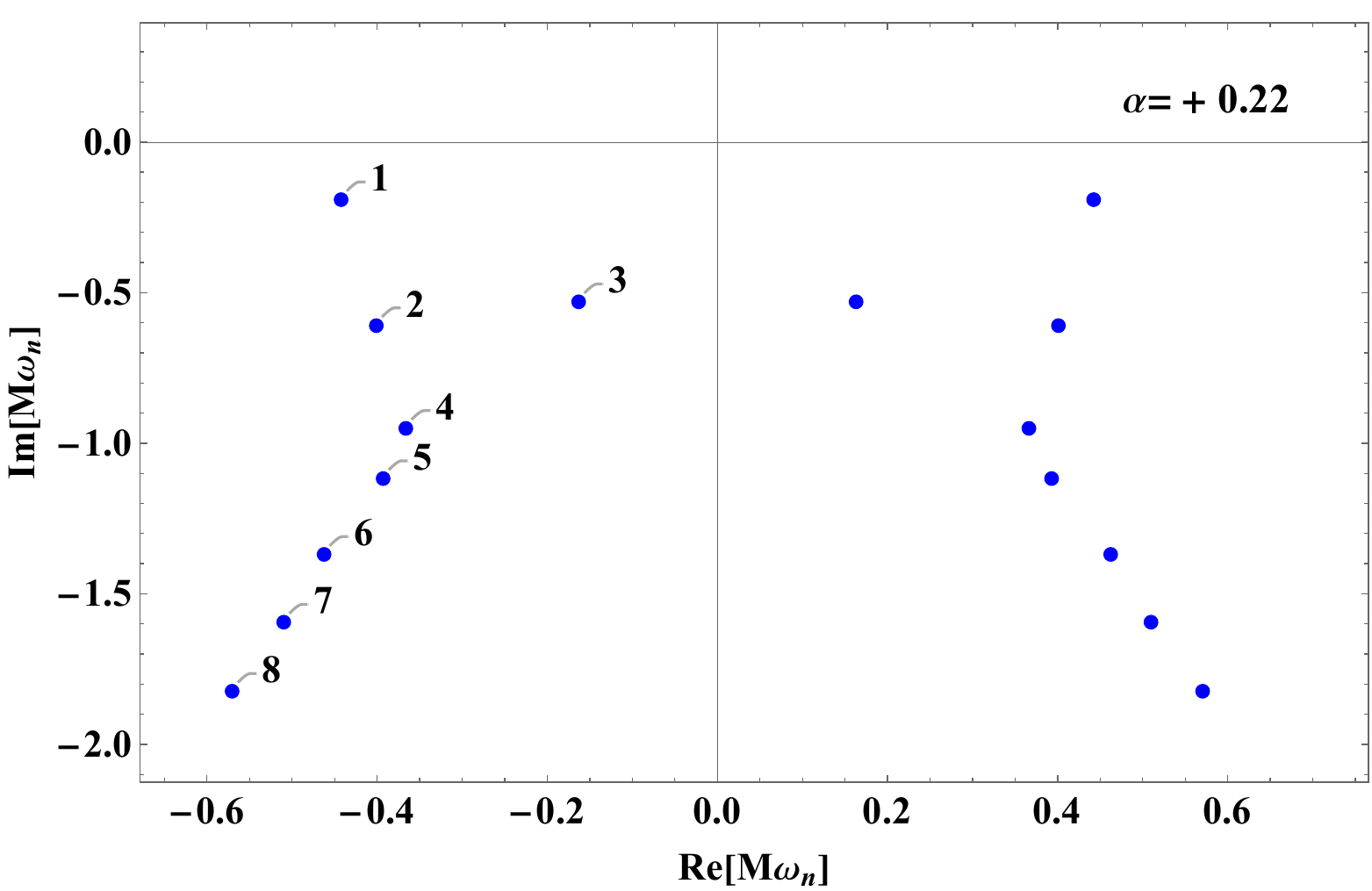}
    \qquad
    \begin{tabular}[b]{c cc}\hline
       & Re[$M\omega_n$] & Im[$M\omega_n$] \\ \hline
      1 & $\pm 0.44235$ & $-0.19132$ \\
      2 & $\pm 0.40095$ & $-0.60965$ \\
      3 & $\pm 0.16308$ & $-0.53055$ \\
      4 & $\pm 0.36626$ & $-0.95082$ \\
      5 & $\pm 0.39292$ & $-1.11752$ \\
      6 & $\pm 0.46238$ & $-1.36942$ \\
      7 & $\pm 0.5098\textcolor{gray}{1}$ & $-1.5944\textcolor{gray}{4}$ \\
      8 & $\pm 0.570\textcolor{gray}{47}$ & $-1.823\textcolor{gray}{79}$ \\ \hline
      \vspace{0.1cm}
    \end{tabular}
    \captionlistentry[table]{A table beside a figure}
    \captionsetup{labelformat=andtable}
    \caption{QNMs with various values of $\alpha$ for $\sigma_h = 0.1$. Blue dots are tolerable less than $10^{-3}$, whereas lighter blue dots have a tolerance of $10^{-2}$.}
    \label{fig:QNMsigmah01}
  \end{figure}

We employed the same strategy used in the symmetric phase to ensure the convergence of the eigenvalues. Calculations were performed on grids of $100$ and $140$, resulting in $100$ and $140$ eigenvalues, respectively. To identify reliable eigenvalues, we considered only those for which both the real and imaginary parts agreed within a tolerance of $10^{-2}$ between the two grid sizes. Here, we only used hairy black hole solutions with a fixed value of $\lambda = 0.1$. To illustrate the behavior of quasinormal frequencies for varying $\alpha$ values, we first computed them for a background solution with $\sigma_h= 0.01$. Representative results are presented for $\alpha = 0.01, 0.2, 0.34$, and $0.38$ in Figure~\ref{fig:QNMsigmah001a}, and for $\alpha = 0.42, 0.46, 0.50, 0.54, 0.58, 0.60, 0.62$, and $0.66$ in Figure~\ref{fig:QNMsigmah001b}. The numerical values for Figure~\ref{fig:QNMsigmah001b} are presented in Appendix A. The QNM spectrum in Figure~\ref{fig:QNMsigmah001b} for increasing positive values of $\alpha$ exhibits behaviors similar to those observed in Figure~\ref{fig:QNMvp001aN}  for increasing negative values of $\alpha$. Different from the symmetric phase with $\varphi_h=0.01$, the imaginary parts of quasinormal frequencies are all negative, which indicates the stability of hairy black holes. We also examined the quasinormal frequencies for $\sigma_h=0.1$ and plotted them for various values of $\alpha$ ($\alpha=0.01,0.1, 0.2$, and $0.22$) in Figure~{\ref{fig:QNMsigmah01}}, which shows all negative values for the imaginary parts of quasinormal frequencies. We further investigated the quasinormal frequencies with the background solutions with $\sigma_h=-0.01$ and $\sigma_h=-0.1$ for various values of $\alpha$. It turns out that they are stable for the linear scalar perturbation and the quasinormal frequencies.

\begin{figure}
	\begin{center}
	\includegraphics[scale=0.28]{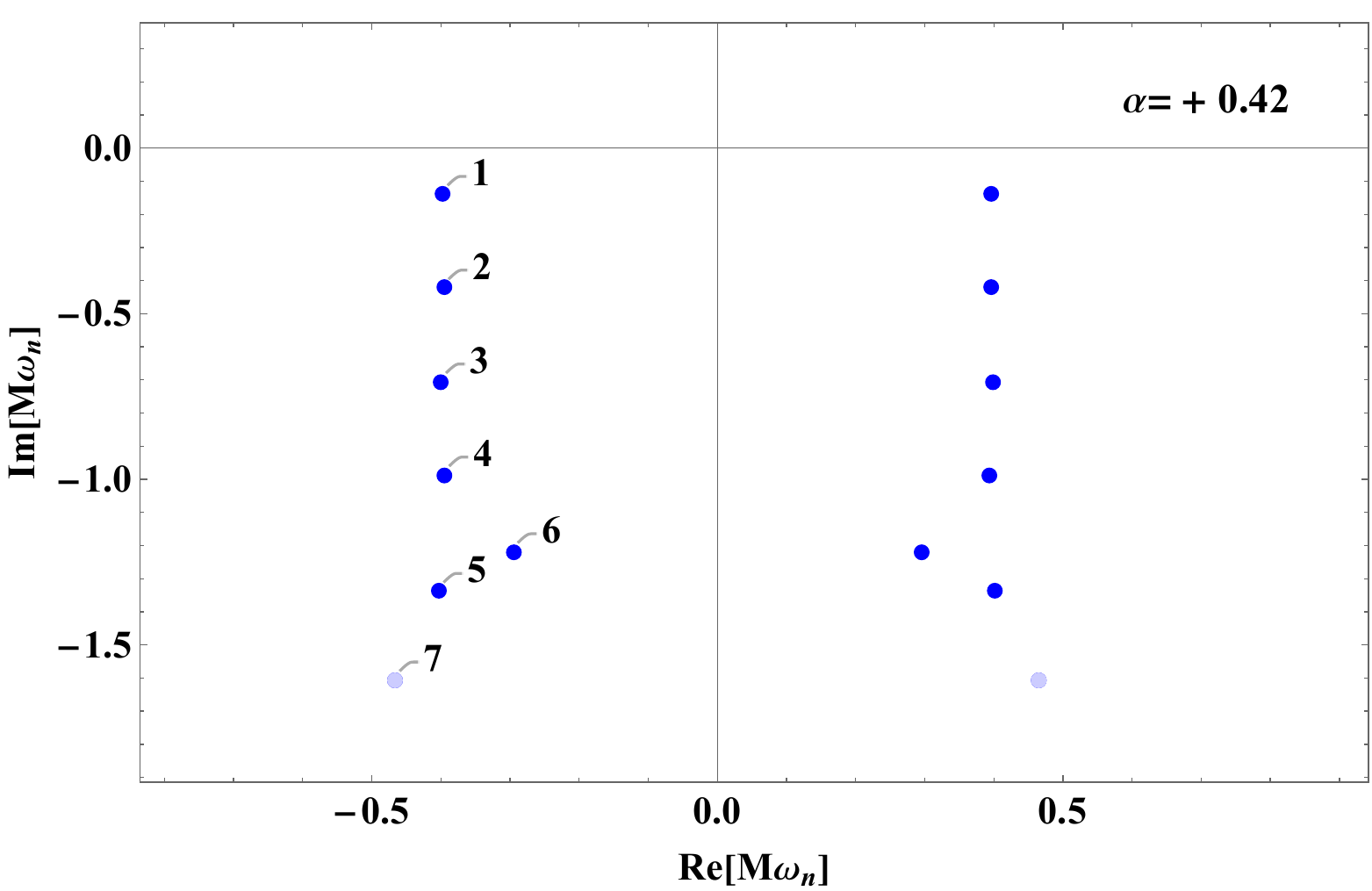} $\; \; \; \;$ \includegraphics[scale=0.28]{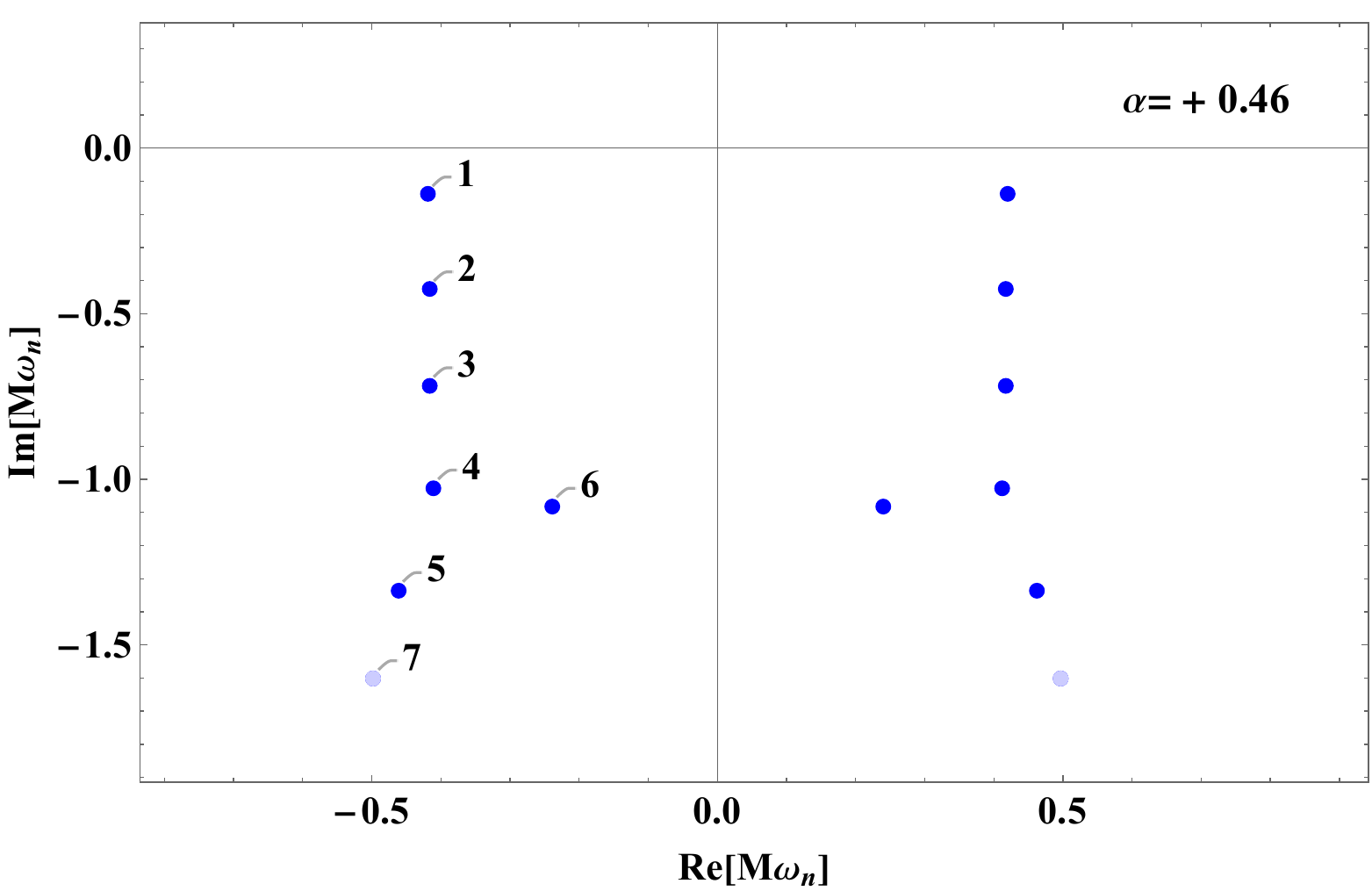}\\
	\includegraphics[scale=0.28]{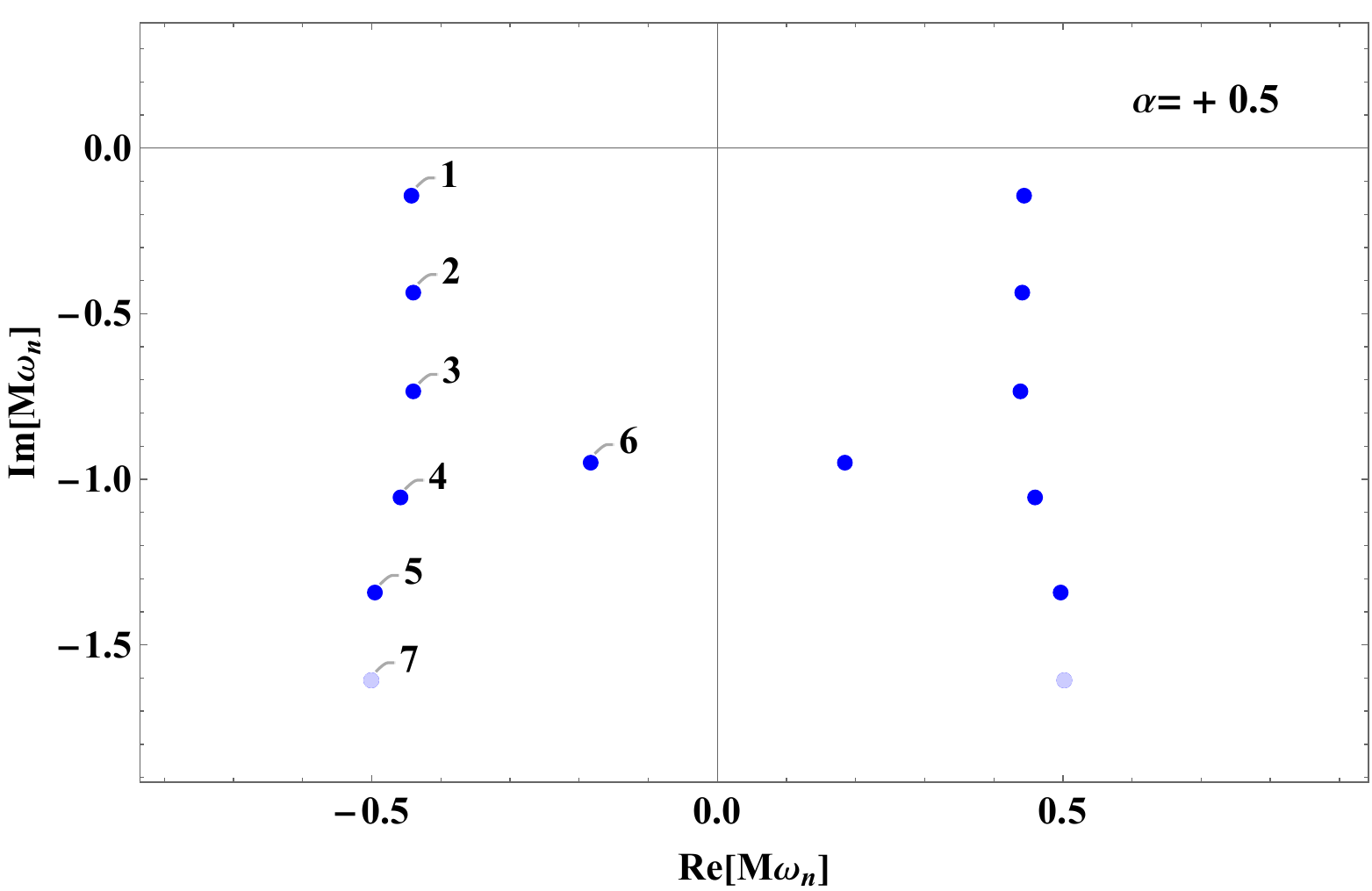} $\; \; \; \;$ \includegraphics[scale=0.28]{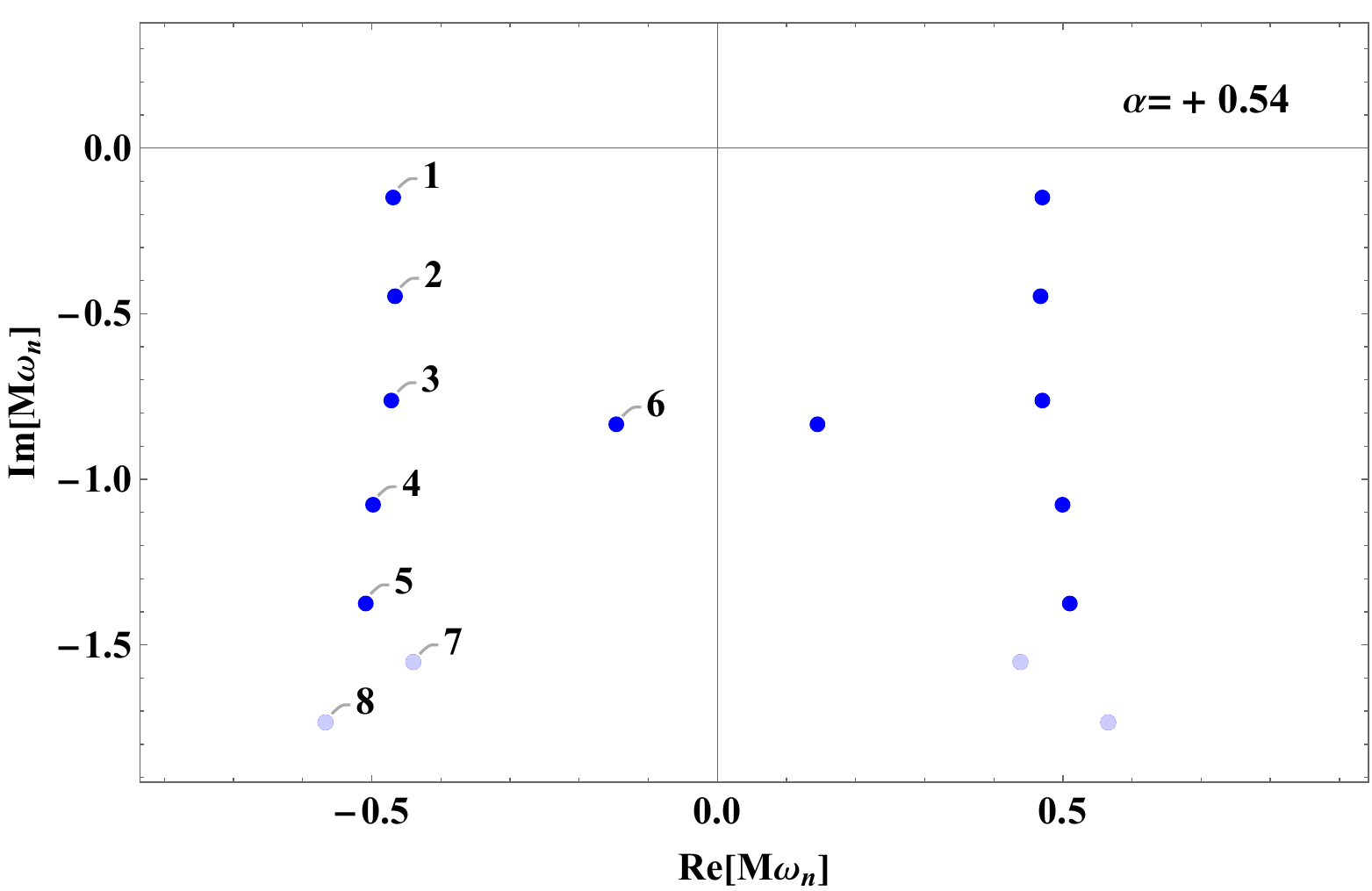}\\
	\includegraphics[scale=0.28]{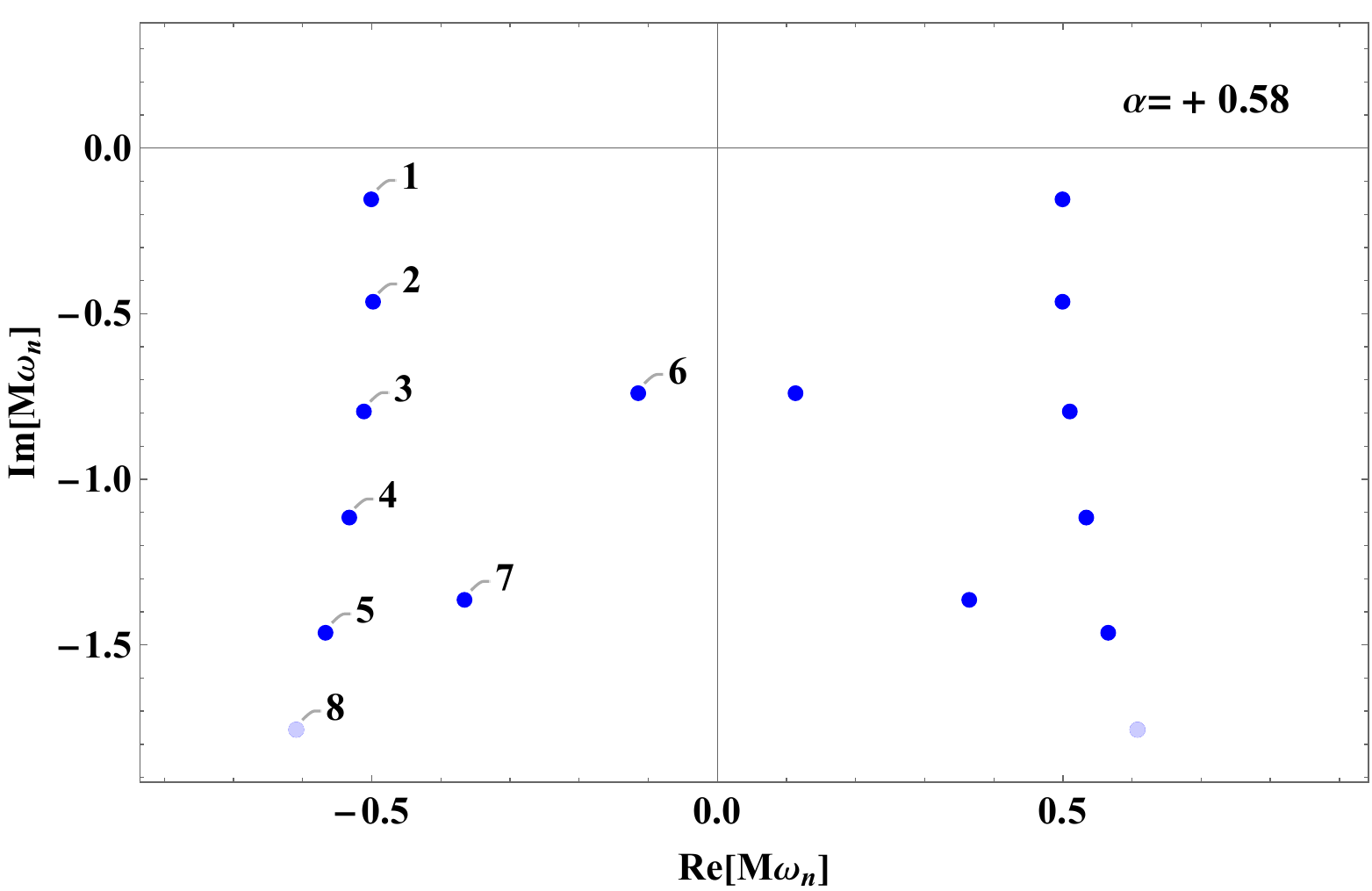} $\; \; \; \;$ \includegraphics[scale=0.28]{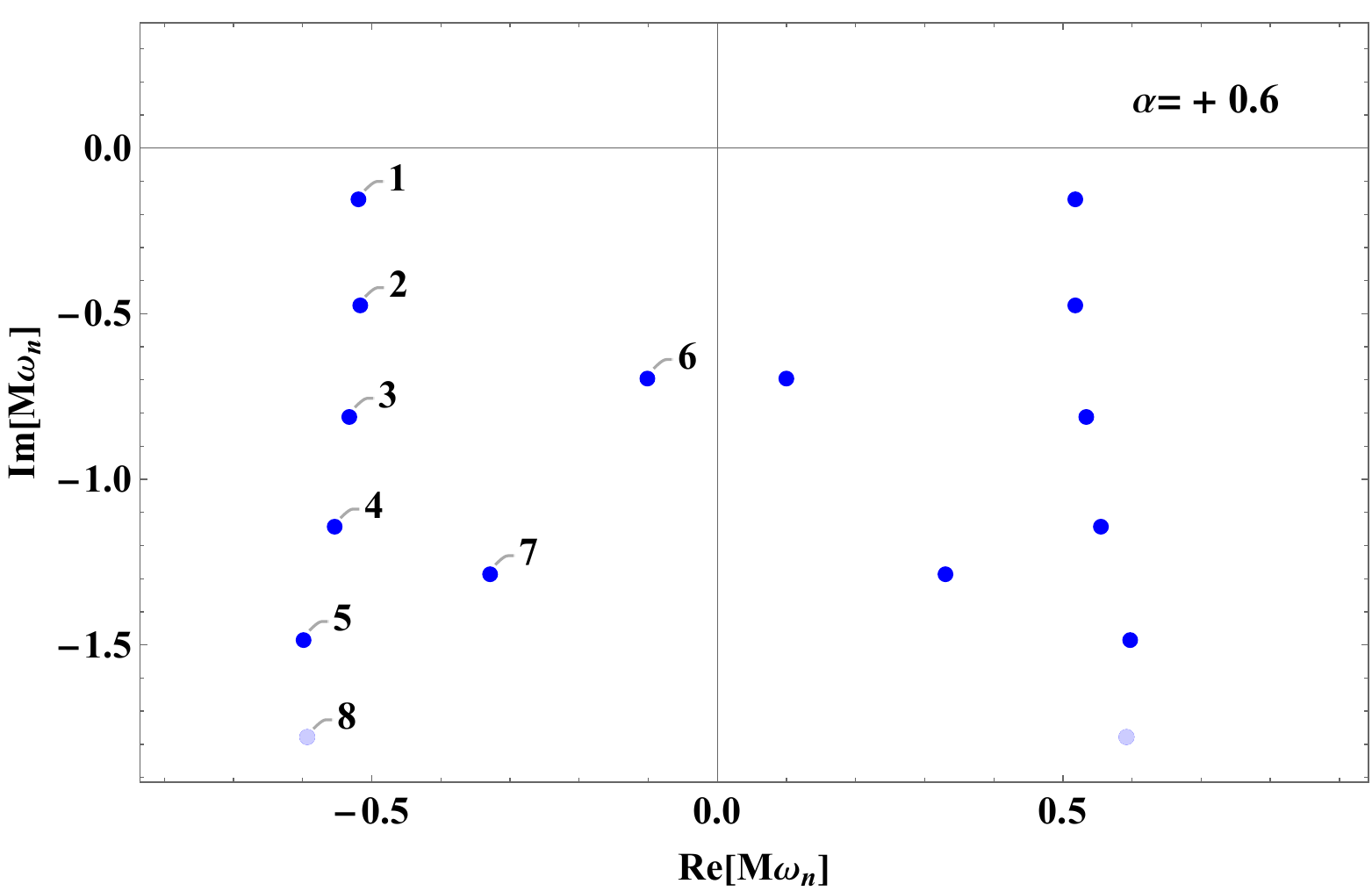}\\
	\includegraphics[scale=0.28]{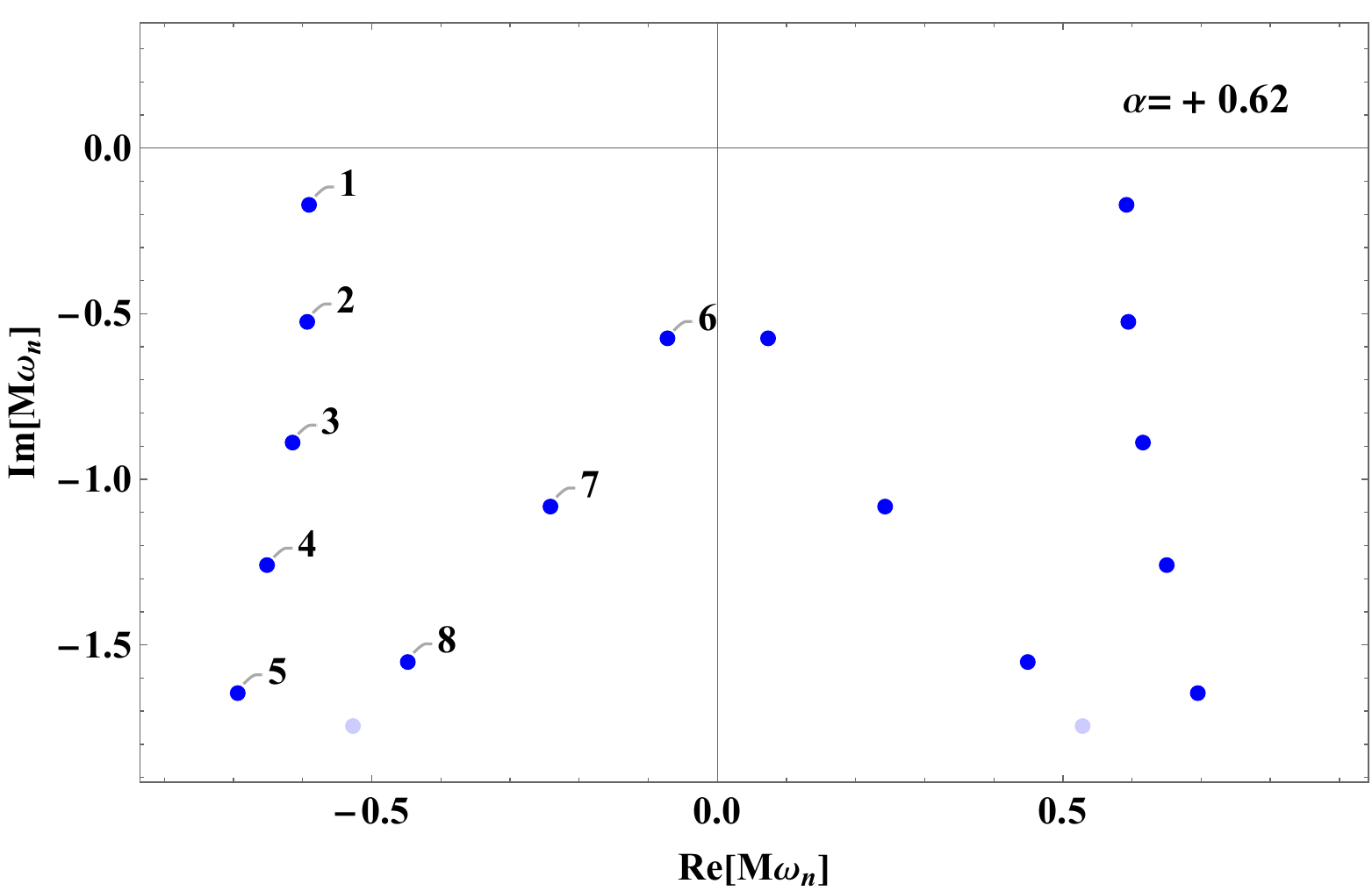} $\; \; \; \;$ \includegraphics[scale=0.28]{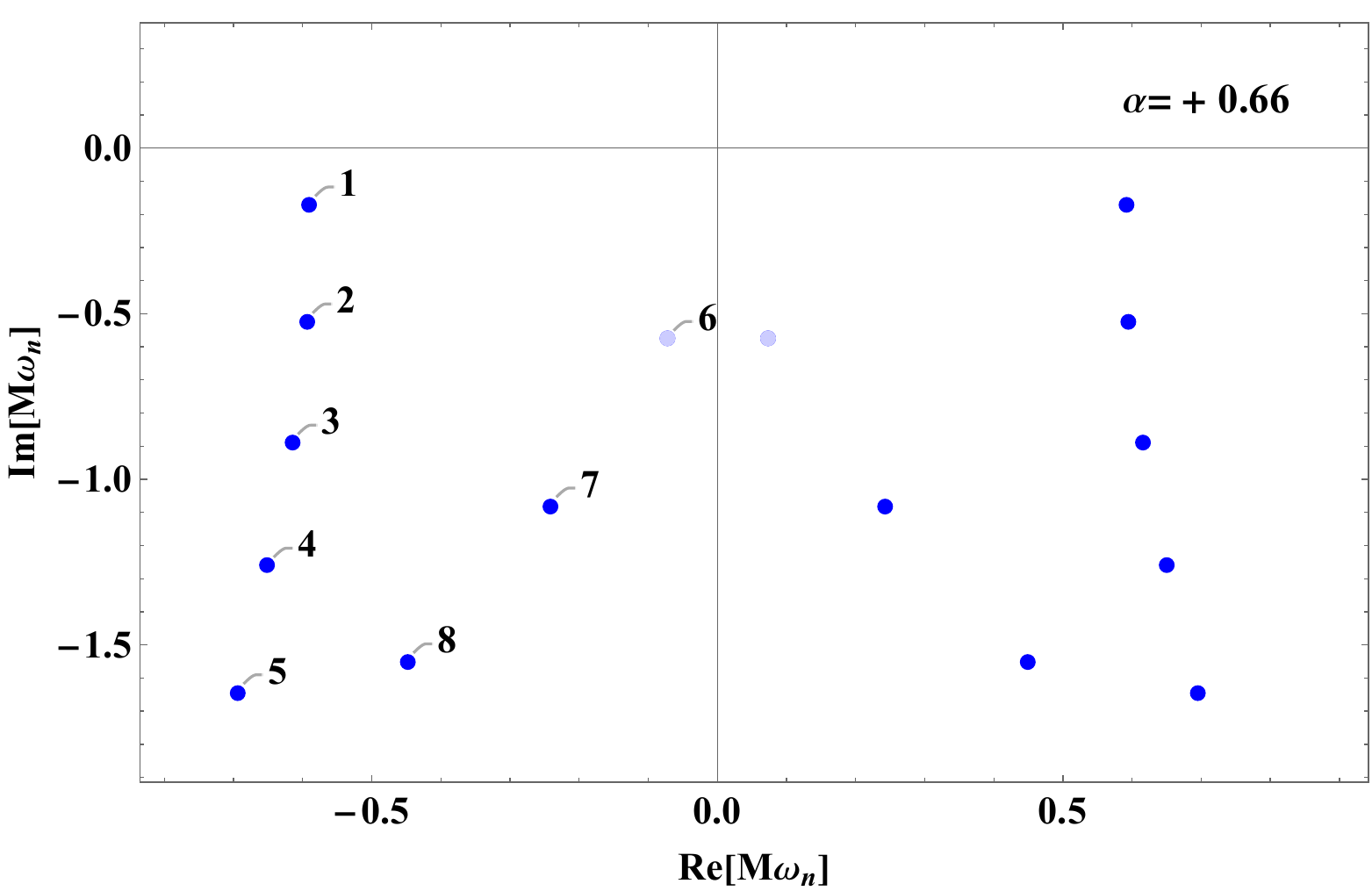}
	    \caption{QNMs with various values of $\alpha$ for $\sigma_h = 0.01$. Blue dots are tolerable less than $10^{-3}$, whereas lighter blue dots have a tolerance of $10^{-2}$.}
    \label{fig:QNMsigmah001b}
	\end{center}
\end{figure}

%%%%%%%%%%%%%%%%%%%%%%%%%%%%%%%%%%%%%%%%%%%%%%%%
\section{Summary}
%%%%%%%%%%%%%%%%%%%%%%%%%%%%%%%%%%%%%%%%%%%%%%%%

We have argued that black holes without hairs grow their scalar hairs by spontaneous symmetry breaking under global U(1) symmetry in Einstein-scalar-Gauss-Bonnet theory, following the suggestion in \cite{Latosh:2023cxm}. We investigated the scalar field perturbation on the hairy black hole spacetimes to provide evidence for this process. We have fully analyzed the solution for the linearized scalar field equation by using Green's function, which consists of three parts ($G_\textrm{F}, G_\textrm{B}$, and $G_{\textrm{QNM}}$). Our main focus was the calculation of QNMs ($G_{\textrm{QNM}}$), which provides a powerful tool to study the stability of concerning systems. The branch cut contribution ($G_\textrm{B}$) differs from the Schwarzschild black hole case only by the mass term. Since our hairy black hole solutions are numerically generated, we employed the numerical methods to compute the QNMs. 

We firstly considered the Schwarzschild black hole, which is the solution without scalar fields, and examined the QNMs of scalar perturbation on this background. We found a critical point at which the imaginary part of the quasinormal frequency becomes positive, indicating the system's instability. Then, we numerically constructed hairy black hole solutions in the symmetric and symmetry-broken phases. We assumed the scalar field always lies near the vacuum in each phase, so their near horizon values are small. Our calculations of the QNMs show that the hairy black holes in the symmetric phase also have a positive imaginary value of the quasinormal frequency at nearly the same critical point as the Schwarzschild black hole. However, the hairy black holes in the symmetry-broken phase always produce a negative imaginary value of the quasinormal frequency in the parameter ranges that we are concerned with. This numerical result indicates that the Schwarzschild black holes undergo a phase transition to the ones in the symmetry-broken phase by spontaneous symmetry breaking. Thus, their final stage through evolution becomes stable. 

Our investigation focused on the scalar field perturbation within the probe limit, neglecting the potential backreaction on the metric. Since the background scalar field has a non-trivial contribution to the spacetime geometry, the perturbation itself will induce a corresponding change in that geometry. This backreaction effect can potentially influence the stability of the system. Thus, it is crucial to consider this effect in future work. Furthermore, studying the full non-linear time evolution of this process would be much more intriguing, but this work presents a significantly greater technical challenge. 

\section*{Acknowledgments}

%We wish to thank mmm for useful discussion. 
B.L. and M.P. were supported by the Institute for Basic Science (Grant No. IBS-R018-Y1). Y.-H.H.  was supported by Basic Science Research Program through the National Research Foundation of Korea (NRF) funded by the Ministry of Education (NRF-2021R1I1A2050775). We appreciate APCTP for its hospitality during completion of this work.

\section*{Appendix A. Numerical data for QNM}

\begin{table}[H]
\begin{center}
    \begin{tabular}{c@{\hspace{2.5em}}c@{\hspace{0.5em}}c@{\hspace{1.5em}}c@{\hspace{0.5em}}c@{\hspace{1.5em}}c@{\hspace{0.5em}}c@{\hspace{1.5em}}c@{\hspace{0.5em}}c}
\toprule
    \multicolumn{3}{c}{$\boldsymbol{\alpha=-0.2}$} &  \multicolumn{2}{c}{$\boldsymbol{\alpha=-0.8}$} &  \multicolumn{2}{c}{$\boldsymbol{\alpha=-0.9}$} &  \multicolumn{2}{c}{$\boldsymbol{\alpha=-1}$} \\ \midrule
      & $\mathrm{Re}[M\omega_n]$ & $\mathrm{Im}[M\omega_n]$  & $\mathrm{Re}[M\omega_n]$ & $\mathrm{Im}[M\omega_n]$ & $\mathrm{Re}[M\omega_n]$ & $\mathrm{Im}[M\omega_n]$ & $\mathrm{Re}[M\omega_n]$ & $\mathrm{Im}[M\omega_n]$\\ \midrule
     1& $\pm 0.26768$ & $-0.13059$ & $\pm 0.38081$ & $-0.14052$  & $\pm 0.40992$ & $-0.14437$  & $\pm 0.44302$ & $-0.15017$  \\
     2& $\pm 0.25648$ & $-0.41005$ & $\pm 0.37650$ & $-0.42891$  & $\pm 0.40404$ & $-0.44013$ & $\pm 0.43517$ & $-0.45885$  \\
     3& $\pm 0.26784$ & $-0.69773$ & $\pm 0.37689$ & $-0.71752$  & $\pm 0.39674$ & $-0.74220$ & $\pm 0.43080$ & $-0.78788$  \\
     4& $\pm 0.28846$ & $-0.96995$ & $\pm 0.35977$ & $-1.00014$  & $\pm 0.40838$ & $-1.08673$ & $\pm 0.1785\textcolor{gray}{3}$ & $-0.8451\textcolor{gray}{4}$  \\
     5&$\pm 0.30676$ & $-1.23278$  & $\pm 0.3033\textcolor{gray}{8}$ & $-1.1743\textcolor{gray}{0}$  & $\pm 0.2384\textcolor{gray}{8}$ & $-0.9822\textcolor{gray}{1}$ & $\pm 0.46553$ & $-1.11114$  \\
     6&$\pm 0.322\textcolor{gray}{21}$ & $-1.490\textcolor{gray}{56}$  & $\pm 0.4079\textcolor{gray}{5}$ & $-1.3732\textcolor{gray}{0}$  & $\pm 0.45992$ & $-1.37158$ & $\pm 0.4743\textcolor{gray}{1}$ & $-1.3931\textcolor{gray}{0}$  \\
     7& --                      & --                  & $\pm 0.45\textcolor{gray}{789}$ & $-1.63\textcolor{gray}{247}$   & $\pm 0.48\textcolor{gray}{216}$ & $-1.62\textcolor{gray}{890}$ & $\pm 0.46\textcolor{gray}{557}$ & $-1.58\textcolor{gray}{835}$ \\
\addlinespace
\toprule
    \multicolumn{3}{c}{$\boldsymbol{\alpha=-1.1}$} &  \multicolumn{2}{c}{$\boldsymbol{\alpha=-1.2}$} &  \multicolumn{2}{c}{$\boldsymbol{\alpha=-1.3}$} &  \multicolumn{2}{c}{$\boldsymbol{\alpha=-1.4}$} \\ \midrule
      & $\mathrm{Re}[M\omega_n]$ & $\mathrm{Im}[M\omega_n]$  & $\mathrm{Re}[M\omega_n]$ & $\mathrm{Im}[M\omega_n]$ & $\mathrm{Re}[M\omega_n]$ & $\mathrm{Im}[M\omega_n]$ & $\mathrm{Re}[M\omega_n]$ & $\mathrm{Im}[M\omega_n]$\\ \midrule
     1& $\pm 0.48383$ & $-0.15920$ & $\pm 0.53874$ & $-0.17361$  & $\pm 0.61923$ & $-0.19750$  & $\pm 0.74663$ & $-0.23984$  \\
     2& $\pm 0.47526$ & $-0.48878$ & $\pm 0.53170$ & $-0.53357$  & $\pm 0.06\textcolor{gray}{431}$ & $-0.53\textcolor{gray}{658}$ & $\pm 0.048\textcolor{gray}{72}$ & $-0.448\textcolor{gray}{24}$  \\
     3& $\pm 0.129\textcolor{gray}{55}$ & $-0.726\textcolor{gray}{87}$ & $\pm 0.092\textcolor{gray}{04}$ & $-0.626\textcolor{gray}{74}$  & $\pm 0.61256$ & $-0.60174$ & $\pm 0.73275$ & $-0.71363$  \\
     4& $\pm 0.48476$ & $-0.83578$ & $\pm 0.54299$ & $-0.90330$  & $\pm 0.62800$ & $-1.01698$ & -- & --  \\
     5&$\pm 0.49622$ & $-1.16638$  & $\pm 0.3133\textcolor{gray}{8}$ & $-1.1647\textcolor{gray}{6}$  & $\pm 0.24\textcolor{gray}{812}$ & $-1.02\textcolor{gray}{833}$ & -- & --  \\
     6&$\pm 0.3944\textcolor{gray}{3}$ & $-1.3448\textcolor{gray}{6}$  & $\pm 0.57314$ & $-1.28380$  & $\pm 0.64\textcolor{gray}{992}$ & $-1.44\textcolor{gray}{187}$ & -- & --  \\
     7& $\pm 0.5515\textcolor{gray}{1}$                      & $-1.5382\textcolor{gray}{2}$                  & $\pm 0.563\textcolor{gray}{13}$ & $-1.609\textcolor{gray}{13}$   & -- & -- & -- & -- \\
     8& --                      & --                  & $\pm 0.58\textcolor{gray}{607}$ & $-1.74\textcolor{gray}{028}$   & -- & -- & -- & --  \\ \bottomrule
    \end{tabular}
         \end{center}
    \caption{$\varphi_h=0.01$ for Figure~\ref{fig:QNMvp001aN}}
\end{table}

\begin{table}[H]
\begin{center}
    \begin{tabular}{c@{\hspace{2.5em}}c@{\hspace{0.5em}}c@{\hspace{1.5em}}c@{\hspace{0.5em}}c@{\hspace{1.5em}}c@{\hspace{0.5em}}c@{\hspace{1.5em}}c@{\hspace{0.5em}}c}
\toprule
    \multicolumn{3}{c}{$\boldsymbol{\alpha=0.12}$} &  \multicolumn{2}{c}{$\boldsymbol{\alpha=0.1815}$} &  \multicolumn{2}{c}{$\boldsymbol{\alpha=0.6}$} &  \multicolumn{2}{c}{$\boldsymbol{\alpha=1}$} \\\midrule
      & $\mathrm{Re}[M\omega_n]$ & $\mathrm{Im}[M\omega_n]$  & $\mathrm{Re}[M\omega_n]$ & $\mathrm{Im}[M\omega_n]$ & $\mathrm{Re}[M\omega_n]$ & $\mathrm{Im}[M\omega_n]$ & $\mathrm{Re}[M\omega_n]$ & $\mathrm{Im}[M\omega_n]$\\ \midrule
     1& $\pm 0.011\textcolor{gray}{34}$ & $-0.058\textcolor{gray}{66}$ & $0$ & $+0.00004$  & $ 0$ & $+0.16688$  & $0$ & $+0.26344$  \\
     2& $\pm 0.039\textcolor{gray}{57}$ & $-0.209\textcolor{gray}{00}$ & $\pm 0.06827$ & $-0.19617$  & $\pm 0.06536$ & $-0.13421$ & $\pm 0.014\textcolor{gray}{47}$ & $-0.061\textcolor{gray}{15}$  \\
     3& $\pm 0.0881\textcolor{gray}{2}$ & $-0.4623\textcolor{gray}{2}$ & $\pm 0.11016$ & $-0.45775$  & $\pm 0.10783$ & $-0.45636$ & $\pm 0.1198\textcolor{gray}{1}$ & $-0.5668\textcolor{gray}{3}$  \\
     4& $\pm 0.114\textcolor{gray}{06}$ & $-0.720\textcolor{gray}{43}$ & $\pm 0.1349\textcolor{gray}{3}$ & $-0.7181\textcolor{gray}{6}$  & $\pm 0.1223\textcolor{gray}{7}$ & $-0.7241\textcolor{gray}{4}$ & $\pm 0.1753\textcolor{gray}{1}$ & $-0.8646\textcolor{gray}{1}$  \\
     5&$\pm 0.133\textcolor{gray}{01}$ & $-0.977\textcolor{gray}{73}$  & $\pm 0.1534\textcolor{gray}{6}$ & $-0.9766\textcolor{gray}{4}$  & $\pm 0.128\textcolor{gray}{74}$ & $-0.976\textcolor{gray}{12}$ & $\pm 0.26764$ & $-1.14678$  \\
     6&$\pm 0.148\textcolor{gray}{21}$ & $-1.233\textcolor{gray}{72}$  & $\pm 0.168\textcolor{gray}{46}$ & $-1.233\textcolor{gray}{48}$  & $\pm 0.162\textcolor{gray}{07}$ & $-1.206\textcolor{gray}{61}$ & $\pm 0.339\textcolor{gray}{84}$ & $-1.400\textcolor{gray}{29}$  \\
     7& $\pm 0.16\textcolor{gray}{124}$                      & $- 1.48\textcolor{gray}{961}$                  & $\pm 0.18\textcolor{gray}{124}$ & $-1.48\textcolor{gray}{954}$   & $\pm 0.22\textcolor{gray}{061}$ & $-1.44\textcolor{gray}{564}$ & -- & -- \\
\addlinespace
\toprule
    \multicolumn{3}{c}{$\boldsymbol{\alpha=1.6}$} &  \multicolumn{2}{c}{$\boldsymbol{\alpha=2.1}$} &  \multicolumn{2}{c}{$\boldsymbol{\alpha=2.4}$} &  \multicolumn{2}{c}{$\boldsymbol{\alpha=2.8}$} \\ \midrule
      & $\mathrm{Re}[M\omega_n]$ & $\mathrm{Im}[M\omega_n]$  & $\mathrm{Re}[M\omega_n]$ & $\mathrm{Im}[M\omega_n]$ & $\mathrm{Re}[M\omega_n]$ & $\mathrm{Im}[M\omega_n]$ & $\mathrm{Re}[M\omega_n]$ & $\mathrm{Im}[M\omega_n]$\\ \midrule
     1& $0$ & $+0.37476$ & $0$ & $+0.45168$  & $0$ & $+0.49338$  & $0$ & $+0.54504$  \\
     2& $0$ & $+0.07902$ & $0$ & $+0.16014$  & $0$ & $+0.20280$ & $0$ & $+0.25509$  \\
     3& $\pm 0.0395\textcolor{gray}{2}$ & $-0.1821\textcolor{gray}{2}$ & $\pm 0.0423\textcolor{gray}{8}$ & $-0.1307\textcolor{gray}{7}$  & $\pm 0.0331\textcolor{gray}{5}$ & $-0.1010\textcolor{gray}{7}$ & $\pm 0.011\textcolor{gray}{46}$ & $-0.054\textcolor{gray}{02}$  \\
     4& $\pm 0.1300\textcolor{gray}{5}$ & $-0.5689\textcolor{gray}{5}$ & $\pm 0.1058\textcolor{gray}{8}$ & $-0.5960\textcolor{gray}{5}$  & $\pm 0.103\textcolor{gray}{94}$ & $-0.602\textcolor{gray}{31}$ & -- & --  \\
     5&$\pm 0.25477$ & $-0.85334$  & $\pm 0.2738\textcolor{gray}{4}$ & $-0.8188\textcolor{gray}{8}$  & $\pm 0.27\textcolor{gray}{571}$ & $-0.79\textcolor{gray}{458}$ & -- & --  \\
     6&$\pm 0.35290$ & $-1.09506$  & $\pm 0.387\textcolor{gray}{09}$ & $-1.039\textcolor{gray}{06}$  & -- & -- & -- & --  \\
     7& $\pm 0.4379\textcolor{gray}{7}$                      & $-1.3286\textcolor{gray}{3}$                  & -- & --   & -- & -- & -- & -- \\
\midrule
    \end{tabular}
         \end{center}
    \caption{$\varphi_h=0.01$ for Figure~\ref{fig:QNMvp001aP}}
\end{table}

\begin{table}[H]
\begin{center}
    \begin{tabular}{c@{\hspace{2.5em}}c@{\hspace{0.5em}}c@{\hspace{1.5em}}c@{\hspace{0.5em}}c@{\hspace{1.5em}}c@{\hspace{0.5em}}c@{\hspace{1.5em}}c@{\hspace{0.5em}}c}
\toprule
    \multicolumn{3}{c}{$\boldsymbol{\alpha=0.42}$} &  \multicolumn{2}{c}{$\boldsymbol{\alpha=0.46}$} &  \multicolumn{2}{c}{$\boldsymbol{\alpha=0.5}$} &  \multicolumn{2}{c}{$\boldsymbol{\alpha=0.54}$} \\ \midrule
      & $\mathrm{Re}[M\omega_n]$ & $\mathrm{Im}[M\omega_n]$  & $\mathrm{Re}[M\omega_n]$ & $\mathrm{Im}[M\omega_n]$ & $\mathrm{Re}[M\omega_n]$ & $\mathrm{Im}[M\omega_n]$ & $\mathrm{Re}[M\omega_n]$ & $\mathrm{Im}[M\omega_n]$\\ \midrule
     1& $\pm 0.39686$  & $-0.13874$ & $\pm 0.41901$ & $-0.14061$  & $\pm 0.44270$ & $-0.14319$  & $\pm 0.46922$ & $-0.14685$  \\
     2& $\pm 0.39535$ & $-0.42179$ & $\pm 0.41703$ & $-0.42707$  & $\pm 0.44016$ & $-0.43520$ & $\pm 0.46650$ & $-0.44737$  \\
     3& $\pm 0.39899$ & $-0.70504$ & $\pm 0.41731$ & $-0.71668$  & $\pm 0.43915$ & $-0.73637$ & $\pm 0.47048$ & $-0.76296$  \\
     4& $\pm 0.39449$ & $-0.98608$ & $\pm 0.41082$ & $-1.02562$  & $\pm 0.45927$ & $-1.05587$ & $\pm 0.49901$ & $-1.07715$  \\
     5&$\pm 0.4026\textcolor{gray}{4}$ & $-1.3372\textcolor{gray}{2}$  & $\pm 0.46181$ & $-1.33502$  & $\pm 0.49551$ & $-1.34339$ & $\pm 0.50952$ & $-1.3728$  \\
     6&$\pm 0.2952\textcolor{gray}{9}$ & $-1.2177\textcolor{gray}{6}$  & $\pm 0.2395\textcolor{gray}{2}$ & $-1.0800\textcolor{gray}{1}$  & $\pm 0.184\textcolor{gray}{62}$ & $-0.948\textcolor{gray}{48}$ & $\pm 0.145\textcolor{gray}{11}$ & $-0.836\textcolor{gray}{37}$  \\
     7&$\pm 0.46\textcolor{gray}{611}$ & $-1.60\textcolor{gray}{476}$  &  $\pm 0.497\textcolor{gray}{27}$ & $-1.601\textcolor{gray}{01}$   & $\pm 0.50\textcolor{gray}{148}$ & $-1.60\textcolor{gray}{679}$ & $\pm 0.43\textcolor{gray}{876}$ & $-1.55\textcolor{gray}{436}$ \\
     8&      --                &       --           &  --  &  --   & -- & -- & $\pm 0.56\textcolor{gray}{580}$ & $-1.73\textcolor{gray}{527}$  \\\addlinespace
\toprule
    \multicolumn{3}{c}{$\boldsymbol{\alpha=0.58}$} &  \multicolumn{2}{c}{$\boldsymbol{\alpha=0.6}$} &  \multicolumn{2}{c}{$\boldsymbol{\alpha=0.62}$} &  \multicolumn{2}{c}{$\boldsymbol{\alpha=0.66}$} \\ \midrule
      & $\mathrm{Re}[M\omega_n]$ & $\mathrm{Im}[M\omega_n]$  & $\mathrm{Re}[M\omega_n]$ & $\mathrm{Im}[M\omega_n]$ & $\mathrm{Re}[M\omega_n]$ & $\mathrm{Im}[M\omega_n]$ & $\mathrm{Re}[M\omega_n]$ & $\mathrm{Im}[M\omega_n]$\\ \midrule
     1& $\pm 0.50045$ & $-0.15212$ & $\pm 0.51870$ & $-0.15559$  & $\pm 0.53936$ & $-0.15979$  & $\pm 0.59066$ & $-0.17118$  \\
     2& $\pm 0.49861$ & $-0.46493$ & $\pm 0.51784$ & $-0.47621$  & $\pm 0.53981$ & $-0.48951$ & $\pm 0.59435$ & $-0.52418$  \\
     3& $\pm 0.51056$ & $-0.79258$ & $\pm 0.53271$ & $-0.81016$  & $\pm 0.55678$ & $-0.83141$ & $\pm 0.61569$ & $-0.89002$  \\
     4& $\pm 0.53295$ & $-1.11412$ & $\pm 0.55363$ & $-1.14460$  & $\pm 0.58299$ & $-1.18030$ & $\pm 0.65080$ & $-1.26134$  \\
     5&$\pm 0.56587$ & $-1.46074$  & $\pm 0.59853$ & $-1.48406$  & $\pm 0.62056$ & $-1.51745$ & $\pm 0.6943\textcolor{gray}{7}$ & $-1.6437\textcolor{gray}{2}$  \\
     6&$\pm 0.113\textcolor{gray}{91}$ & $-0.738\textcolor{gray}{06}$  & $\pm 0.100\textcolor{gray}{44}$ & $-0.693\textcolor{gray}{27}$  & $\pm 0.090\textcolor{gray}{07}$ & $-0.652\textcolor{gray}{34}$ & $\pm 0.07\textcolor{gray}{204}$ & $-0.57\textcolor{gray}{704}$  \\
     7&$\pm 0.3646\textcolor{gray}{0}$ & $-1.3626\textcolor{gray}{4}$ & $\pm 0.3295\textcolor{gray}{5}$ & $-1.2854\textcolor{gray}{8}$   & $\pm 0.2974\textcolor{gray}{3}$ & $-1.2101\textcolor{gray}{3}$ & $\pm 0.241\textcolor{gray}{81}$ & $-1.079\textcolor{gray}{90}$ \\
     8&$\pm 0.60\textcolor{gray}{883}$ & $-1.75\textcolor{gray}{396}$ & $\pm 0.59\textcolor{gray}{270}$ & $-1.78\textcolor{gray}{070}$   & $\pm 0.52\textcolor{gray}{726}$ & $-1.74\textcolor{gray}{332}$ & $\pm 0.4485\textcolor{gray}{8}$ & $-1.5502\textcolor{gray}{0}$  \\ \bottomrule
    \end{tabular}
         \end{center}
    \caption{$\sigma_h=\frac{1}{100}$ for Figure~\ref{fig:QNMsigmah001b}}
\end{table}  

\section*{Appendix B. Eigenfunctions}

\begin{figure}[h]
\begin{center}
    \includegraphics[scale=0.115]{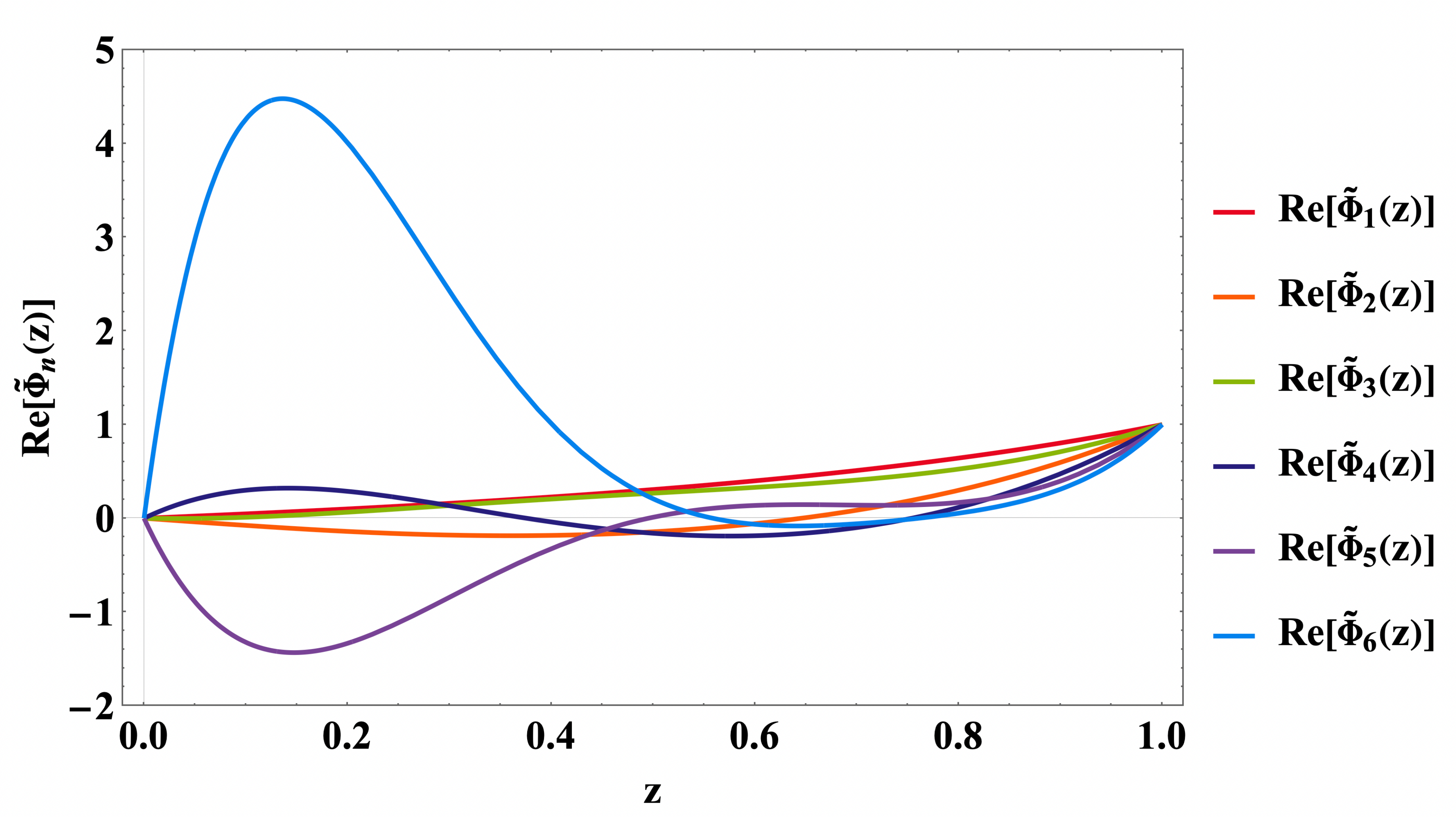}     \includegraphics[scale=0.115]{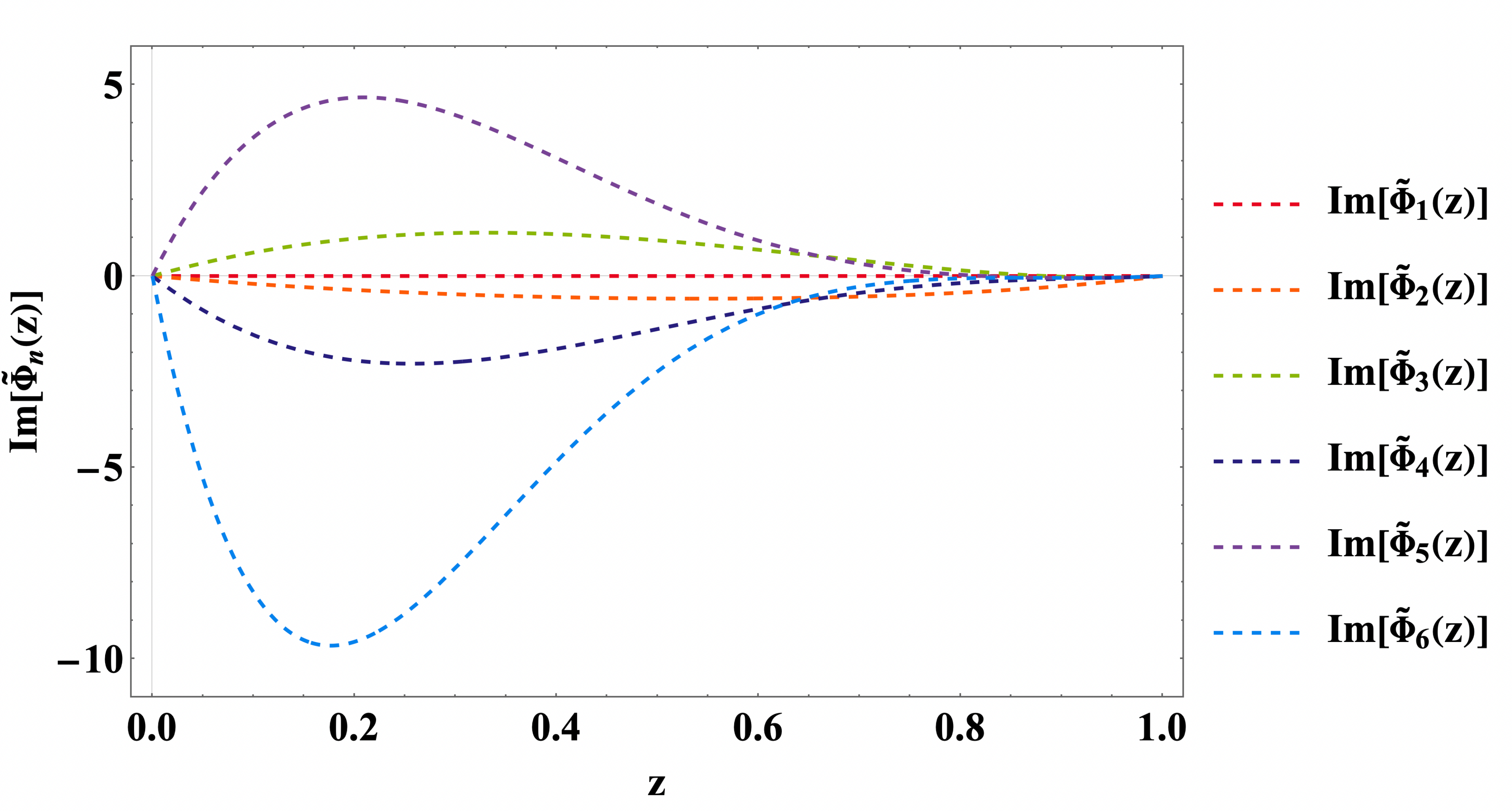}
    \end{center}
    \caption{The eigenfunctions corresponding to the quasinormal frequencies of the hairy black hole with $\varphi_h=0.01$ at $\alpha=0.1815$ and $\lambda=0.1$. The subscript of $\tilde{\Phi}(z)$ shares the same number of the QNM data presented in Table 6.}
    \label{fig:eigenfunctions}
\end{figure}

By examining the corresponding eigenfunctions, we confirm the physical validity of the QNM frequencies from our numerical data. We here demonstrate the eigenfunctions associated with the QNM frequencies which are computed on the hairy black holes with $\varphi_h=0.01$ at $\alpha=0.1815$ in Figure~\ref{fig:eigenfunctions}. We used the QNM frequencies with the positive real values and normalized the eigenfunctions to be $1$ at the horizon. As we expected, all eigenfunctions are smooth and finite.

\section*{Appendix C. QNM with angular momentum}

\begin{figure}[H]
	\begin{center}
	\includegraphics[scale=0.2]{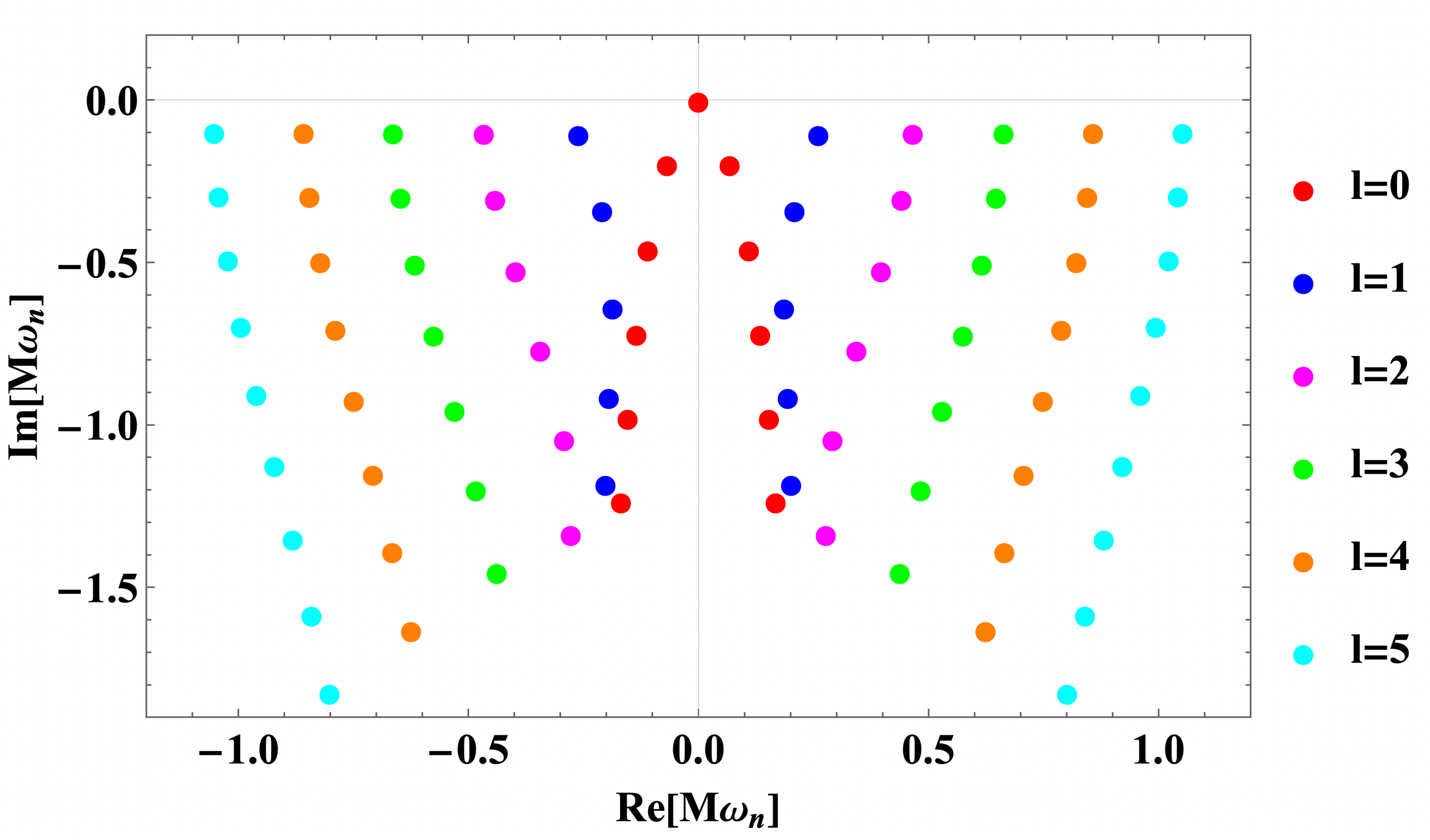}    \\
	\vspace{0.5cm}
	\begin{tabular}{c@{\hspace{2.5em}}c@{\hspace{0.5em}}c@{\hspace{1.5em}}c@{\hspace{0.5em}}c@{\hspace{1.5em}}c@{\hspace{0.5em}}c}
\toprule

 & \multicolumn{2}{c}{$\boldsymbol{l=0}$} & \multicolumn{2}{c}{$\boldsymbol{l=1}$} & \multicolumn{2}{c}{$\boldsymbol{l=2}$}  \\
\midrule
 & $\text{Re}[M \omega_{n}]$ & $\text{Im}[M \omega_{n}]$ & $\text{Re}[M \omega_{n}]$ & $\text{Im}[M \omega_{n}]$ &  $\text{Re}[M \omega_{n}]$ & $\text{Im}[M \omega_{n}]$ \\
%\addlinespace
\midrule
     1& $0$ & $+0.00004$   & $\pm   0.26085$ & $-  0.10328$  & $\pm   0.46605$ & $-  0.09877$  \\
     2& $\pm 0.06827$ & $-0.19617$   & $\pm   0.20933$ & $-  0.33765$  & $\pm   0.44129$ & $-  0.30235$  \\
     3& $\pm   0.11016$ & $-  0.45775$   & $\pm   0.18652$ & $-  0.63711$  & $\pm   0.39743$ & $-  0.52268$  \\
     4& $\pm   0.13493$ & $-  0.71816$   & $\pm   0.19404$ & $-  0.91276$  & $\pm   0.34352$ & $-  0.76635$  \\
     5& $\pm   0.1534\textcolor{gray}{6}$ & $-  0.9766\textcolor{gray}{4}$   & $\pm   0.2018\textcolor{gray}{3}$ & $-  1.1804\textcolor{gray}{6}$  & $\pm   0.29168$ & $-  1.04271$  \\
     6& $\pm   0.168\textcolor{gray}{46}$ & $-  1.233\textcolor{gray}{48}$   & -- & --  & $\pm   0.2769\textcolor{gray}{7}$ & $-  1.3340\textcolor{gray}{3}$  \\
\addlinespace
\toprule
 & \multicolumn{2}{c}{$\boldsymbol{l=3}$} & \multicolumn{2}{c}{$\boldsymbol{l=4}$} & \multicolumn{2}{c}{$\boldsymbol{l=5}$}  \\
\midrule
 & $\text{Re}[M \omega_{n}]$ & $\text{Im}[M \omega_{n}]$ & $\text{Re}[M \omega_{n}]$ & $\text{Im}[M \omega_{n}]$ &  $\text{Re}[M \omega_{n}]$ & $\text{Im}[M \omega_{n}]$ \\
%\addlinespace
\midrule
     1& $\pm   0.66314$ & $-  0.09755$   & $\pm   0.85802$ & $-  0.09704$  & $\pm   1.05198$ & $-  0.09677$  \\
     2& $\pm   0.64659$ & $-  0.29556$   & $\pm   0.84552$ & $-  0.29285$  & $\pm   1.04191$ & $-  0.29148$  \\
     3& $\pm   0.61574$ & $-  0.50191$   & $\pm   0.82160$ & $-  0.49373$  & $\pm   1.02237$ & $-  0.48959$  \\
     4& $\pm   0.57502$ & $-  0.72051$   & $\pm   0.78845$ & $-  0.70251$  & $\pm   0.99456$ & $-  0.69316$  \\
     5& $\pm   0.52974$ & $-  0.95246$   & $\pm   0.74919$ & $-  0.92095$  & $\pm   0.96030$ & $-  0.90380$  \\
     6& $\pm   0.48374$ & $-  1.19641$   & $\pm   0.70719$ & $-  1.14931$  & $\pm   0.92182$ & $-  1.12241$  \\
     7& $\pm   0.4382\textcolor{gray}{5}$ & $-  1.4511\textcolor{gray}{1}$   & $\pm   0.66516$ & $-  1.38650$  & $\pm   0.88141$ & $-  1.34903$  \\
     8& -- & --   & $\pm   0.6246\textcolor{gray}{5}$ & $-  1.6307\textcolor{gray}{6}$  & $\pm   0.84101$ & $-  1.58292$  \\
     9& -- & --   & -- & --  & $\pm   0.801\textcolor{gray}{94}$ & $-  1.822\textcolor{gray}{89}$  \\ 
\bottomrule
\end{tabular}
    \captionlistentry[table]{A table beside a figure}
    \captionsetup{labelformat=andtable}
\caption{The QNM frequencies for various values of $l$. The background hairy black hole solution is employed for the case of $\varphi_h=0.01, \alpha=0.1815$ and $\lambda = 0.1$.}
\label{fig:QNMwithL}
\end{center}
\end{figure}

To investigate the dependence on angular momentum $l$ of the perturbed scalar field, we calculated quasinormal frequencies for various values $l$. We employed a hairy black hole solution generated with specific parameters: $\varphi_h = 0.01$, $\alpha = 0.1815$, and $\lambda = 0.1$. The resulting QNM spectrum for $l = 0$ to $5$ and their numerical data are presented in Figure~\ref{fig:QNMwithL} and Table 8 respectively.

\bibliography{hairyQNMbib}
\bibliographystyle{ieeetr}

\end{document}